\documentclass[a4paper,11pt]{article}
\usepackage{jheppub} 
\usepackage{lineno}

\usepackage{hyperref}
\usepackage{array}
\usepackage{mwe}

\usepackage[utf8]{inputenc}
\usepackage[normalem]{ulem}

\usepackage{lipsum}
\usepackage{xcolor}
\usepackage{etoolbox}
\usepackage{graphicx}
\usepackage{wrapfig}
\usepackage{amsmath}
\usepackage{amsfonts}
\usepackage{amssymb}
\usepackage{multirow}
\usepackage{slashed}
\usepackage{physics}
\usepackage{diagbox}
\usepackage{float}
\usepackage{cleveref}
\usepackage{bbm}
\usepackage{tikz-cd}

\DeclareRobustCommand{\Eq}[1]{Eq.~\eqref{eq:#1}}

\newcommand\bets{\begin{table*}}
\newcommand\eets[1]{\label{tb:#1}\end{table*}}

\newcommand{\MS}{{\overline{\mathrm{MS}}}}

\newcommand{\nn}{\nonumber}

\newcommand{\ripmom}{RI$^\prime$-MOM}

\Crefname{equation}{Eq.}{Eqs.}
\Crefname{figure}{Fig.}{Figs.}
\Crefname{table}{Table}{Tables}
\Crefname{section}{Sec.}{Secs.}

\title{\boldmath Nucleon Parton Distribution Functions from Boosted Correlations in the Coulomb gauge}

\author[a]{Xiang Gao,}
\author[b,c,*]{Jinchen He,}
\author[c]{Joshua Lin,}
\author[a]{Swagato Mukherjee,}
\author[a]{Peter Petreczky,}
\author[d,c]{Rui Zhang,}
\author[c]{and Yong Zhao}

\affiliation[a]{Physics Department, Brookhaven National Laboratory, Upton, New York 11973, USA}
\affiliation[b]{Maryland Center for Fundamental Physics, University of Maryland, College Park, MD 20742, USA}
\affiliation[c]{Physics Division, Argonne National Laboratory, Lemont, IL 60439, USA}
\affiliation[d]{Center for Theoretical Physics - a Leinweber Institute, Massachusetts Institute of Technology, Cambridge, MA 02139, USA}

\emailAdd{xgao@bnl.gov}
\emailAdd{jinchen@umd.edu}
\emailAdd{joshua.lin@anl.gov}
\emailAdd{swagato@bnl.gov}
\emailAdd{petreczk@bnl.gov}
\emailAdd{rzhang93@mit.edu}
\emailAdd{yong.zhao@anl.gov}

\abstract{Recently, a novel approach has been proposed to compute parton distributions through the use of boosted correlators fixed in the Coulomb gauge from lattice QCD, within the framework of Large-Momentum Effective Theory (LaMET). This approach circumvents the need for Wilson lines, potentially enhancing the efficiency and accuracy of lattice calculations. In this work, we present the first exploratory implementation of the Coulomb gauge method for calculating nucleon unpolarized, helicity, and transversity parton distribution functions (PDFs). 
The calculations are performed on a Highly-Improved-Staggered-Quark
ensemble with lattice spacing $a = 0.06$ fm, volume $L_s^3 \times L_t=48^3\times 64$, and valence pion mass $m_\pi=300$ MeV, employing boosted nucleon states with momenta up to 3.04 GeV.
Our lattice predictions for the valence-quark PDFs---extracted from the real part of the correlators---show good convergence with increasing nucleon momentum and are compatible with the most recent global analyses for all spin structures. On the other hand, the full-quark-channel PDFs obtained from the imaginary part of the correlators exhibit discrepancies between the two large nucleon momenta considered, although the results at the higher momentum are consistent with phenomenology. The discrepancies are likely driven by stronger excited-state contamination in the imaginary matrix elements, which is consistent with the observation in the literature. Overall, this work demonstrates the efficacy of the Coulomb gauge approach for nucleon PDFs and serves as a benchmark for its broader applications. 
}

\begin{document}
\maketitle
\flushbottom

\section{Introduction}
\label{sec:intro}

Achieving a quantitative description of the internal structure of the proton and other hadrons at the partonic level remains a central challenge in hadron physics. Such a description is essential not only for providing precise inputs to standard-model predictions at collider experiments, but also for uncovering the origin of the hadron mass and spin and clarifying the mechanism of color confinement, which are central to our understanding of the visible universe. The parton distribution functions (PDFs), which describe the one-dimensional momentum densities of quarks and gluons inside hadrons, are the simplest and most fundamental quantities in this context. Because of their universality in QCD factorization, PDFs extracted from a subset of processes can be consistently employed as nonperturbative inputs for the calculation of cross sections in a broad class of hard-scattering reactions. During recent decades, high-energy scattering experiments at facilities such as HERA, COMPASS, SLAC, Fermilab, and LHC have provided state-of-the-art measurements of PDFs, leading to {global analyses} that determine the collinear structure of unpolarized nucleons and polarized nucleons, with some reaching percent-level uncertainties in the well-constrained kinematic regions~\cite{Alekhin:2017kpj,Hou:2019efy,Bailey:2020ooq,NNPDF:2021njg,Cruz-Martinez:2025ahf}. The future Electron-Ion Collider~\cite{Accardi:2012qut,AbdulKhalek:2021gbh} will probe the PDFs and other parton distributions with unprecedented precision, offering opportunities to map the multidimensional structure of hadrons.

Nevertheless, significant challenges remain in the determination of parton distributions, particularly in kinematic regions where experimental data provide insufficient constraints on phenomenological parameterizations, as well as in understanding spin-dependent effects and the multidimensional structure of hadrons~\cite{Aidala:2012mv,Amoroso:2022eow}. Over the past decades, lattice QCD has emerged as a promising first-principles framework for parton physics, with a variety of approaches developed to access partonic observables from Euclidean correlation functions~\cite{Kronfeld:1984zv,Martinelli:1987si,Liu:1993cv,Detmold:2005gg,Davoudi:2012ya,Ji:2013dva,Chambers:2017dov,Radyushkin:2017cyf,Ma:2017pxb,Detmold:2021uru,Shindler:2023xpd}. Among these approaches, large-momentum effective theory (LaMET)~\cite{Ji:2013dva,Ji:2014gla,Ji:2020ect,Ji:2024oka} enables precision-controlled calculations of the $x$-dependence of parton distributions. Since its proposal, LaMET has been extensively used in the computation of pion and nucleon PDFs~\cite{Lin:2014zya,Alexandrou:2016tjj,Alexandrou:2016jqi,Alexandrou:2016eyt,Chen:2016utp,Ishikawa:2017iym,Lin:2017ani,Chen:2017mzz,Alexandrou:2017dzj,Alexandrou:2017qpu,Alexandrou:2018eet,Alexandrou:2018pbm,Alexandrou:2018yuy,Chen:2018xof,Fan:2018dxu,LatticeParton:2018gjr,Lin:2018pvv,Liu:2018hxv,Zhang:2018nsy,Alexandrou:2019lfo,Izubuchi:2019lyk,Alexandrou:2020qtt,Bhattacharya:2020cen,Fan:2020nzz,Gao:2020ito,Lin:2020fsj,Shugert:2020tgq,Alexandrou:2021oih,Bhattacharya:2021moj,Constantinou:2021nbn,Gao:2021dbh,Salas-Chavira:2021wui,Gao:2022iex,Gao:2022uhg,LatticeParton:2022xsd,Gao:2023ktu,Good:2023ecp,Good:2023gai,Good:2024iur,Holligan:2024umc,Holligan:2024wpv,Ji:2024hit,Good:2025daz,NieMiera:2025mwj,Chen:2025xww,NieMiera:2025vcx,Miller:2025wgr}.
Notably, recent LaMET computations of the pion valence quark PDFs~\cite{Gao:2021dbh,Gao:2022iex,Holligan:2024umc,Ji:2024hit} have reached NNLO accuracy~\cite{Li:2020xml,Chen:2020ody} with systematic improvements and demonstrated consistency with phenomenological extractions in the intermediate-$x$ region~\cite{Zhao:2025oto}. Moreover, the recent work on the transversity PDF~\cite{LatticeParton:2022xsd} represents the first nucleon PDF calculation within this framework that incorporates controlled systematic improvements together with physical extrapolations.

Within the LaMET framework, most lattice calculations so far have employed gauge-invariant (GI) quasi-PDF operators defined with spatially separated quark bilinears connected by Wilson lines. This formulation preserves gauge invariance while introducing several features that require careful treatment. In particular, the linear ultraviolet (UV) divergence and the associated renormalon arising from Wilson-line self-energy call for nonperturbative renormalization strategies such as hybrid renormalization~\cite{Ji:2020brr,LatticePartonLPC:2021gpi} and leading-renormalon resummation (LRR)~\cite{Zhang:2023bxs,Holligan:2023rex}. In addition, the signal-to-noise ratio (SNR) of quasi-PDF matrix elements decreases exponentially with the length of the Wilson line, which can affect the precision at large distances. While the uncertainties from this region can be controlled and quantified using asymptotic extrapolation~\cite{Chen:2025cxr,Ji:2026vir}, further improvements in numerical precision and computational efficiency {are} still desirable. Furthermore, the presence of Wilson lines tends to restrict practical simulations to on-axis directions, thereby limiting flexibility in accessing large off-axis hadron momenta.

An alternative formulation of the GI quasi-PDF operators are quark bilinears fixed in the axial gauge, which approaches the light-cone gauge under an infinite Lorentz boost. However, the linear divergences remain in the renormalization, so the SNR still decreases exponentially with the separation of quark fields.
Recently, a Coulomb gauge (CG) formulation of quasi-PDFs within the LaMET framework has been proposed~\cite{Gao:2023lny,Zhao:2023ptv}. In this approach, quasi-PDFs are constructed from nonlocal quark correlators in Coulomb gauge without Wilson lines. It has been shown that, in this formulation, the linear divergences and associated linear renormalon effects stemming from Wilson-line self-energy are absent~\cite{Gao:2023lny,Liu:2023onm,Zhang:2024omt}, which substantially simplifies the renormalization procedure. This absence of linear divergences also leads to an improved SNR in the large-distance region and reduces uncertainties associated with the Fourier transform. The preservation of 3D rotational symmetry in the CG also allows for more flexible choices of off-axis momenta with fewer oscillatory modes~\cite{Gao:2023lny}. Despite the difference in operator construction, CG quasi-PDFs belong to the same universality class as their GI counterparts~\cite{Hatta:2013gta,Ji:2020ect}, implying that they can be matched to light-cone PDFs through LaMET, albeit with different power-suppressed corrections. Therefore, calculations based on the CG quasi-PDFs provide a complementary cross check of the GI approach and can lead to better precision in some cases. To date, exploratory studies employing the CG approach have focused primarily on collinear and transverse-momentum-dependent distributions (TMDs) of the pion~\cite{Gao:2023lny,Bollweg:2024zet,Mukherjee:2024xie,Bollweg:2025iol}. Beyond the pion, applications of the CG framework remain very limited, with the proton isovector helicity transverse-momentum-dependent parton distribution function (TMDPDF) and flavor-dependent unpolarized TMDPDFs~\cite{Bollweg:2025ecn} standing as the only nucleon-related study.

In this work, we present the first comprehensive benchmark of the CG method for nucleon PDFs through a lattice QCD calculation of the isovector unpolarized, helicity, and transversity distributions. By boosting the nucleon along an off-axis direction $\vec{n}=(1,1,0)$, we are able to reach momenta of $P^{z}=2.43$ and $3.04$ GeV at a pion mass of $m_{\pi}=300$ MeV, which are among the largest boosts achieved in nucleon PDF calculations at a comparable pion mass. Nevertheless, the CG method faces similar systematic uncertainties as the GI approach, including excited-state contamination and power-suppressed corrections. A detailed analysis is carried out to address excited-state contamination in the extraction of matrix elements, where we find that imaginary parts of matrix elements are significantly more sensitive to excited-state contamination, consistent with previous observations for GI operators~\cite{Alexandrou:2020qtt}.
Besides excited-state contamination, we also perform a systematic investigation of other sources of uncertainty in the CG framework, including the choice of Dirac gamma structure in quasi-PDF operators, the renormalization schemes, uncertainties associated with the Fourier transform, and power-suppressed corrections in the LaMET factorization. 
For the real parts of the quasi-PDF matrix elements, we utilize a hybrid scheme~\cite{Ji:2020brr} to renormalize the matrix elements. To renormalize the imaginary part of the quasi-PDF matrix elements, we argue that the hybrid scheme is not optimal and instead propose a static version of the \ripmom~scheme~\cite{Martinelli:1994ty}, where we find that the systematic uncertainties are under control. 
While the CG formulation does not eliminate the intrinsic challenges of nucleon PDF calculations, such as excited-state contamination due to the dense excited-state spectra, our results demonstrate reasonable agreement with existing lattice and phenomenological studies. 
This work thus demonstrates the CG method as a promising approach for precision studies of the nucleon partonic structure and provides a solid foundation for its future extensions, including the computation of TMDs.

This paper is organized as follows: in Sec.~\ref{sec:theory}, we introduce the theoretical framework for extracting nucleon PDFs from CG quasi-PDFs, including discussions of renormalization strategies and perturbative matching. The lattice setup is described in Sec.~\ref{sec:lattice_setup}. In Sec.~\ref{sec:quasi_pdf}, we present a detailed analysis of the quasi-PDFs. The resulting nucleon PDFs in all three polarization channels are presented and compared with the phenomenological determinations in Sec.~\ref{sec:pdfs}. The systematic uncertainties arising from the renormalization procedures and power corrections are discussed in Sec.~\ref{sec:sys}. Finally, our conclusions are given in Sec.~\ref{sec:conclusion}.

\section{Theoretical Framework}
\label{sec:theory}

\subsection{Definitions}

The light-cone PDF is defined as
\begin{align}
    f_\Gamma(x; \mu) &= \int_{-\infty}^{+\infty} \frac{\mathrm{d} \lambda}{2 \pi} e^{-i \lambda x} h_\Gamma(\lambda, \mu) ~,\\
    h_\Gamma(\lambda; \mu) &\equiv \frac{1}{2 P^{+}}\left\langle PS\left|\bar{\psi}\left(\xi^{-}\right) W\left(\xi^{-}, 0\right) \Gamma \psi(0)\right| PS\right\rangle ~,
\end{align}
where $|PS\rangle$ is a proton plane wave state with momentum $P^\mu = (P^t, 0, 0, P^z)$ and spin $S$, and $W\left(\xi^{-}, 0\right)$ is a Wilson  line. The light-cone coordinates and variables are $\xi^\pm = (t \pm z) / \sqrt{2}$ and $\lambda= P^+ \xi^-$. The PDF is renormalized in the $\MS$ scheme at scale $\mu$. For the unpolarized, helicity, and transversity PDFs, the Dirac structures $\Gamma$ are chosen as $\gamma^+$, $\gamma^+ \gamma^5$, and $i\gamma^+ \gamma^\perp$, respectively.

The quasi PDFs in the CG are defined as~\cite{Gao:2023lny}
\begin{align}
    \tilde{f}_{\tilde \Gamma}(x, P^z; \mu) &= P^z \int \frac{d z}{2 \pi} e^{i z (x P^z)} \tilde{h}_{\tilde\Gamma}(z, P^z, \mu) ~,
    \label{eq:quasi_operator_momentum}\\
    \tilde{h}_{\tilde\Gamma}(z, P^z; \mu) &\equiv \frac{1}{2 N_{\tilde\Gamma}} \bra{PS} \left. \bar{\psi}(z) \tilde\Gamma \psi (0) \right|_{\vec{\nabla} \cdot \vec{A} = 0} \ket{PS} ~,
    \label{eq:quasi_operator_coordinate}
\end{align}
where the normalization factor $N_{\tilde{\Gamma}}$ is given by:
\begin{align}
    N_{\tilde\Gamma} &=\left\{ \begin{array}{cc}
    {P^t} : & \tilde\Gamma \in \{ \gamma^t, \gamma^z\gamma_5  , \gamma^t\gamma_\perp^\alpha\gamma_5 \}\,, \\
    {P^z} : & \tilde\Gamma\in \{ \gamma^z, \gamma^t\gamma_5, \gamma^z\gamma_\perp^\alpha\gamma_5 \} \,.
    \end{array} \right.
\end{align}
The Dirac matrices $\tilde\Gamma = \{\gamma^t/\gamma^z,\gamma^t\gamma_5/\gamma^z\gamma_5, \gamma^t\gamma_\perp^\alpha\gamma_5/\gamma^z\gamma_\perp^\alpha\gamma_5\}$ in the quasi-PDF\footnote{{In this work, we use off-axis momenta with direction $\vec{n} = (n^x, n^y, 0)$. Accordingly, the Dirac structure $\gamma^z$ is replaced by $(\gamma^x+\gamma^y)$, and the transverse structure $\gamma_\perp^\alpha$ is defined with respect to this direction. In particular, for $n^x = n^y$, the transverse directions can be chosen as $(\gamma^x-\gamma^y)$, which is orthogonal to $(\gamma^x+\gamma^y)$.}} corresponds to $\Gamma=\{\gamma^+, \gamma^+\gamma_5, \gamma^+\gamma_\perp^\alpha \gamma_5\}$ in the PDF. 

\subsection{Renormalization}
\label{sec:renormalization}

\subsubsection{Hybrid renormalization of the real part}\label{ss:hybrid}

In the LaMET framework, the bare quasi distributions must be renormalized and matched to the $\overline{\mathrm{MS}}$ PDF. Several nonperturbative renormalization schemes are used in practice to renormalize the gauge-invariant quasi-PDF matrix element, including ratio-type schemes~\cite{Constantinou:2017sej,Stewart:2017tvs,Alexandrou:2017huk,Chen:2017mzz,Orginos:2017kos,Braun:2018brg,Li:2020xml}, regularization independent momentum subtraction schemes~\cite{Constantinou:2017sej,Stewart:2017tvs,Alexandrou:2017huk,Chen:2017mzz}, and the hybrid scheme~\cite{Ji:2020brr,LatticePartonLPC:2021gpi}. In this section, we adapt the hybrid scheme to renormalize the Coulomb-gauge fixed quasi-PDF matrix elements. 
The main advantage of the hybrid scheme is that discretization effects are cancelled by taking appropriate ratios at short separations, whilst avoiding  introducing any nonperturbative artifacts in the long-distance matrix elements.

{The proofs of renormalizability of CG QCD to all orders in perturbation theory have been made in the literature~\cite{Zwanziger:1998ez,Baulieu:1998kx,Niegawa:2006ey,Niegawa:2006hg}, though some still consider them incomplete~\cite{Lichtenegger:2010uda}. After all, the renormalizability of equal-time quark correlators have been validated on the lattice~\cite{Burgio:2012ph,Gao:2023lny,Zhang:2024omt} and at one loop~\cite{Gao:2023lny}, with the absence of linear power divergences and the associated renormalon~\cite{Liu:2023onm}}. As a result, the renormalization of the nonlocal CG operator is purely multiplicative and is given by
\begin{align}\label{eq:renorm}
\bar{\psi}_0 (z) \tilde\Gamma \psi_0 (0) = Z^R_\psi (a) \left[ \bar{\psi}_R (z) \tilde\Gamma \psi_R(0) \right] ~,
\end{align}
where $Z^R_\psi (a)$ is the renormalization factor of the quark wave-function into some renormalization scheme $R$ as a function of the lattice spacing $a$. $\psi_0$ and $\psi_R$ represent the bare and renormalized quark fields, respectively.

In the standard definition of the hybrid scheme~\cite{Gao:2021dbh,LatticeParton:2022xsd,Holligan:2024wpv}, the renormalized quasi-PDF matrix element is defined by the following ratios: 
\begin{align}
\begin{aligned}
    \tilde{h}_{\tilde\Gamma}^{\mathrm{hyb.}} \left(z, P^z; z_s \right)&=N \frac{\tilde{h}^0_{\tilde\Gamma} \left(z, P^z; a\right)} {\tilde{h}^0_{\tilde\Gamma} (z, 0; a)} \theta\left(z_s-|z|\right) +N \frac{\tilde{h}^0_{\tilde\Gamma} \left(z, P^z; a \right)} {\tilde{h}^0_{\tilde\Gamma} \left(z_s, 0; a \right)} \theta\left(|z|-z_s\right) ~,
    \label{eq:hybrid_renorm_norm}
\end{aligned}
\end{align}
where $\tilde{h}^0_{\tilde{\Gamma}}/\tilde{h}^{\mathrm{hyb.}}_{\tilde{\Gamma}}$ denotes the bare/hybrid-renormalized matrix element respectively, the normalization factor is defined by $N = \tilde{h}^0_{\tilde\Gamma} \left(0, 0; a\right) / \tilde{h}^0_{\tilde\Gamma} (0, P^z; a)$, and $z_s$ ($a\ll z_s \ll \Lambda_{\rm QCD}^{-1}$) denotes the point that separates the short- and long-distance regimes. Since the matrix elements are usually strongly correlated at neighboring lattice sites, this normalization can reduce their statistical uncertainties at small $z$. 
However, a drawback of this treatment is that the discretization errors of $z=0$ matrix elements are unnecessarily propagated to all $z\neq0$, where such effects are expected to be suppressed as powers of $a/|z|$. An ideal solution would be a $z$-dependent normalization that accurately removes the discretization effects mostly at small $z$, which might be achieved with lattice perturbation theory. However, it is beyond the scope of this work.
After all, the discretization effects can be systematically eliminated through continuum extrapolation, so this normalization may still be favored for the purpose of reducing statistical uncertainties.

On a single lattice ensemble, there is no clear preference for applying or omitting the above normalization, since the discretization effects cannot be quantified. However, since the $z=0$ matrix elements are expected to be independent of $P^z$ due to Lorentz covariance, a significant deviation of $N$ from unity signals sizable lattice artifacts; in this case, a blind normalization would rescale the quasi-PDF by the same factor. Therefore, in this work we adopt a hybrid scheme without the normalization at $z=0$, namely,
\begin{align}
    \tilde{h}^{\mathrm{hyb.}}_{\tilde \Gamma} \left(z, P^z; z_s \right)&= \frac{\tilde{h}^0_{\tilde \Gamma} \left(z, P^z; a\right)} {\tilde{h}^0_{\tilde \Gamma} (z, 0; a)} \theta\left(z_s-|z|\right) + \frac{\tilde{h}^0_{\tilde \Gamma} \left(z, P^z; a \right)} {\tilde{h}^0_{\tilde \Gamma} \left(z_s, 0; a \right)} \theta\left(|z|-z_s\right) ~.
    \label{eq:hybrid_renorm_no_norm}
\end{align}
As a result, the discretization effects at $z\sim a$ remain uncorrected and propagate into momentum space as artifacts of order $(P^z a)^n$. Although this treatment spoils the normalization of the quasi-PDF and PDF, the effect vanishes in the continuum limit. {The comparison between two strategies is shown in Fig.~\ref{fig:PDF_scheme_real}, where the double ratio corresponds to including $N$, while the single ratio corresponds to omitting $N$.}

The hybrid renormalization is applied to the real part of the quasi-PDF matrix elements with the separation point chosen as $z_s = 3 \sqrt{2}a$, where the factor of $\sqrt{2}$ originates from the off-axis momentum setup. 

\subsubsection{S\ripmom \ renormalization of the imaginary part}\label{ss:srip}

For the imaginary part of the quasi-PDF matrix elements, however, it is likely that the ratios employed in the hybrid scheme do \textit{not} remove the leading lattice artifacts for $z \le z_s$, as the zero-momentum matrix elements in the denominator of the ratio are purely real. To address this issue, we instead employ a modification of the \ripmom~\cite{Martinelli:1994ty} scheme for the quark wave function renormalization factor $Z_R$ in Eq.~(\ref{eq:renorm}). Because the Coulomb gauge is not a complete gauge fixing due to the residual gauge freedom in the timelike direction, such a scheme based on the quark propagator either requires introducing \textit{additional} gauge-fixing conditions in the timelike direction~\cite{Burgio:2008jr,Burgio:2012ph}, or one must only use spatial correlation functions. 
Choosing the latter route leads to a Static \ripmom  \ scheme (S\ripmom) which we define below. Given the CG static propagator:
\begin{equation}
S(\vec{k})  = \int \mathrm{d}^3 \vec{x}\  e^{i \vec{x} \cdot \vec{k}} \left\langle \psi(\vec{0},t_E) \overline{\psi}(\vec{x},t_E) \right\rangle,
\end{equation}
computed from position-space propagators at a single Euclidean time $t_E$, we can define a renormalization constant such that the renormalized static propagator is equal to the tree-level value at $\vec{k}$:
\begin{align}
\psi_{0} &= Z^{1/2}_{\mathrm{SRI'}}(|\vec{k}|) \psi_{\mathrm{SRI'}},
\qquad Z_{\mathrm{SRI'}}(|\vec{k}|) =  \lim_{m \to 0} \frac{ \mathrm{Tr}[\slashed{\vec{k}} S_0(\vec{k})]}{ \mathrm{Tr}[\slashed{\vec{k}} S_\mathrm{tree}(\vec{k})]} , \label{eq:Zdef} 
\end{align}
where the chiral limit is taken in order to define a mass-independent renormalization scheme to match to $\overline{\mathrm{MS}}$.  
{
Due to the fact that we only have a single ensemble, we cannot actually take this chiral limit. Studies of similar massless schemes have typically found that the wavefunction renormalisation factor varies only very mildly with quark mass \cite{He:2022lse,Blossier:2014kta,Shanahan:2019zcq,Bi:2017ybi,Blossier:2010vt} for even heavier pion masses than the $m_\pi = 300 \mathrm{MeV}$ used in this work. 
It is however not clear if this observation should generalise to SRI'-MOM, and further studies with ensembles including various quark masses are needed to properly take a chiral limit. 

In principle, with a single ensemble it is possible to perform a massive variant of the scheme (similar to RI/mSMOM \cite{Boyle:2016wis}) by enforcing the renormalization conditions at a fixed renormalized quark mass. 
The difference between such a massive scheme and the chiral scheme proposed in this section scales with $m_q/|\vec{k}|$, where $m_q$ is the renormalised quark mass, and given that we consider scales up to $|\vec{k}| = 6~\mathrm{GeV}$ (as further detailed in Appendix~\ref{sec:RIMOM}), the systematic error due to finite quark mass is likely small. 
Properly accounting for the finite quark mass however will require multiple ensembles with multiple values of the quark mass. 
}

Using leading-logarithmic running and $O(\alpha_S)$ matching between S\ripmom \ and $\mathrm{\overline{MS}}$, we find that the renormalization constant for the ensemble used in this work (described in \Cref{sec:lattice_setup}) is:
\begin{align}\label{eq:finalresult}
Z_{\overline{\mathrm{MS}}}(a,\mu = 2~\mathrm{GeV}) = 0.914(5).
\end{align}
Details of the calculation, including treatment of lattice artifacts, breakup of error into statistical, perturbative truncation, and systematic errors are described in Appendix~\ref{sec:RIMOM}. 

\subsection{Large-momentum expansion and perturbative matching}

According to LaMET~\cite{Ji:2020ect,Ji:2024oka}, when $P^z\gg \Lambda_{\rm QCD}$ the PDF can be obtained from the CG quasi-PDF through a perturbative matching~\cite{Izubuchi:2018srq,Gao:2023lny},
\begin{align}\label{eq:fact}
	f_{\Gamma}^{\overline{\mathrm{MS}}}(x,\mu) &= \int_{-\infty}^{+\infty} {\mathrm{d}y\over |y|}\ C_{\tilde \Gamma}^R\left({x\over y}, {\mu\over |y|P^z}, z_sP^z\right) \tilde f_{\tilde\Gamma}^R(y,P^z,z_s)  + {\cal O}\left({\Lambda_{\rm QCD}^2\over x^2P_z^2},{\Lambda_{\rm QCD}^2\over(1-x)^2P_z^2}\right)\,,
\end{align}
where $C_{\tilde \Gamma}^R$ is the matching coefficient corresponding to the quasi-PDF renormalized in a renormalization scheme labelled $R$, and the power corrections are suppressed by the active and spectator parton momenta. 
Due to the absence of linear renormalon in the CG method~\cite{Liu:2023onm}, the power corrections in \Eq{fact} start from the quadratic order.

The above matching formula can be derived following similar procedures to those in Refs.~\cite{Zhao:2023ptv,Holligan:2025baj} with Soft-Collinear Effective Theory~\cite{Bauer:2002uv,Bauer:2001yt,Beneke:2002ph,Bauer:2003mga}. This approach has been successfully applied to derive the factorization formulas for the CG quasi-TMD~\cite{Zhao:2023ptv} and for the GI quasi-PDF and generalized parton distribution~\cite{Holligan:2025baj}. Alternatively, the factorization can be established through the lead-region analysis of Feynman diagrams~\cite{Libby:1978bx}, as has been done for the GI quasi-PDF and quasi-TMD~\cite{Ji:2020ect,Liu:2022nnk}. Although extending this proof to the CG case requires additional care because of the Lorentz noncovariance, the renormalizability of CG QCD guarantees that no new counterterms arise at higher loop orders, allowing the proof to be carried out iteratively. Since the derivation of the factorization formula largely follows the procedures in Refs.~\cite{Zhao:2023ptv,Holligan:2025baj}, we do not repeat it here.

\subsubsection{One-loop matching kernels in the Coulomb gauge}

By calculating the next-to-leading order (NLO) corrections to the quark CG quasi-PDF and PDF in a free quark state, we find out that their collinear divergences are identical~\cite{Xiong:2013bka}. The $\MS$ matching coefficient is obtained as a series in the strong coupling $\alpha_s$,
\begin{align}\label{eq:quasi-PDF_match}
	C_{\tilde \Gamma}^{\overline{\mathrm{MS}}}\big(\xi, {\mu\over p^z}\big) &= \delta(\xi-1) {-} {\alpha_sC_F\over 2\pi}C_{\tilde \Gamma(1)}^{\overline{\mathrm{MS}}}\big(\xi,{\mu\over p^z}\big) + {\cal O}(\alpha_s^2)\,,
\end{align}
where $C_F=4/3$.
At NLO the matching kernels are given by:
\begin{align}
	C_{\gamma^t(1)}^{\overline{\mathrm{MS}}}\big(\xi,{\mu\over p^z}\big) &= C_{\gamma^t\gamma_5(1)}^{\overline{\mathrm{MS}}}\big(\xi,{\mu\over p^z}\big) = C^{(1)}_r\big(\xi,{\mu\over p^z}\big) + {1\over 2|1-\xi|} \nn\\
    &\qquad \qquad \qquad \qquad  \qquad  + {1\over2}\delta(1-\xi) \left[ 1 + \ln { 4p_z^2\over\mu^2} - \int_0^2 d\xi' {1\over |1-\xi'|}\right]\,,\\
	C_{\gamma^z(1)}^{\overline{\mathrm{MS}}}\big(\xi,{\mu\over p^z}\big) &= C_{\gamma^z\gamma_5(1)}^{\overline{\mathrm{MS}}}\big(\xi,{\mu\over p^z}\big) = C_{\gamma^t(1)}^{\overline{\mathrm{MS}}}\big(\xi,{\mu\over p^z}\big) + 2(1-\xi)_{+(1)}^{[0,1]} + \delta(1-\xi)\,,
\end{align}
where
\begin{align}
    C^{(1)}_r\big(\xi,{\mu\over p^z}\big) &= \left[\frac{1+\xi^2}{1-\xi}\ln{4p_z^2\over\mu^2} + \xi - 1\right]_{+(1)}^{[0,1]}  + \Bigg\{\frac{1+\xi^2}{1-\xi} \Big[\text{sgn}(\xi)\ln|\xi| + \text{sgn}(1-\xi)\ln|1-\xi|\Big] \nn\\
    &\qquad + \text{sgn}(\xi) + {3 \xi-1\over \xi-1} \frac{\tan^{-1}\left(\sqrt{1-2 \xi}/|\xi|\right)}{\sqrt{1-2 \xi}} - {3\over 2|1-\xi|} \Bigg\}_{+(1)}^{(-\infty, \infty)}\nn
\end{align}
satisfies particle number conservation. Here the plus functions on a domain $D$ are defined as
\begin{align}
	\left[g(x)\right]^D_{+(x_0)} &= g(x) - \delta(x-x_0)\int_D dx'\ g(x')\,.
\end{align}
Note that $C^{(1)}_r$ is analytical at $\xi=1/2$ despite its form.

For the transversity PDF case,
\begin{align}
	C_{\gamma^t\gamma_\perp^\alpha\gamma_5(1)}^{\overline{\mathrm{MS}}}\big(\xi,{\mu\over p^z}\big) &= C_{\gamma^z\gamma_\perp^\alpha\gamma_5(1)}^{\overline{\mathrm{MS}}}\big(\xi,{\mu\over p^z}\big) = C^{\perp (1)}_r\big(\xi,{\mu\over p^z}\big) \,,
\end{align}
where
\begin{align}
    C^{\perp (1)}_r\big(\xi,{\mu\over p^z}\big)&= \left[{2\xi \over 1-\xi}\ln{4p_z^2\over\mu^2} \right]_{+(1)}^{[0,1]}  + \Bigg\{{2\xi \over 1-\xi} \Big[\text{sgn}(\xi)\ln|\xi| + \text{sgn}(1-\xi)\ln|1-\xi|\Big] \nn\\
    &\qquad + {3 \xi-1\over \xi-1} \frac{\tan^{-1}\left(\sqrt{1-2 \xi}/|\xi|\right)}{\sqrt{1-2 \xi}} - {1\over |1-\xi|} \Bigg\}_{+(1)}^{(-\infty, \infty)}\,.
\end{align}

Finally, the hybrid scheme matching coefficients are
\begin{align}
	C_{(1)}^{\mathrm{hyb.}}\left(\xi,{\mu\over p^z},z_sp^z\right) &= C^{(1)}_r\left(\xi,{\mu\over p^z}\right) + \delta C_{(1)}^{\mathrm{hyb.}}\left(\xi,z_sp^z\right)\,,
\end{align}
where
\begin{align}
	\delta C_{\gamma^t(1)}^{\mathrm{hyb.}}\big(\xi,z_sp^z\big) &= \delta C_{\gamma^z(1)}^{\mathrm{hyb.}}\big(\xi,z_sp^z\big) =  \delta C_{\gamma^t\gamma_5(1)}^{\mathrm{hyb.}}\big(\xi,z_sp^z\big) =  \delta C_{\gamma^z\gamma_5(1)}^{\mathrm{hyb.}}\big(\xi,z_sp^z\big) \nonumber\\
    &= {1\over2}\left[{1\over |1-\xi|} - {2{\rm Si}[(1-\xi)z_sp^z]\over \pi(1-\xi)}\right]_{+(1)}^{(-\infty, \infty)}\,,\\
    \delta C_{\gamma^t\gamma_\perp^\alpha\gamma_5(1)}^{\mathrm{hyb.}}\big(\xi,z_sp^z\big) &= \delta C_{\gamma^z\gamma_\perp^\alpha\gamma_5(1)}^{\mathrm{hyb.}}\big(\xi,z_sp^z\big)= 0\,.
\end{align}

\subsubsection{Renormalization group evolution}

The matching coefficient in \Eq{fact} includes the logarithm $\ln {\mu^2\over (2yP^z)^2}$ which follows the DGLAP evolution equation. However, since $y$ is an integration variable, the matching would be ill defined for all $x$ if we naively resum such logarithms. Instead, the physically meaningful logarithms should be given by the external scales after the convolution integral. For \Eq{fact}, they are identified as $\ln {\mu^2\over (2xP^z)^2}$~\cite{Su:2022fiu}, which indicates that the resummation becomes significant at small $x$. When $\alpha_s(2xP^z)\gtrsim 1$, the matching loses perturbative control, which provides an estimate of the smallest $x_{\rm min}$ that LaMET can reach.

The procedure for resumming small-$x$ logarithm is as follows: first, for each $x$ we match the quasi-PDF to the PDF at the initial scale $\mu_0(x)=2\kappa xP^z$ where $\kappa\sim 1$. Then, we evolve the PDF $f(x, \mu_0(x))$ to a fixed $\MS$ scale $\mu=2$ GeV for $x\in [x_{\rm min}, 1]$. For matching accuracy at NLO, the two-loop DGLAP evolution kernel~\cite{Curci:1980uw} is needed, which makes the resummation at next-to-leading logarithmic (NLL) order. Since the DGLAP evolution is closed for $x\in [x_{\rm min}, 1]$, such an evolution kernel can be calculated uniquely following the strategies in Refs.~\cite{Su:2022fiu,Ding:2024saz}.

Apart from the small-$x$ logarithms, the matching coefficient also contains the singular terms $\ln(1-\xi)/(1-\xi)$ and $1/(1-\xi)$~\cite{Gao:2021hxl}, which become important as $\xi\to1$. These are the standard threshold logarithms that also appear in the matching kernel for the GI quasi-PDFs~\cite{Ji:2023pba,Ji:2024hit}. The factorization and resummation of such logarithms have been derived for the GI quasi-PDFs~\cite{Ji:2023pba,Ji:2024hit} and for the quasi generalized parton distributions~\cite{Holligan:2025baj}, and the same methodology can be applied to the CG quasi distributions, although this has not yet been carried out in the literature. According to Ref.~\cite{Ji:2024hit}, the relevant external scale is $2(1-x)P^z$, implying that threshold resummation becomes significant---and eventually loses perturbative control---at large $x$. For this reason, we postpone its implementation in the present work and instead estimate the maximum value $x_{\rm max}$ allowed by LaMET by imposing the condition $\alpha_s(2(1-x_{\rm max})P^z)\sim 1$.

\section{Lattice Setup}
\label{sec:lattice_setup}

\begin{table}[]
    \centering
    \begin{tabular}{|c|c|c|c|c|}
    \hline \hline$\abs{\vec{p}}(\mathrm{GeV})$ & $\vec{n}$ & $\vec{k}$ & $t_s / a$ & $(\#$ ex, \#sl) \\
    \hline 0 & $(0,0,0)$ & $(0,0,0)$ & $8,10,12$ & $(1,16)$ \\
    \hline \multirow{5}{*}{2.43} & \multirow{5}{*}{$(4,4,0)$} & \multirow{5}{*}{$(2,2,0)$} & 8 & $(1,112)$ \\
    & & & 9 & $(3,128)$ \\
    & & & 10 & $(3,176)$ \\
    & & & 11 & $(3,128)$ \\
    & & & 12 & $(3,176)$ \\
    \hline \multirow{5}{*}{3.04} & \multirow{5}{*}{$(5,5,0)$} & \multirow{5}{*}{$(2,2,0)$} & 8 & $(1,112)$ \\
    & & & 9 & $(3,128)$ \\
    & & & 10 & $(3,176)$ \\
    & & & 11 & $(3,128)$ \\
    & & & 12 & $(3,176)$ \\
    \hline \hline
    \end{tabular}
    \caption{
    Summary of the lattice setup and statistics for the nucleon two-point and three-point correlation functions used in this work. 
    For each momentum $|\vec{p}|$, we list the corresponding boost direction $\vec{n}$, the quark smearing momentum $\vec{k}$, the source sink separations $t_s/a$, and the statistics in terms of the numbers of exact and sloppy inversions $(\#\,\mathrm{ex}, \#\,\mathrm{sl})$.
    }
    \label{tb:lattice_setup}
\end{table}

We perform a numerical lattice QCD calculation on a 2+1 flavor gauge ensemble generated by the HotQCD Collaboration~\cite{HotQCD:2014kol} with Highly Improved Staggered Quarks (HISQ)~\cite{Follana:2006rc}. The ensemble has a lattice spacing of $a=0.0601(5)$ fm and a volume of $L_s^3 \times L_t = 48^3 \times 64$. We used tadpole-improved clover Wilson valence fermions on a hypercubic (HYP) smeared~\cite{Hasenfratz:2001hp} gauge background. We set the clover coefficient $c_{\rm sw} = u_0^{-3/4}$, in which $u_0$ is the average plaquette after HYP smearing; the parameter is set as $c_{\rm sw} = 1.0336$ for both time and spatial directions. For the masses of the valence quarks, we used hopping parameter $\kappa = 0.12623$ for the light quarks $u$ and $d$, corresponding to a valence pion mass $m_\pi = 300$ MeV. {To implement the CG method, we fix the lattice ensemble in the CG via finding the gauge transformation $\Omega$ of link variables $U_i(t, \vec{x})$ that minimizes the functional
\begin{align}
    F\left[U^{\Omega}\right]=\frac{1}{9 V} \sum_{\vec{x}} \sum_{i=1,2,3}\left[-\operatorname{Re} \operatorname{Tr} U_i^{\Omega}(t, \vec{x})\right] 
\end{align}
on each time slice. Each gauge fixing is performed with 600 sweeps and reaches a precision of $\delta F / F < 10^{-7}$, the first successful gauge fixed instance is kept. Although imperfect gauge fixing and the presence of Gribov copies~\cite{Gribov:1977wm,Singer:1978dk} can, in principle, affect gauge-variant correlators~\cite{Zhang:2024omt}, numerical studies in Ref.~\cite{Gao:2024fbh} indicate that the resulting distortions from Gribov copies are negligible compared to the statistical uncertainties in the CG matrix elements of quasi-PDFs. 
After all, such effects will be power suppressed in the large-momentum expansion of the quasi-PDF, as the Gribov ambiguities vanish in the collinear modes or light-cone gauge.}

As shown in Eq.~\ref{eq:fact}, a sufficiently large momentum of the hadron is necessary to suppress the power corrections. In order to attain a larger boost factor, we leverage the three-dimensional rotational symmetry inherent in the CG methodology by using off-axis momentum directions, which are aligned with $\vec{n} = (n^x, n^y, 0)$. Although we use off-axis momentum directions, for simplicity, we will still refer to the momentum direction as the $z$ direction in the following discussions. The hadron momenta on the lattice are given by $P^z = \frac{2\pi |\vec{n}|}{L_s a}$. In this study, we consider $ n^x = n^y \in \{4, 5\} $, which allows us to reach nucleon momentum $ P^z =\{2.43, 3.04\} $ GeV for the quasi-PDF calculations. To optimize the SNR and enhance overlap with the ground states of large-momentum hadrons, we employ boosted Gaussian smearing~\cite{Bali:2016lva}, using the same setup as in Ref.~\cite{Gao:2023lny}. To extract the ground-state contribution, we compute three-point functions of quasi-PDFs for multiple source-sink separations, choosing $ t_{\rm sep} / a \in \{8, 9, 10, 11, 12\} $. Calculations are performed on 553 gauge configurations and we apply the All-Mode Averaging (AMA) technique~\cite{Shintani:2014vja} to further improve the signal. The stopping criteria for the sloppy and exact inversions are set to $10^{-4}$ and $10^{-10}$, respectively, aligning with the settings in Ref.~\cite{Gao:2020ito}. In addition, the quasi-PDF matrix elements at zero momentum are computed for the purpose of hybrid renormalization, with the source sink separations chosen as $t_{\rm sep}/a \in \{8, 10, 12\}$. The statistics of all datasets used in this work, including both the zero-momentum and the nonzero-momentum correlators entering the quasi-PDF analysis, are summarized in Table~\ref{tb:lattice_setup}.

\section{Quasi Distributions}
\label{sec:quasi_pdf}

\subsection{Two-point function and dispersion relation}

\begin{figure}[th!]
    \centering
    \includegraphics[width=.49\linewidth]{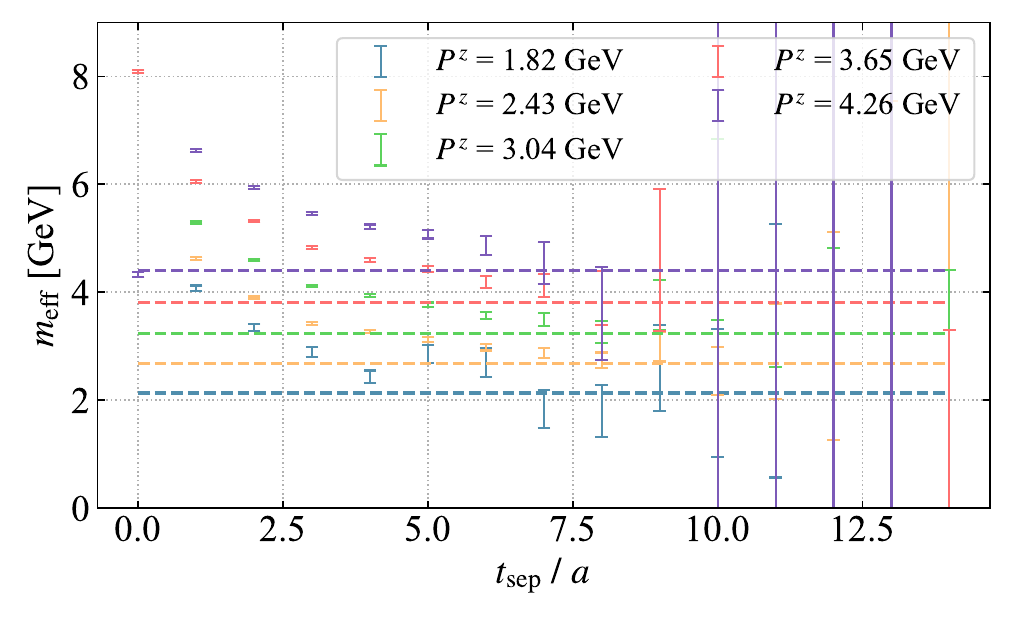}
    \includegraphics[width=.49\linewidth]{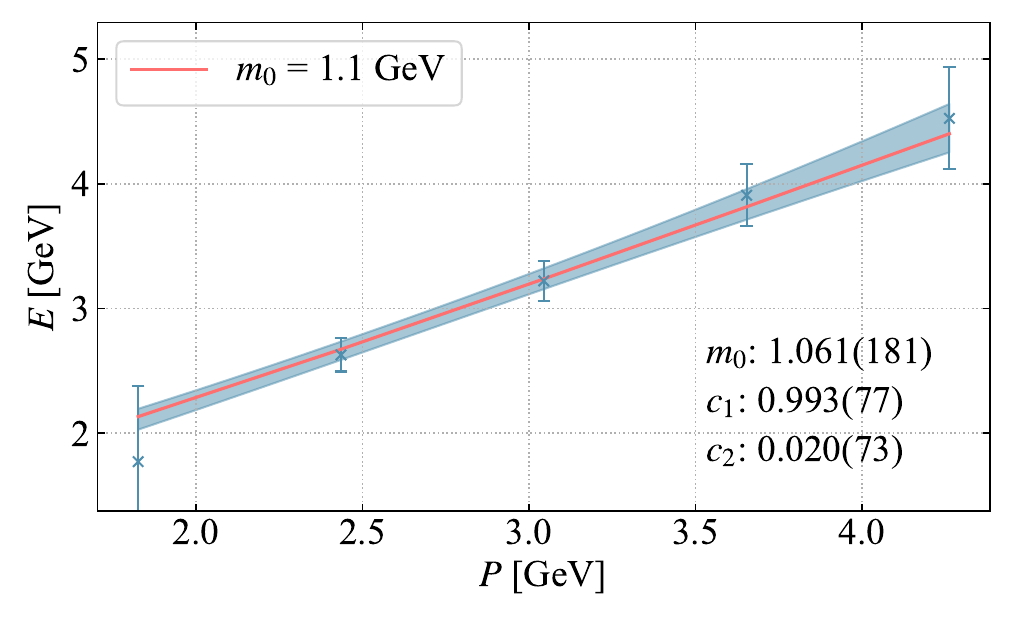}
    \caption{Effective mass and dispersion relation of the nucleon. The left panel shows the effective masses extracted from the two-point correlators as functions of the source-sink separation $t_{\mathrm{sep}}$ for different hadron momenta. The dashed horizontal lines correspond to the energies computed from the relativistic dispersion relation $m_{\rm eff} = \sqrt{m_0^2 + (P^z)^2}$ with the static mass fixed to $m_0 = 1.1$ GeV. The right panel displays the ground-state energies obtained from two-state fits as functions of the hadron momentum, along with the fit to the dispersion relation $E = \sqrt{m_0^2 + c_1 P^2 + c_2 a^2 P^4}$. The shaded band denotes the fit uncertainty, and the fitted values of the parameters are indicated in the figure.
    \label{fig:disp}}
\end{figure}

The nucleon interpolating operator used in this work is given by
\begin{align}
    N^s_\alpha(\vec{x},t) = \epsilon_{abc} ~ u_{a \alpha}^s (\vec{x}, t)  \left( u^{s}_b (\vec{x}, t)^T ~ C \gamma_5 ~ d_c^s (\vec{x}, t) \right) ~,
    \label{eq:interpolating}
\end{align}
where $C=\gamma^t \gamma^y$ is the charge conjugation operator, the subscript $a, b, c$ are the color indices, $\alpha$ is the spin index, and the superscript $s$ indicates whether the quark field is smeared ($s=S$) or not ($s=P$).

The two-point correlator with hadron momentum $\vec{P}$ is defined as  
\begin{align}
    C^{s s'}_{\text{2pt}} \left(t_{\rm sep} \right) = \sum_{\vec{x}} e^{-i \vec{P} \cdot \vec{x}} ~ \mathcal{P}^{\mathrm{2pt}}_{\alpha  \beta} \langle N^{s}_\alpha (\vec{x}, t_{\rm sep}) \bar{N}^{s'}_\beta (\vec{0}, 0) \rangle ~,
\end{align}
where the source is positioned at the origin for simplicity, and the positive parity is projected using $\mathcal{P}^{\mathrm{2pt}} = \frac{1}{2}(1 + \gamma^t)$. In this work, we use a Gaussian-smeared source and sink ($ s = s' = S $), following the setup in Ref.~\cite{Izubuchi:2019lyk}. The superscript $SS$ will be removed in the following text for simplicity. By inserting a complete set of states, the two-point correlator can be expressed as a sum over energy eigenstates:
\begin{align}
    C_{\text{2pt}}\left(t_{\rm sep}\right) = \sum_{n=0}^{n_{\mathrm{s}}-1} \frac{\left|z_n\right|^2}{2 E_n} \left(e^{-E_n t_{\rm sep}}+e^{-E_n\left(L_t-t_{\rm sep}\right)}\right) ~.
    \label{eq:2pt}
\end{align}  
The overlap amplitude $ z_n = \langle \Omega | N_\alpha \mathcal{P}^{\mathrm{2pt}}_{\alpha  \beta} | n \rangle $ quantifies the projection of the nucleon interpolator onto the $ n $th energy eigenstate, where $|\Omega \rangle$ denotes the QCD vacuum state and $ n_s $ denotes the number of excited states with the same quantum numbers as the nucleon, which are considered within the fit function.

To investigate the asymptotic behavior of $ C_{\text{2pt}} $ at large $ t_{\rm sep} $, we define the effective mass as  
\begin{align}
    m_{\rm eff} (t_{\rm sep} ) = \ln\left( \frac{C_{\text{2pt}}\left(t_{\rm sep}\right)}{C_{\text{2pt}}\left(t_{\rm sep} + a\right)} \right) ~.
\end{align}  
The effective masses for different momenta are shown in the left panel of Fig.~\ref{fig:disp}. As $ t_{\rm sep} $ increases, the effective masses approach plateaus, which align with the dashed lines computed from the relativistic dispersion relation $m_{\rm eff} = \sqrt{m_0^2 + (P^z)^2}$, where the static mass of nucleon is set to be $m_0 = 1.1$ GeV. This agreement indicates that the ground state is effectively isolated from the excited-state tower for $ t_{\rm sep} \gtrsim 8a $.

In practical analysis, the upper bound $ n_s $ must be truncated, as higher excited-state contributions decay rapidly with increasing $ t_{\rm sep} $. In this work, we perform a two-state fit by setting $ n_s = 2 $, which allows us to efficiently extract the ground-state contribution while accounting for the leading excited-state contamination. The right panel of Fig.~\ref{fig:disp} presents the extracted ground-state energies as a function of the hadron momentum. To test the validity of the relativistic dispersion relation, we fit the data points using the functional form  
\begin{align}
    E = \sqrt{m_0^2 + c_1 P^2 + c_2 a^2 P^4} ~,
\end{align}  
where $m_0$ is the static mass, $ E $ is the ground-state energy, $ P $ is the hadron momentum, and $ a $ is the lattice spacing. The coefficients $ c_1 $ and $ c_2 $ parameterize the possible discretization effects.  
As shown in Fig.~\ref{fig:disp}, the fit band describes the data well, the fitted coefficients are found to be $c_1 = 0.993(77)$, consistent with unity, and $c_2 = 0.020(73)$, consistent with zero within uncertainties. This indicates that the extracted energy spectrum is consistent with the continuum relativistic dispersion relation up to hadron momenta $ P \approx 4.3 $ GeV.
This agreement shows that discretization effects remain small in the dispersion relation within the momentum range studied in this work.

\subsection{Bare quasi-PDFs}

\begin{table}[h]
\renewcommand{\arraystretch}{1.2} 
    \centering
    \begin{tabular}{|c|c|c|c|}
        \hline\hline
        Fit & $n_s$ & $t_{\rm sep}$ range & $\tau$ range \\
        \hline
        $C_{\rm 2pt}(t_{\rm sep})$ & 2 & $t_{\rm sep} \in [4, 12]$ & / \\
        \hline
        $R(t_{\rm sep}, \tau)$ & 2 & $t_{\rm sep} \in [8, 12]$ & $\tau \in [3, t_{\rm sep} - 3]$ \\
        \hline
        $\mathrm{FH}(t_{\rm sep}, \tau_{\rm cut} = 3)$ & 1 & $t_{\rm sep} \in [8, 12]$ & $\tau \in [3, t_{\rm sep} - 3]$ \\
        \hline\hline
    \end{tabular}
    \caption{Collection of ground state fit settings. \#state = 1 means that only one ground state is included in the fit functions.}
    \label{tb:gsfit}
\end{table}

The bare quasi-PDFs can be extracted from the three-point function with hadron momentum $\vec{P}$
\begin{align}
    C_{\mathrm{3pt}} (t_{\mathrm{sep}}, \tau) = \sum_{\vec{x}, \vec{z}_0} e^{-i \vec{P} \cdot \vec{x}} ~ \mathcal{P}^{\mathrm{3pt}}_{\alpha \beta} \left\langle N_\alpha  (\vec{x}, t_{\mathrm{sep}}) ~ \mathcal{O}_{\tilde \Gamma} (\vec{z}_0, z, \tau) ~ \bar{N}_\beta (\vec{0}, 0) \right\rangle ~,
\end{align}
where nucleon polarization projection operator $\mathcal{P}^{\mathrm{3pt}}$ is chosen to be $\frac{1}{2}(1+\gamma_t)$, $\frac{1}{2}(1+\gamma_t)(-i\gamma_5\gamma_z)$, and $\frac{1}{2}(1+\gamma_t)(-i\gamma_5 \gamma_\perp)$
for the unpolarized, helicity and transversity quasi-PDF respectively. Throughout, we take the hadron momentum direction as $\hat{z}$ and the transverse spin direction as $\hat{x}$. The insertion operator is given in \Eq{quasi_operator_coordinate} with isovector combination
\begin{align}
    \mathcal{O}_{\tilde \Gamma} (\vec{z}_0, z, \tau) = \bar{u}(\vec{z}_0 + z \hat{z}, \tau) \tilde \Gamma u (\vec{z}_0, \tau) - \bar{d}(\vec{z}_0 + z \hat{z}, \tau) \tilde \Gamma d (\vec{z}_0, \tau) ~,
\end{align}
where the corresponding Dirac structures $\tilde \Gamma$ appropriate for the three channels are $\gamma^t$, $i \gamma^z \gamma^5$ and $-i \gamma^z \gamma^y$.  
In terms of the spectral decomposition, it can be expanded as
\begin{align}
\begin{aligned}
    C_{\mathrm{3pt}}\left(t_{\mathrm{sep}}, \tau \right) = \sum^{n_s -1}_{n, m = 0} \frac{z_n^\dagger O_{n m} z_m}{4 E_n E_m} e^{-E_n\left(t_{\mathrm{sep}}- \tau \right)} e^{-E_m \tau} ~,
    \label{eq:qtmdpdf_3pt}
\end{aligned}
\end{align}  
where $ O_{n m}=\langle n| \mathcal{O}_{\tilde \Gamma} |m \rangle$ are the matrix elements of the insertion operator. The ground state matrix element $O_{00}$, determined by our selection of $\tilde \Gamma$, is related to the bare quasi-PDFs $\tilde{h}^0_{\tilde \Gamma}$ through the relationship $O_{00} = 2 E_0 \tilde{h}^0_{\tilde \Gamma}$.

We employ the Feynman–Hellmann (FH) method~\cite{Bouchard:2016heu,He:2021yvm} to suppress excited-state contamination in the extraction of the ground-state matrix elements. The analysis is carried out using a fully correlated Bayesian least-squares procedure. We first fit the two-point correlator $C_{\rm 2pt}$ from which the ground- and excited-state energies and amplitudes ($E_0$, $E_1$, $z_0$, $z_1$) are obtained. The posterior distributions of these parameters are then used as Bayesian priors in a subsequent joint fit of the ratio
\begin{align}
R\left(t_{\mathrm{sep}}, \tau \right) = \frac{C_{\mathrm{3pt}}\left(t_{\mathrm{sep}}, \tau \right)}{C_{\rm 2pt} \left(t_{\mathrm{sep}} \right)} ~,
\end{align}
together with the FH correlators. The FH correlator is defined as
\begin{align}
\begin{aligned}
    S\left(t_{\mathrm{sep}}, \tau_{\mathrm{cut}} \right) &\equiv \sum_{t=\tau_{\text {cut }}}^{t=t_{\text {sep }}-\tau_{\text {cut }}} R\left(t_{\text {sep}}, \tau\right) \\
    \mathrm{FH}\left(t_{\mathrm{sep}}, \tau_{\mathrm{cut}}, d\tau \right) &\equiv \frac{ S\left(t_{\mathrm{sep}} + d\tau, \tau_{\mathrm{cut}} \right) - S\left(t_{\mathrm{sep}}, \tau_{\mathrm{cut}} \right) }{d\tau} ~,
\end{aligned}
\end{align}
where $d\tau$ is set to equal 1. In the large-source–sink separation limit $t_{\rm sep} \to \infty$, both $R$ and $\mathrm{FH}$ converge asymptotically to the same bare quasi-PDF matrix element, $\tilde{h}^0_{\tilde \Gamma}$. The entire chained procedure—first fitting $C_{\rm 2pt}$ and then performing the joint fit of $R$ and FH with the two-point posteriors as priors—is repeated on each jackknife sample to ensure robust and reliable uncertainty estimates. The detailed setting can be found in Table~\ref{tb:gsfit}.

\begin{figure}[th!]
    \centering
    \includegraphics[width=0.49\linewidth]{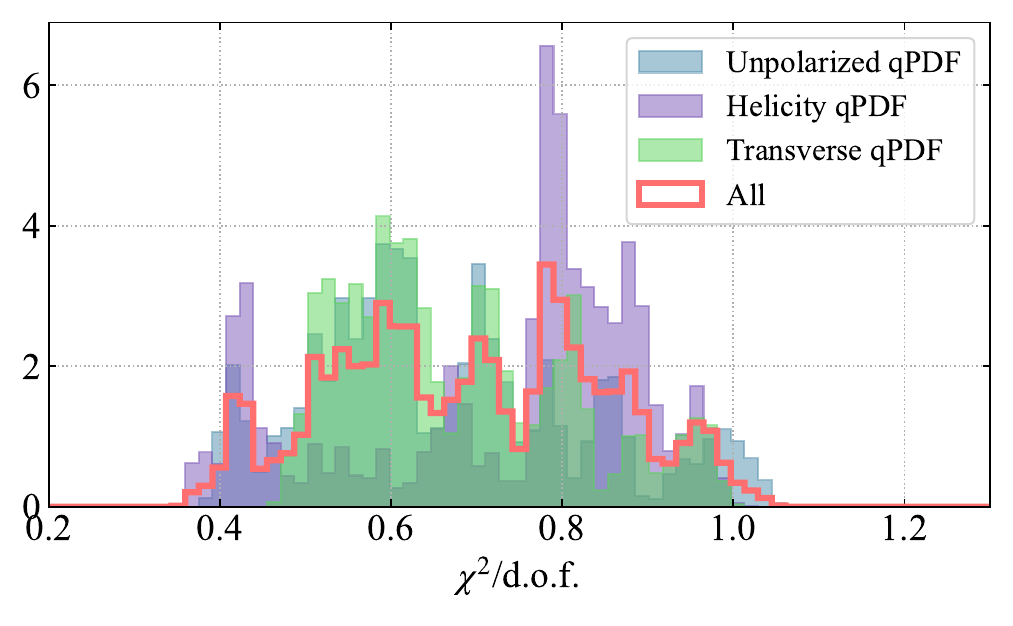}
    \includegraphics[width=0.49\linewidth]{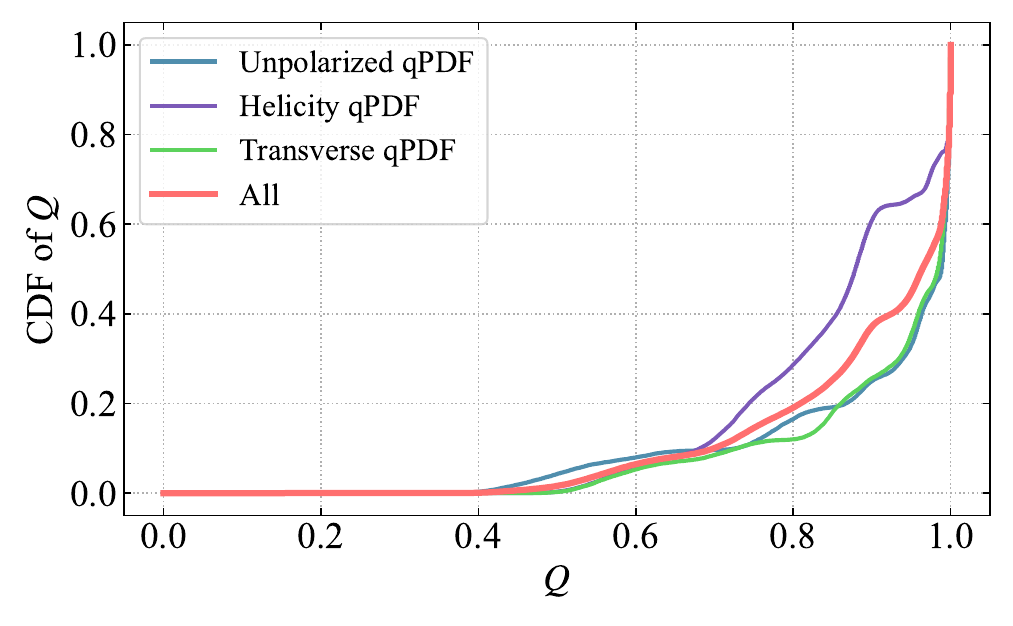}
    \caption{Statistical summary of the ground-state fits. The left panel shows the density distribution of $\chi^2/{\rm d.o.f.}$ for the unpolarized, helicity, and transversity quasi-PDFs (qPDF), as well as the combined distribution. The right panel displays the cumulative distribution function (CDF) of the corresponding p-values. }
    \label{fig:gsfit_stat}
\end{figure}

\begin{figure}[th!]
    \centering
    \includegraphics[width=0.32\linewidth]{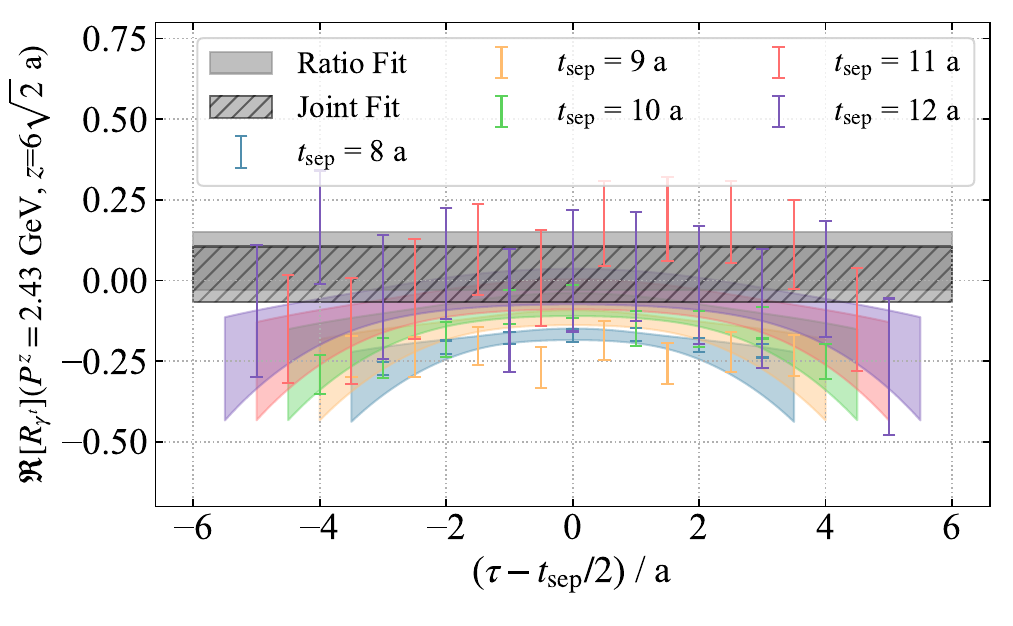}
    \includegraphics[width=0.32\linewidth]{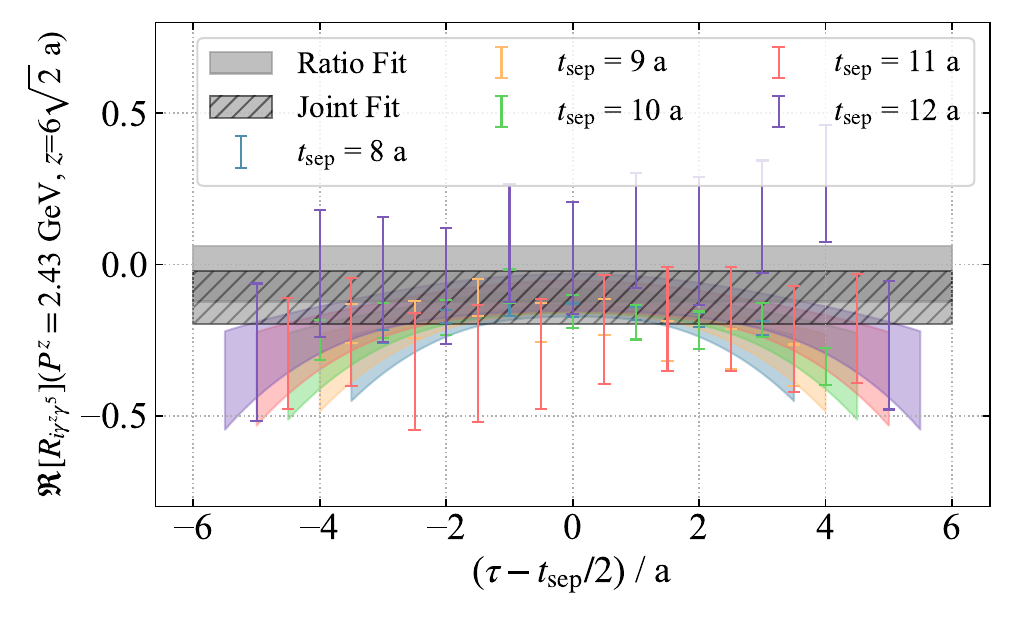}
    \includegraphics[width=0.32\linewidth]{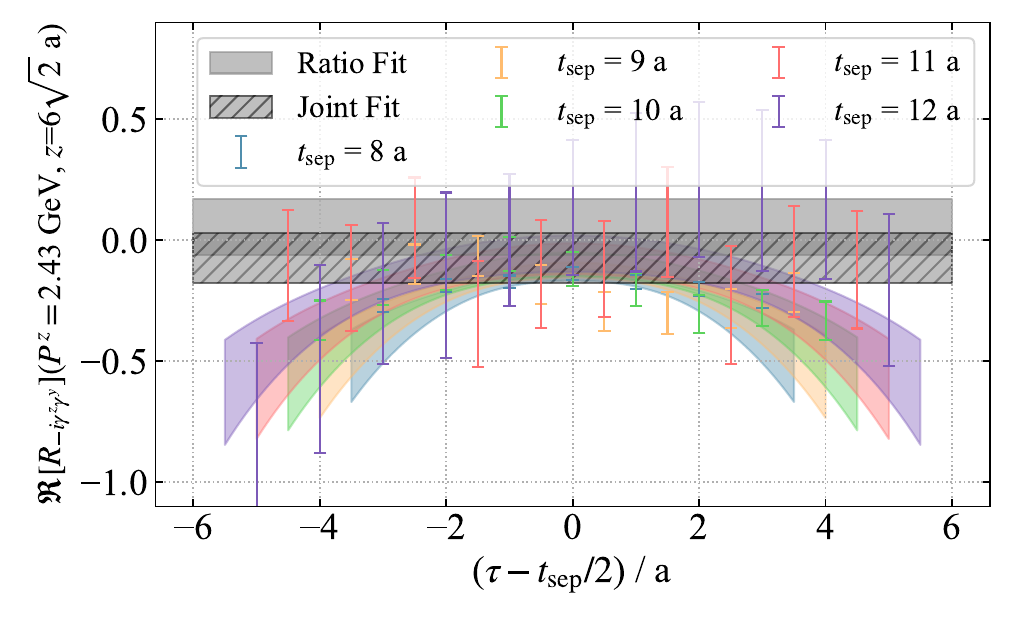}
    \includegraphics[width=0.32\linewidth]{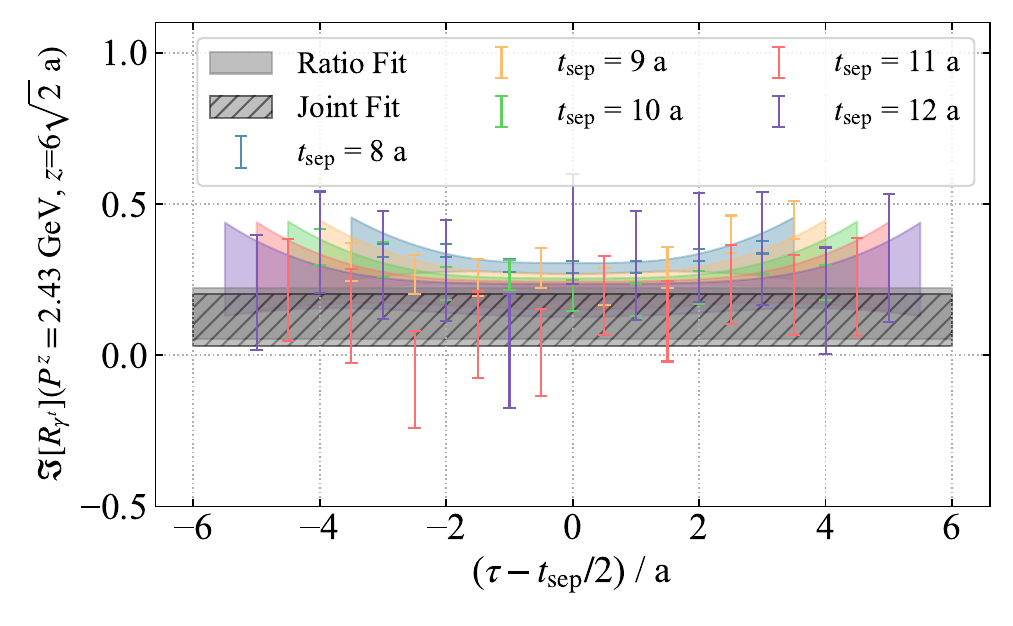}
    \includegraphics[width=0.32\linewidth]{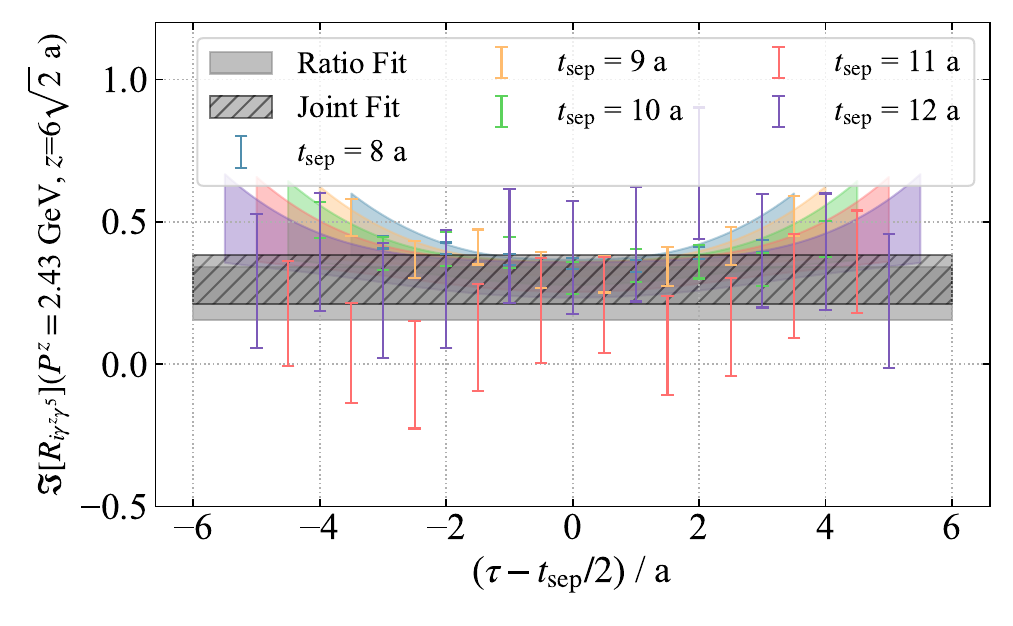}
    \includegraphics[width=0.32\linewidth]{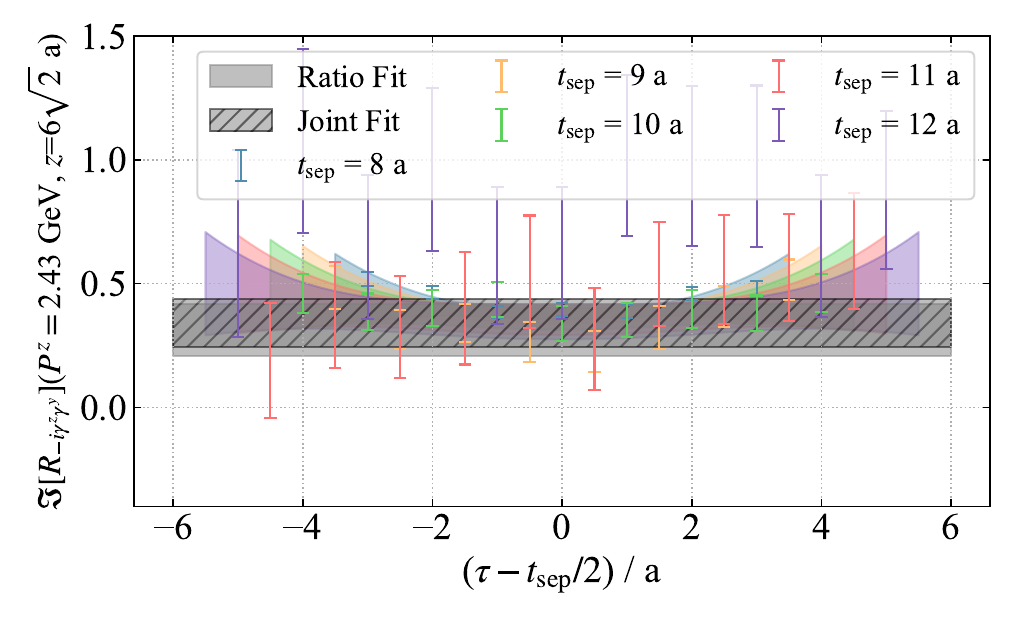}
    \caption{Examples of ground-state fits to the matrix elements of the unpolarized, helicity, and transversity quasi-PDFs at $z = 6\sqrt{2}~a$ and $P^z = 2.43$ GeV. The three columns from left to right correspond to the unpolarized, helicity, and transversity channels, respectively. The upper (lower) row shows the real (imaginary) parts of the matrix elements. The error bars denote the ratio data R. The colored bands represent the results of the joint fit over multiple source-sink separations, while the hatched gray bands indicate the joint-fit results and the solid gray bands show the corresponding ratio-only fit for comparison.}
    \label{fig:gsfit_examples}
\end{figure}

To evaluate the overall quality of the ground-state fits, we examine the density distribution of $\chi^2/\rm{d.o.f.}$ and the cumulative distribution function of the p-value for the quasi-PDF matrix elements, as shown in Fig.~\ref{fig:gsfit_stat}. The figures reveal that $\chi^2/\rm{d.o.f.}$ is predominantly distributed within the expected range, with p-values exceeding $0.05$ for the majority of fits, thereby suggesting a high level of overall quality. To provide a representative illustration, six examples of the joint fit applied to three quasi-PDFs are shown in Fig.~\ref{fig:gsfit_examples}. The three columns from left to right correspond to the ground-state fits of the matrix elements for the unpolarized, helicity, and transversity quasi-PDFs, respectively. The upper row shows the real parts of the matrix elements, while the lower row presents the corresponding imaginary parts. The error bars correspond to the data points of the ratio $R$, while the colored bands denote the result of the joint fit. The hatched gray bands show the joint-fit results, and the solid gray bands without hatching represent the corresponding ratio-only fit for comparison.

\begin{figure}[th!]
    \centering
    \includegraphics[width=0.32\linewidth]{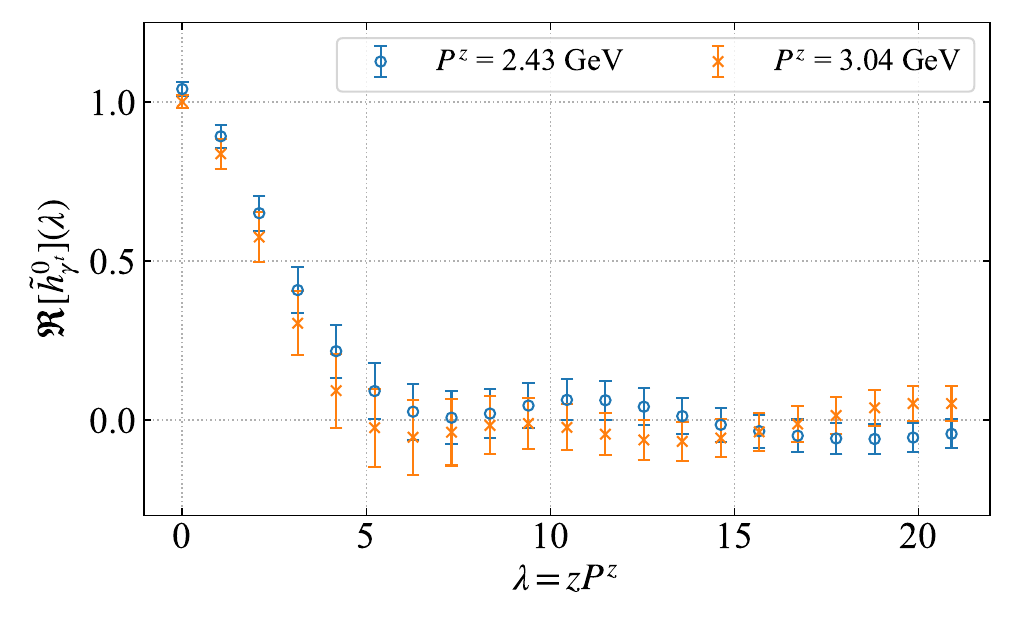}
    \includegraphics[width=0.32\linewidth]{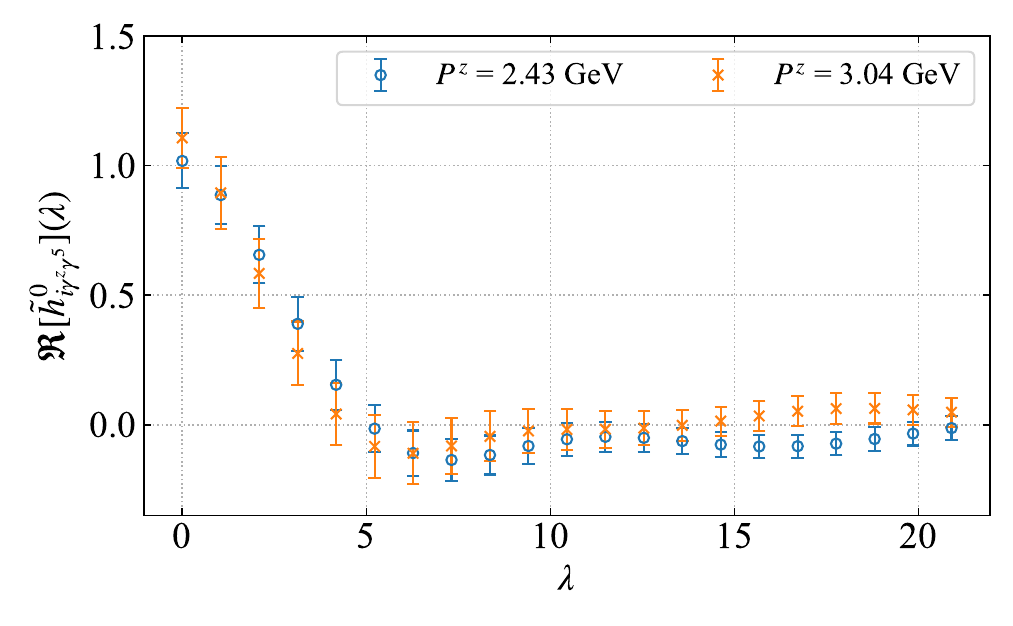}
    \includegraphics[width=0.32\linewidth]{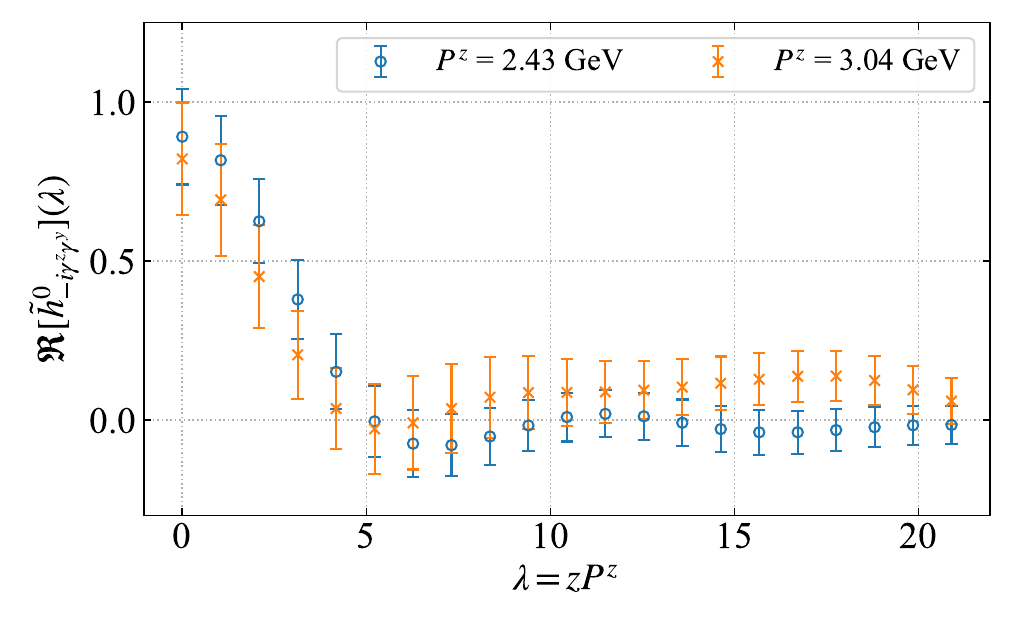}
    \includegraphics[width=0.32\linewidth]{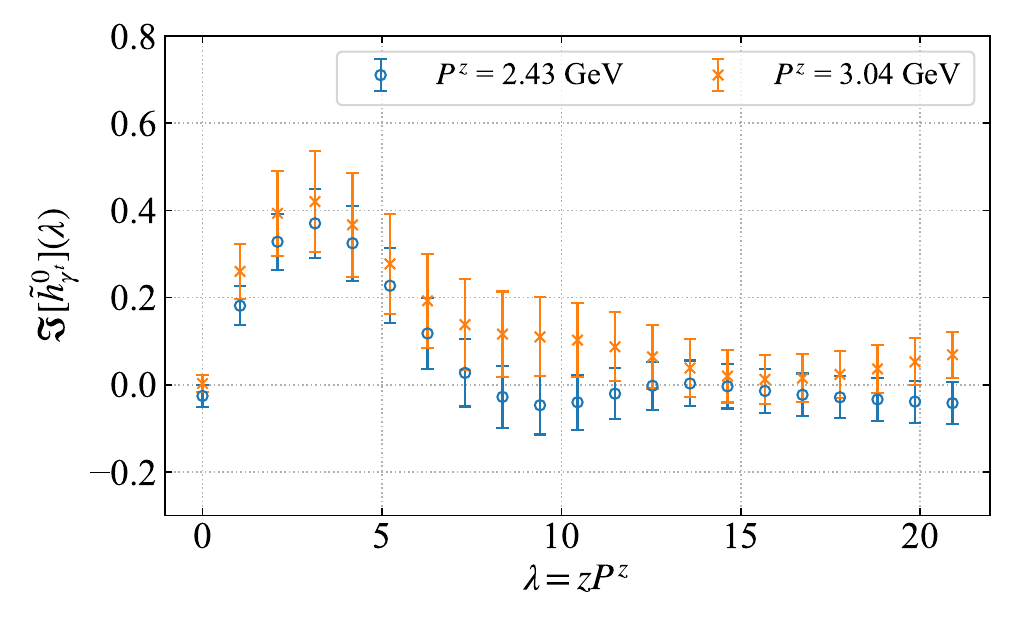}
    \includegraphics[width=0.32\linewidth]{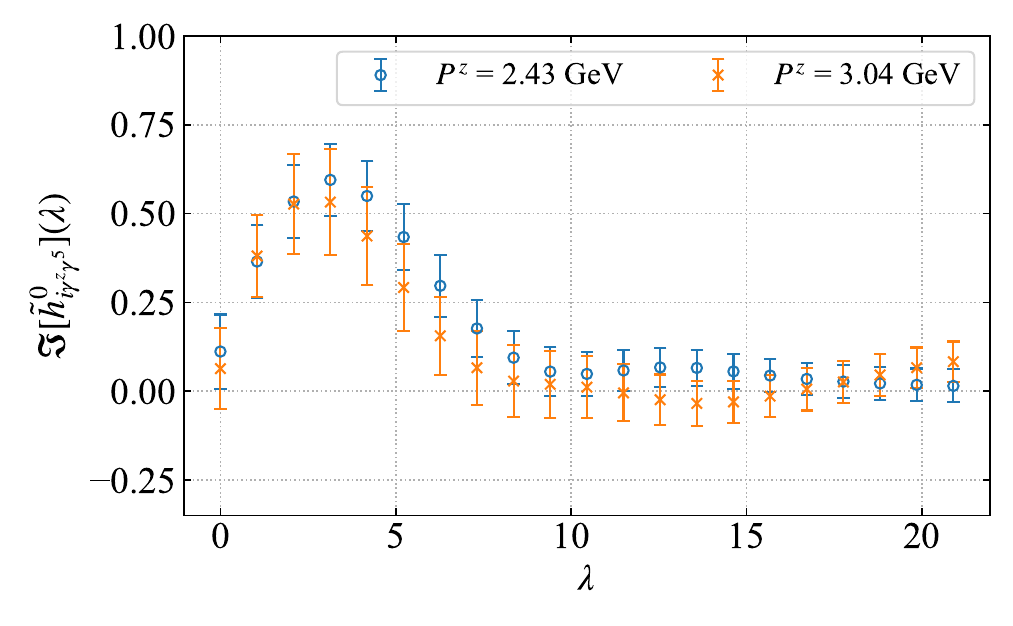}
    \includegraphics[width=0.32\linewidth]{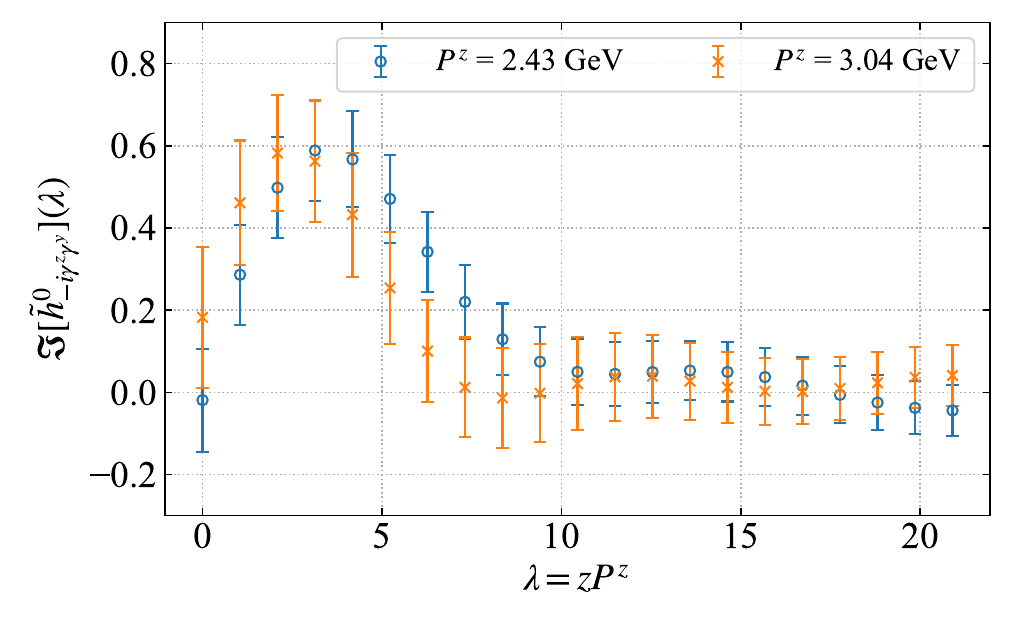}
    \caption{Bare matrix elements of the unpolarized, helicity, and transversity quasi-PDFs in coordinate space. The three columns from left to right correspond to the unpolarized, helicity, and transversity channels, respectively. The upper (lower) row shows the real (imaginary) parts of the matrix elements. Results at two hadron momenta, $P^z=2.43$ GeV and $3.04$ GeV, are shown for comparison.}
    \label{fig:bare_quasi_PDFs}
\end{figure}

Finally, the extracted bare matrix elements of the three quasi-PDFs in coordinate space are presented in Fig.~\ref{fig:bare_quasi_PDFs}, where results for the unpolarized, helicity, and transversity channels are shown from left to right. Both real and imaginary parts are displayed, and a direct comparison between the two momenta, $P^z = 2.43$ GeV and 3.04 GeV, is provided.

\subsection{Renormalization and extrapolation}

\begin{figure*}[th!]
    \centering
    \includegraphics[width=.32\linewidth]{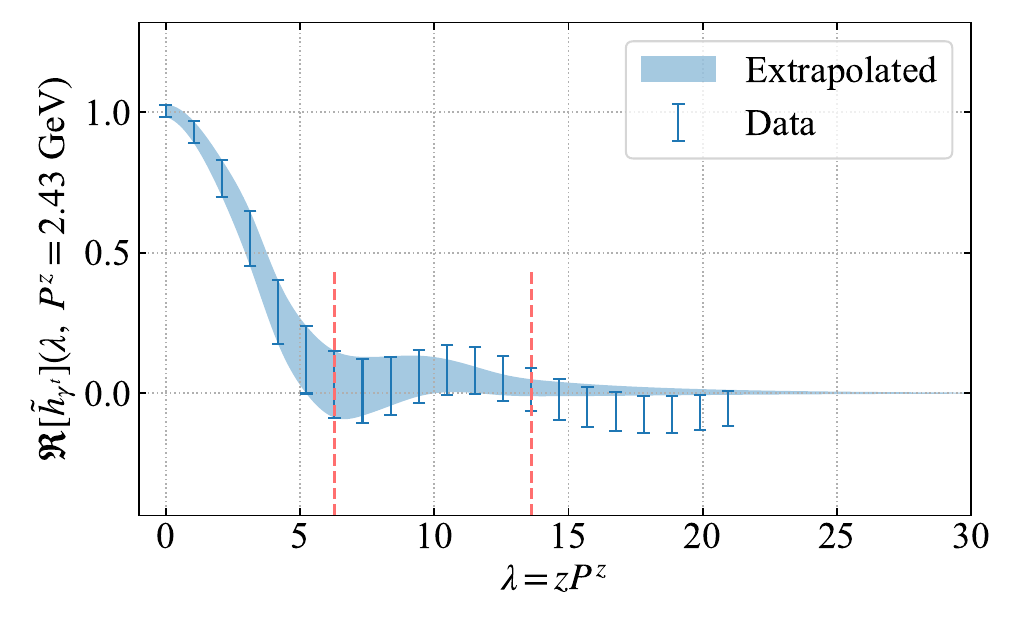}
    \includegraphics[width=.32\linewidth]{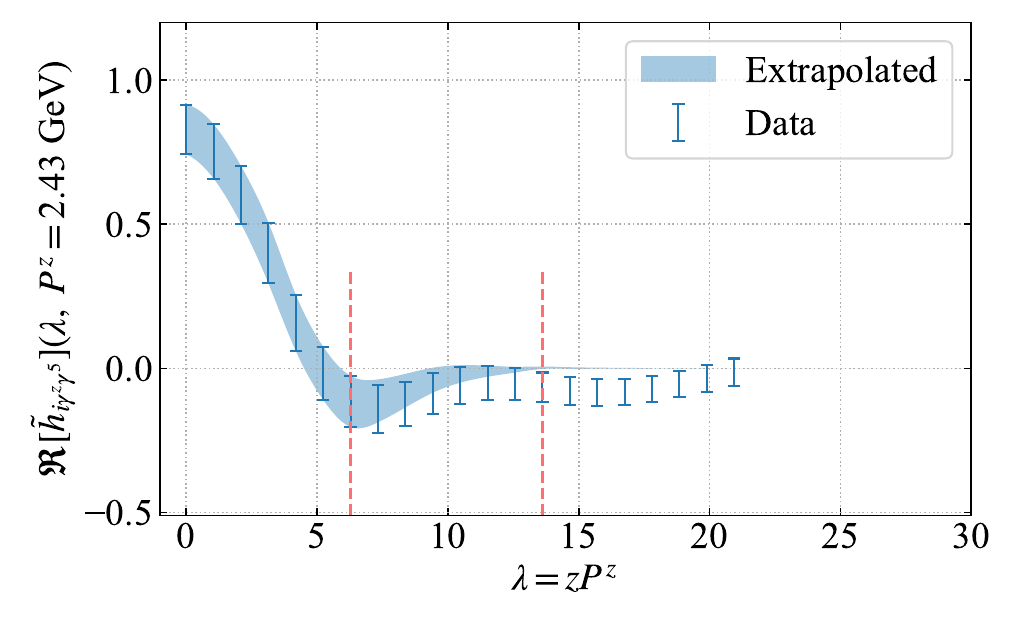}
    \includegraphics[width=.32\linewidth]{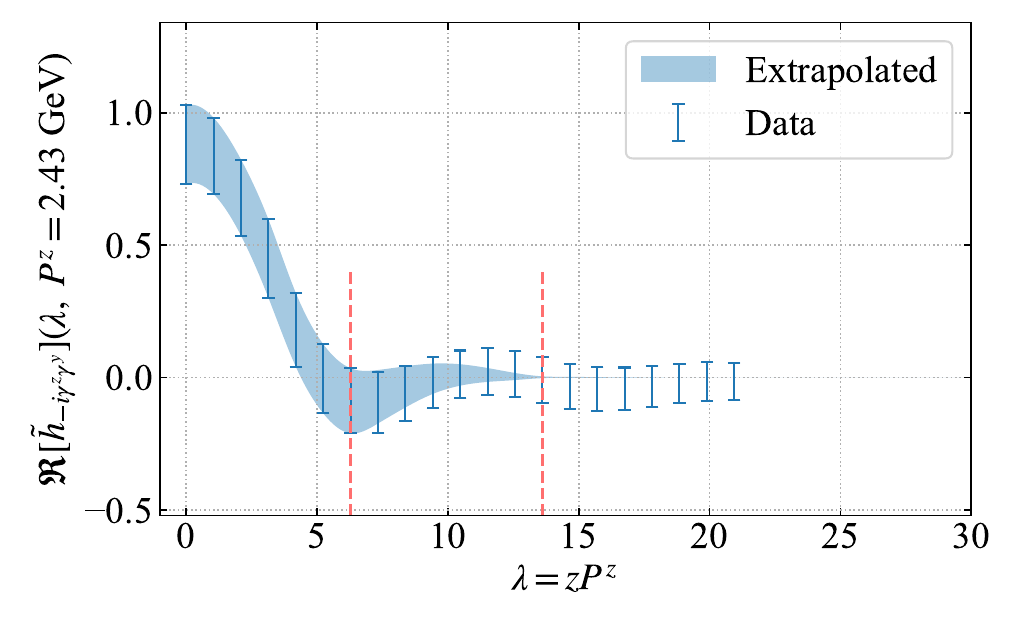}
    \includegraphics[width=.32\linewidth]{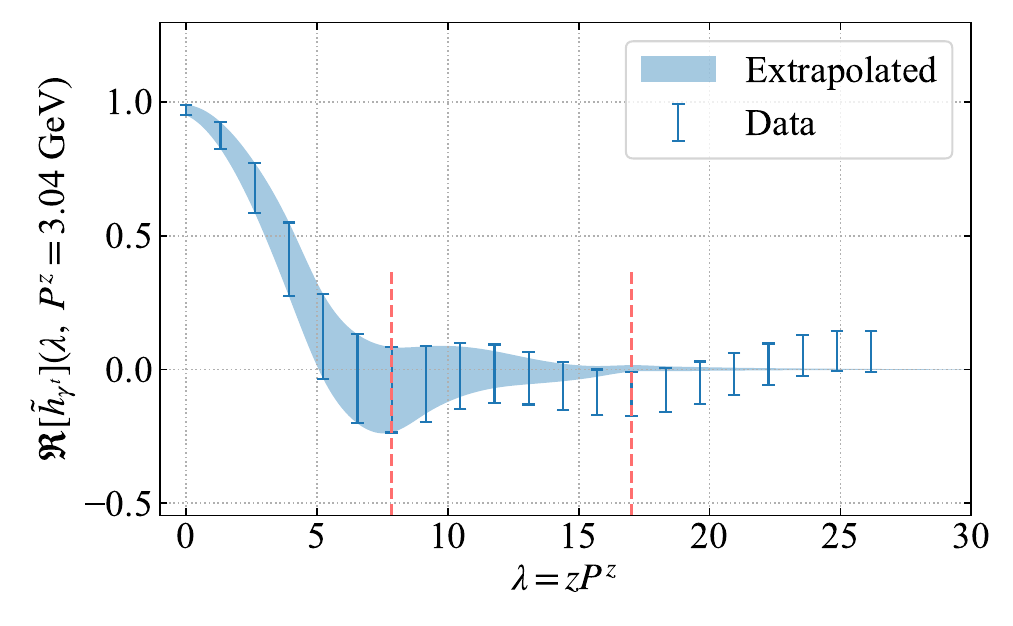}
    \includegraphics[width=.32\linewidth]{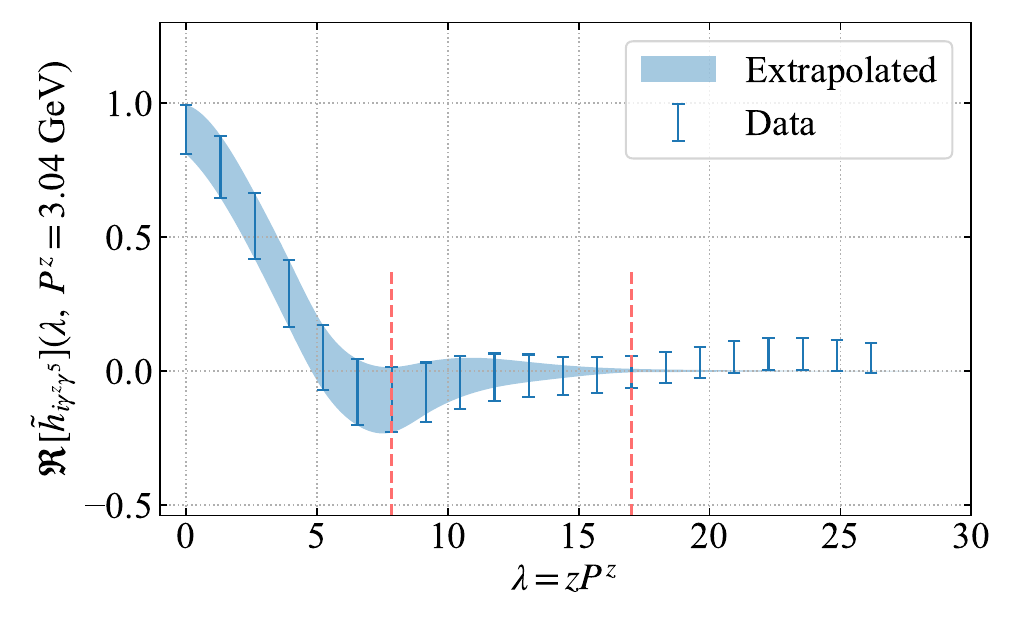}
    \includegraphics[width=.32\linewidth]{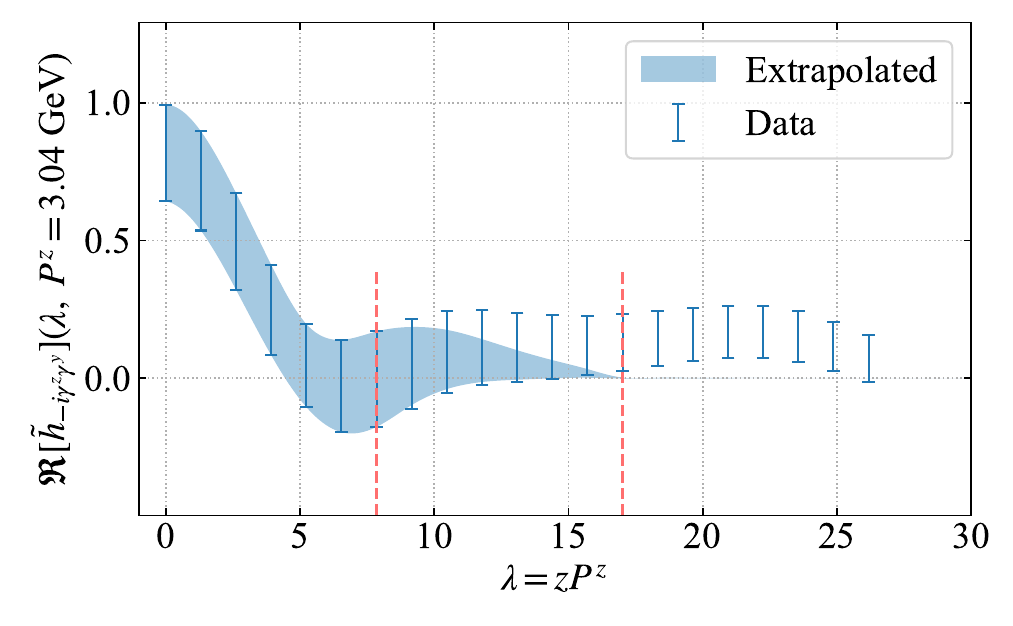}
    \caption{Extrapolation of the real parts of the renormalized quasi-PDF matrix elements in coordinate space as functions of $\lambda = z P^z$. The three columns from left to right correspond to the unpolarized, helicity, and transversity channels, while the upper (lower) row shows the results at $P^z = 2.43$ GeV ($P^z = 3.04$ GeV). The data points are shown with error bars, and the shaded bands represent the extrapolation using the asymptotic form. The two red dashed vertical lines indicate the sub-asymptotic region used in the fit.}
    \label{fig:extrapolation_real}
\end{figure*}

\begin{figure*}[th!]
    \centering
    \includegraphics[width=.32\linewidth]{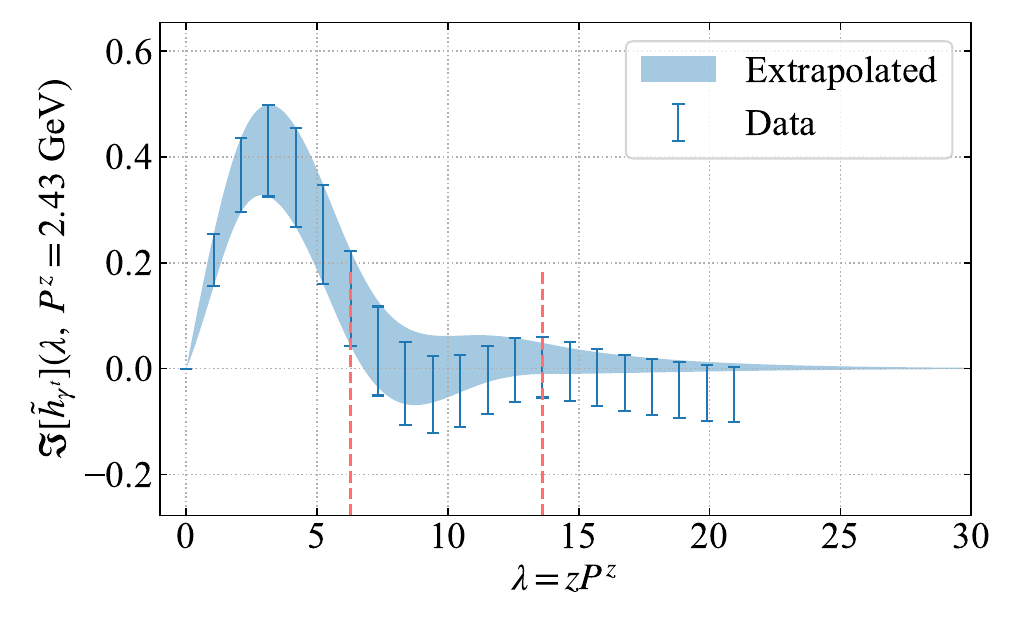}
    \includegraphics[width=.32\linewidth]{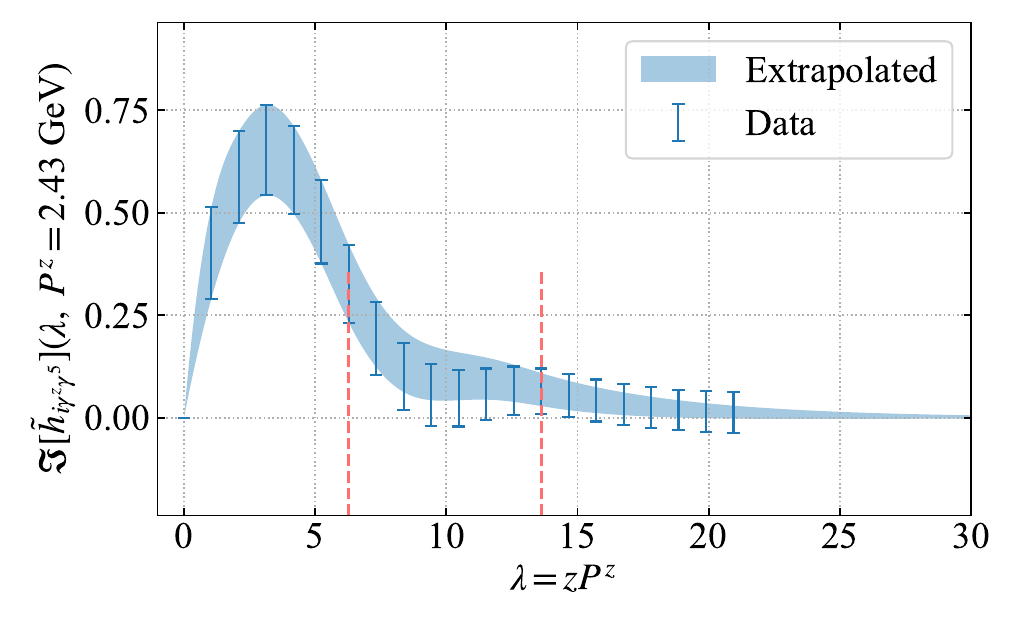}
    \includegraphics[width=.32\linewidth]{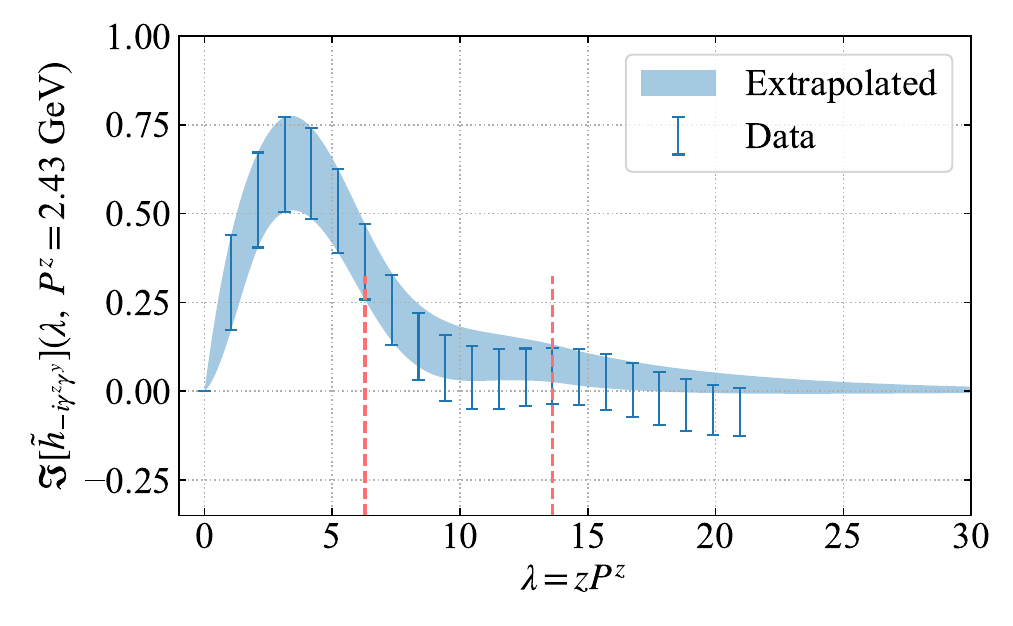}
    \includegraphics[width=.32\linewidth]{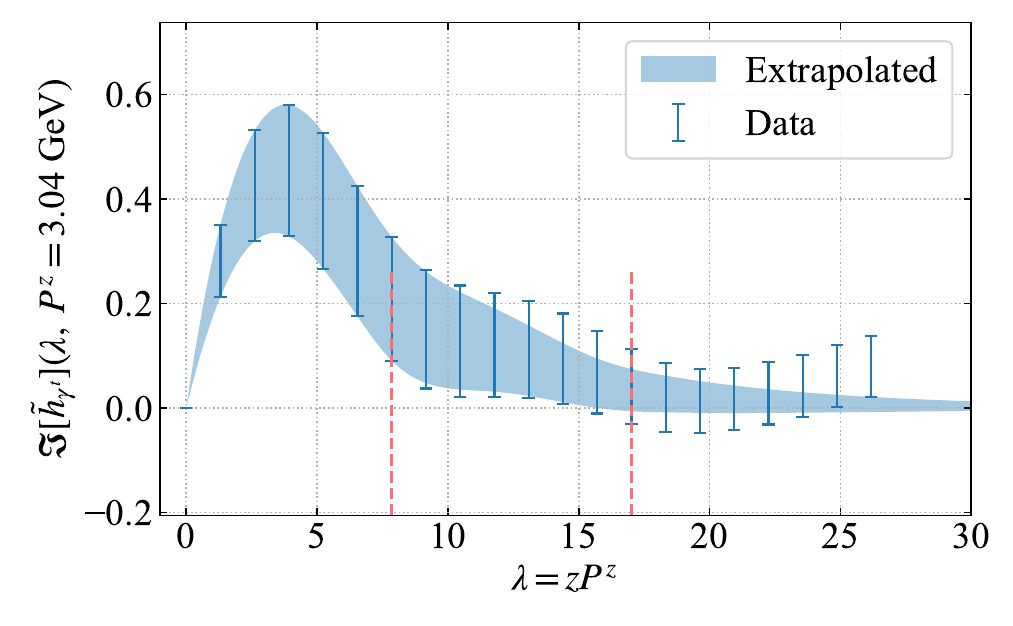}
    \includegraphics[width=.32\linewidth]{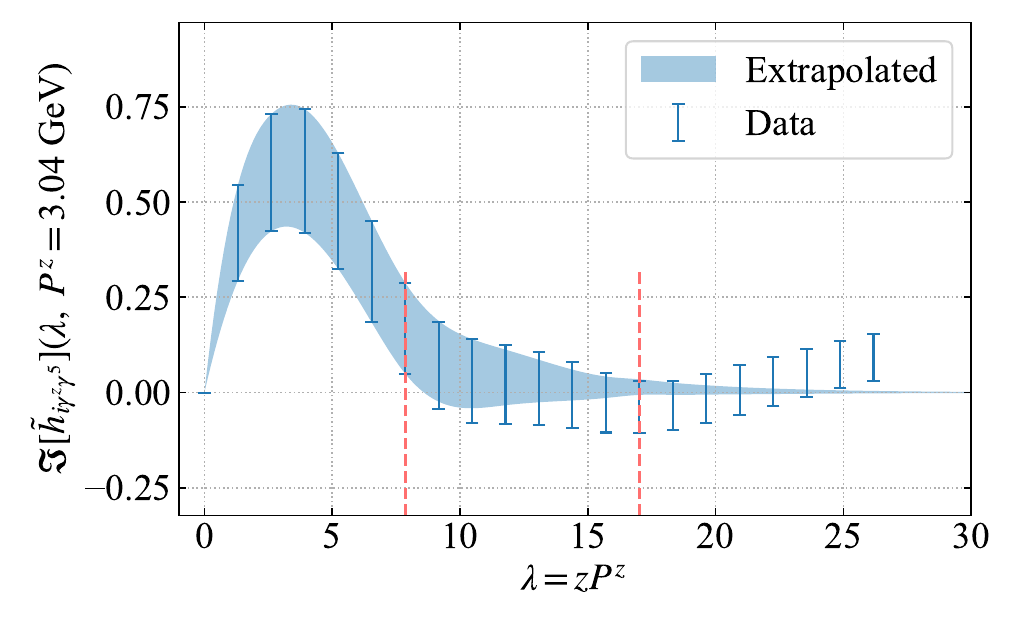}
    \includegraphics[width=.32\linewidth]{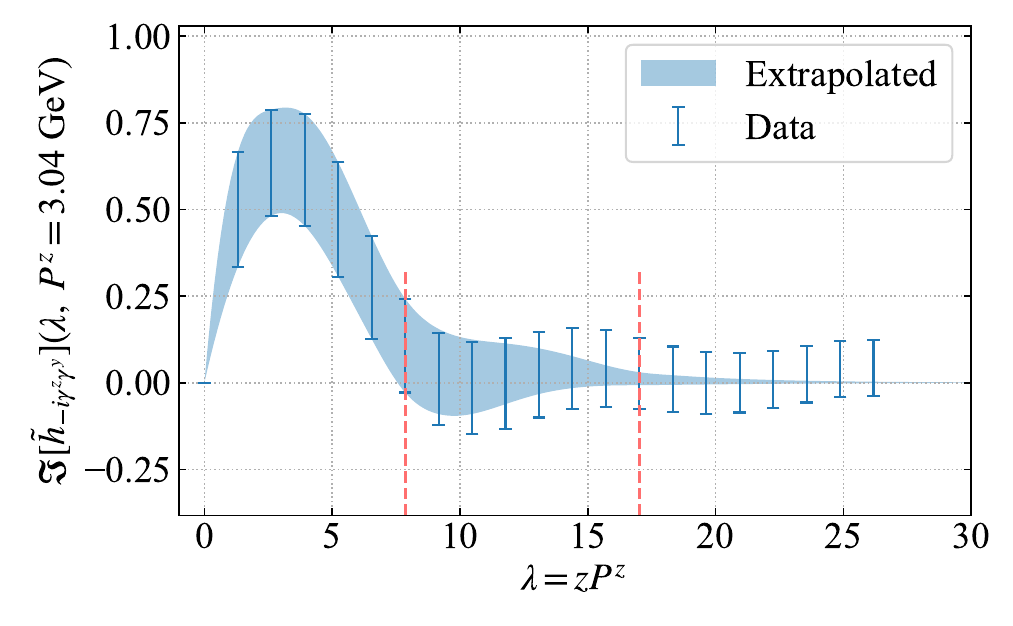}
    \caption{Same as Fig.~\ref{fig:extrapolation_real}, but for the imaginary parts of the renormalized quasi-PDF matrix elements. The local contribution at $z = 0$ is explicitly set to zero to preserve antisymmetry. The shaded bands show the extrapolated results obtained from the fit in the sub-asymptotic region indicated by the red dashed lines.}
    \label{fig:extrapolation_imag}
\end{figure*}
 
As discussed in Sec.~\ref{sec:renormalization}, we employ the hybrid renormalization scheme in \Eq{hybrid_renorm_no_norm} to renormalize the real part of the matrix elements. The separation point is chosen as $z_s = 3 \sqrt{2}~a$. The imaginary part of the matrix elements is instead renormalized in S\ripmom \ before matching to $\overline{\mathrm{MS}}$ as detailed in \Cref{ss:srip} such that $Z_\psi^{\overline{\rm MS}}(a=0.06~\mathrm{fm},~\mu= 2~\mathrm{GeV}) = 0.915(2)$. 

In addition, to preserve the antisymmetry of the imaginary part, its local contribution at $z=0$ is explicitly set to zero. The renormalized matrix elements in the coordinate space are presented in Figs.~\ref{fig:extrapolation_real} and \ref{fig:extrapolation_imag}. Due to the absence of normalization, the local point at $z = 0$ is particularly sensitive to discretization effects. Quantitatively, for the unpolarized case, we find
$N = 0.992(21)$ at $P^z = 2.43~\mathrm{GeV}$ and
$N = 1.032(21)$ at $P^z = 3.04~\mathrm{GeV}$,
both consistent with unity within uncertainties.
For the helicity case, however, larger deviations are observed,
with $N = 1.22(13)$ at $P^z = 2.43~\mathrm{GeV}$ and
$N = 1.12(12)$ at $P^z = 3.04~\mathrm{GeV}$,
indicating sizable lattice artifacts at $z=0$.
The transversity case shows a similar trend,
with $N = 1.14(19)$ at $P^z = 2.43~\mathrm{GeV}$ and
$N = 1.24(27)$ at $P^z = 3.04~\mathrm{GeV}$,
although with larger statistical uncertainties. As discussed in Sec.~\ref{sec:renormalization}, a significant deviation of $N$ from unity signals substantial discretization effects in the local matrix element at $z=0$. 
In such cases, applying a blind normalization would rescale the quasi-PDF by the same factor and propagate these lattice artifacts to all $z\neq 0$. 
This effect is particularly pronounced in the helicity channel and motivates our choice to abandon the normalization at $z=0$ in the hybrid renormalization scheme.
Although this treatment spoils the normalization of the quasi-PDF at finite lattice spacing, the associated effects vanish in the continuum limit and avoid contaminating the long-distance behavior that is essential for the PDF reconstruction.

The quasi-PDF matrix elements decay rapidly as a function of $\lambda = zP^z$, reaching approximately zero for $\lambda \gtrsim 5$. However, at large distances, while the values remain statistically consistent with zero, the statistical errors persist at a constant level, which will lead to non-physical fluctuations of the uncertainty band in a discrete Fourier transform. Due to the finite correlation length of spatial correlators in QCD~\cite{Gao:2021dbh}, the quasi-PDF matrix elements in the coordinate space are expected to exhibit exponential decay when the coordinate separation $z$ is large. Moreover, as demonstrated in Ref.~\cite{Gao:2021dbh}, the extracted quasi-distributions in momentum space within the moderate $x$ region remain largely insensitive to the choice of extrapolation strategy. In this work, we follow the asymptotic analysis strategy proposed in Ref.~\cite{Chen:2025cxr}. 
A sub-asymptotic form for {nucleon quasi-PDF} matrix elements in GI formulations has been derived in Ref.~\cite{Ji:2026vir}, 
where the large-$z$ behavior is shown to consist of an exponential decay governed by the lightest intermediate state, multiplied by a power-law dependence, {and a phase factor that depends on the sign of $\lambda$.}

While the exponential decay structure is robust, the precise power-law behavior depends on the details of the theoretical formulation. In particular, for the CG formalism, the corresponding algebraic dependence has not yet been fully understood from first principles. Motivated by these considerations, we adopt the following parameterization for the sub-asymptotic behavior in the CG,
{
\begin{align}
   \tilde{h} (\lambda) =
   \left[
      A_2 \cdot e^{i \mathrm{sign}(\lambda)\phi_2}
      + \frac{A'_2}{|\lambda|}
      \cdot e^{i \mathrm{sign}(\lambda)\phi'_2}
   \right]
   \frac{e^{-m |\lambda|}}{|\lambda|^n} ,
   \label{extrapolation_fcn}
\end{align}
}
where the overall factor {$1/|\lambda|^n$} is introduced to parametrize the unknown algebraic decay.
This form is used to fit the data in the sub-asymptotic region
$0.5~\mathrm{fm} \leq z \leq 1.1~\mathrm{fm}${, corresponding to $z/a \in [6,14] \sqrt{2}$},
as indicated by the two red dashed lines in
Figs.~\ref{fig:extrapolation_real} and \ref{fig:extrapolation_imag}.
In the fit function above, there are six parameters: $A_2$, $A'_2$, $\phi_2$, $\phi'_2$, $m$ and $n$, where $m$ is set to be larger than $0.1$ GeV / $P^z$~\cite{Gao:2021dbh}. To get a smooth change of the error band, the extrapolated results are given in the form as
 \begin{align}
     \tilde{h}^{\rm ext} = w \cdot \tilde{h}^{\mathrm{data}} + (1 - w) \cdot \tilde{h}^{\rm fit} ~,
    \label{eq:extrapolation_weight}
 \end{align}
where $w$ is a weight function that transitions linearly from 1 to 0 within the fit range, $\tilde{h}^{\mathrm{data}}$ is the renormalized quasi-distribution, and $\tilde{h}^{\rm fit}$ refers to the fit results of the asymptotic fit. 
{Variations of the fit ranges and choices of weight functions were tested and found to have only a minor impact on the quasi-PDFs in $x$ space within the current uncertainties. The systematic uncertainty associated with the fit-range variation is incorporated into the error budget presented in Appendix~\ref{app:error}.}
After applying this extrapolation, the uncertainty bands smoothly converge to zero, mitigating non-physical fluctuations. The model uncertainty of this asymptotic extrapolation is bounded from above and can be improved systematically~\cite{Gao:2021dbh,Chen:2025cxr}.

\section{Nucleon Quark PDFs}
\label{sec:pdfs}

\subsection{Unpolarized PDF}

\begin{figure}[th!]
    \centering
    \includegraphics[width=0.49\linewidth]{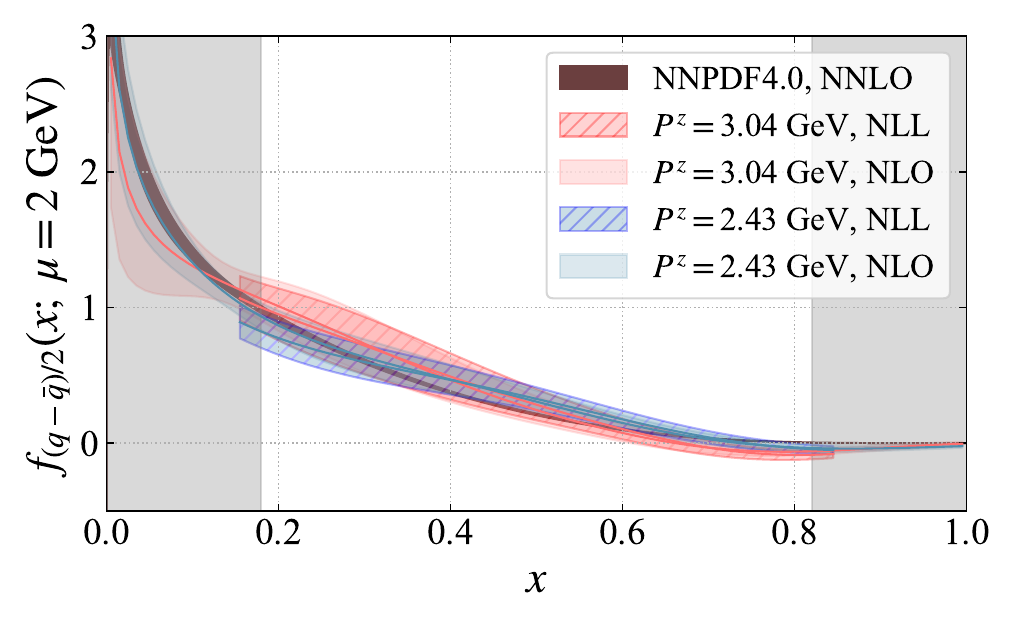}
    \includegraphics[width=0.49\linewidth]{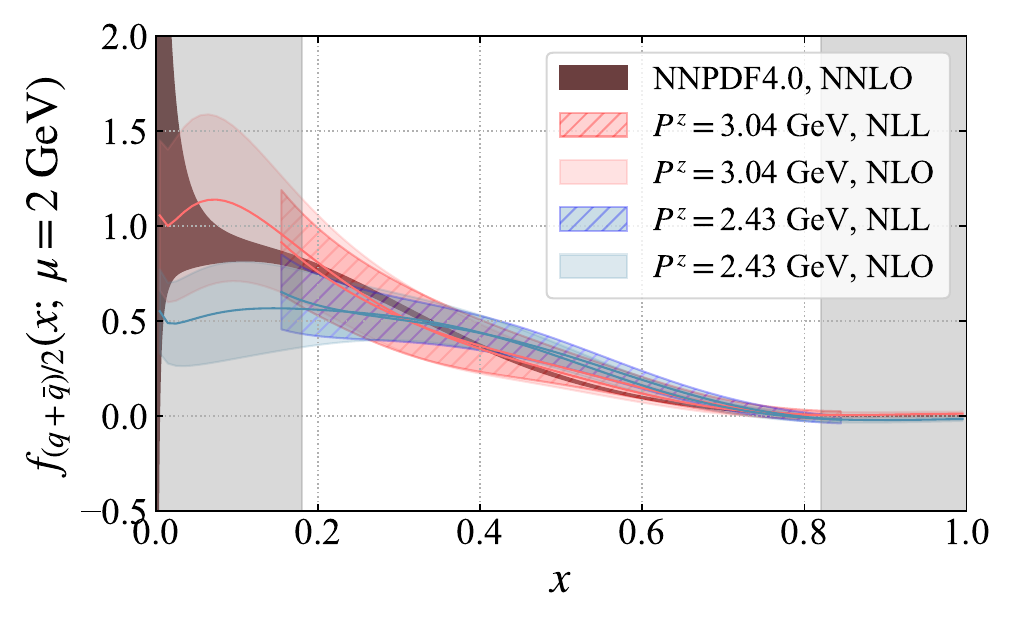}
    \caption{Unpolarized quark PDFs of the nucleon at momenta $P^z = 2.43$ and $3.04$~GeV, extracted from lattice QCD and compared with the NNLO phenomenological results from NNPDF4.0~\cite{NNPDF:2021njg}. The NLO and NLL labels indicate the perturbative accuracy of matching coefficients. The left and right panels show the valence contribution $(q-\bar q)/2$ and the full distribution $(q+\bar q)/2$, respectively. Results obtained using matching coefficients at NLO and NLL are shown to illustrate the perturbative convergence. The shaded gray regions ($x < 0.18$ or $x > 0.82$) denote kinematic domains where the strong coupling $\alpha_s(\mu = 2 x P^z)$ or $\alpha_s(\mu = 2 (1-x) P^z)$ at NLO exceeds 0.5, indicating a poor convergence of perturbation theory.}
    \label{fig:unpol_xdep_sep}
\end{figure}

As discussed above, the real and imaginary parts of the matrix elements are renormalized in different schemes, and their contributions correspond to the valence and full quark channels of the nucleon PDFs, respectively. Therefore, these two contributions can be independently compared with existing results in the literature. After performing the large-momentum expansion in \Eq{fact}, the matched results from the real and imaginary parts are shown in the left and right panels of Fig.~\ref{fig:unpol_xdep_sep}, respectively. The unpolarized nucleon PDFs at hadron momenta $P^z = 2.43$ and $3.04$~GeV are presented together with the NNLO phenomenological determination from NNPDF4.0~\cite{NNPDF:2021njg}. By employing matching coefficients at different perturbative accuracies, we further obtain results at next-to-leading order (NLO) as well as at next-to-leading logarithmic (NLL) accuracy with DGLAP evolution. {However, it should be noted that the lattice calculations in this work are performed on a single ensemble with a fixed lattice spacing and unphysical pion mass, without taking the physical limits such as the continuum and chiral limits. Therefore, the comparison with phenomenological results should be interpreted with caution. The same caveat also applies to the helicity and transversity PDFs presented below.} The shaded gray regions in Fig.~\ref{fig:unpol_xdep_sep} indicate kinematic domains where the strong coupling $\alpha_s(\mu = 2 x P^z)$ evaluated at NLO exceeds 0.5. In these regions, the perturbative expansion underlying the matching procedure is expected to have poor convergence, and the corresponding lattice results should therefore be interpreted with caution.

\begin{figure}[th!]
    \centering
    \includegraphics[width=0.49\linewidth]{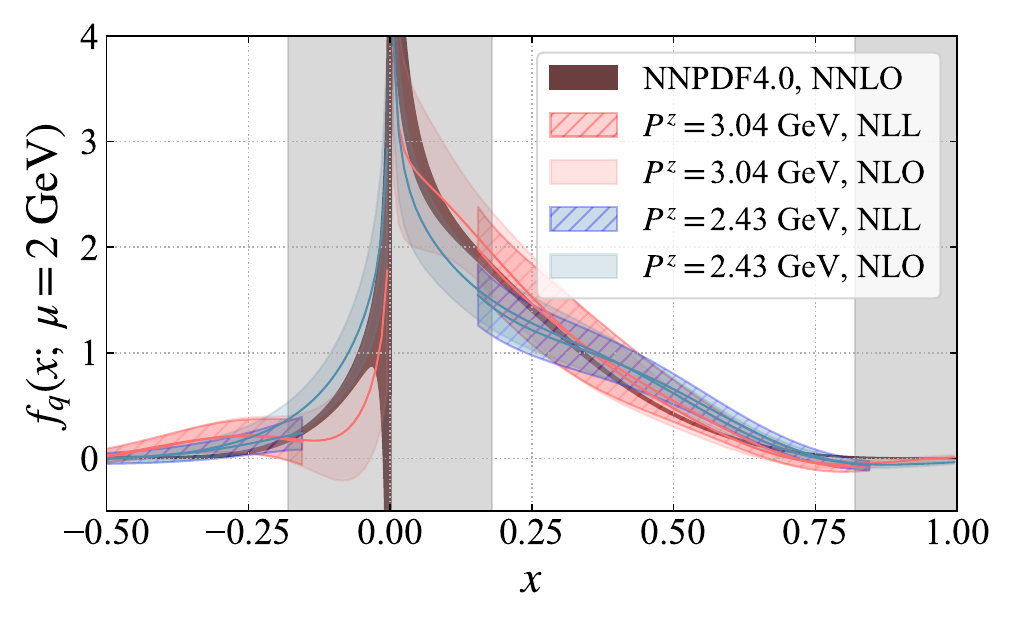}
    \caption{Unpolarized quark PDFs of the nucleon at momenta $P^z = 2.43$ and $P^z = 3.04$~GeV. The lattice results are shown together with the NNLO phenomenological determination from NNPDF4.0 for comparison.}
    \label{fig:unpol_complete}
\end{figure}

The results in Fig.~\ref{fig:unpol_xdep_sep} show that the lattice calculations at hadron momentum $P^z = 3.04$~GeV are in good agreement with the determination of NNPDFs4.0 for both the valence and the full components of the unpolarized PDFs. The comparison between the NLO and NLL cases further indicates a stable perturbative behavior and supports the convergence of the matching expansion. In the right panel, corresponding to $(q + \bar q)/2$, the difference between the two lattice momenta is more pronounced than in the $(q - \bar q)/2$ channel, which is likely caused by residual excited-state contamination in the extraction of the bare quasi-matrix elements. By combining the valence and sea contributions shown in Fig.~\ref{fig:unpol_xdep_sep}, we present the resulting unpolarized nucleon PDFs in Fig.~\ref{fig:unpol_complete}, together with the results of NNPDF4.0~\cite{NNPDF:2021njg} for comparison. 

\subsection{Helicity PDF}

\begin{figure}[th!]
    \centering
    \includegraphics[width=0.49\linewidth]{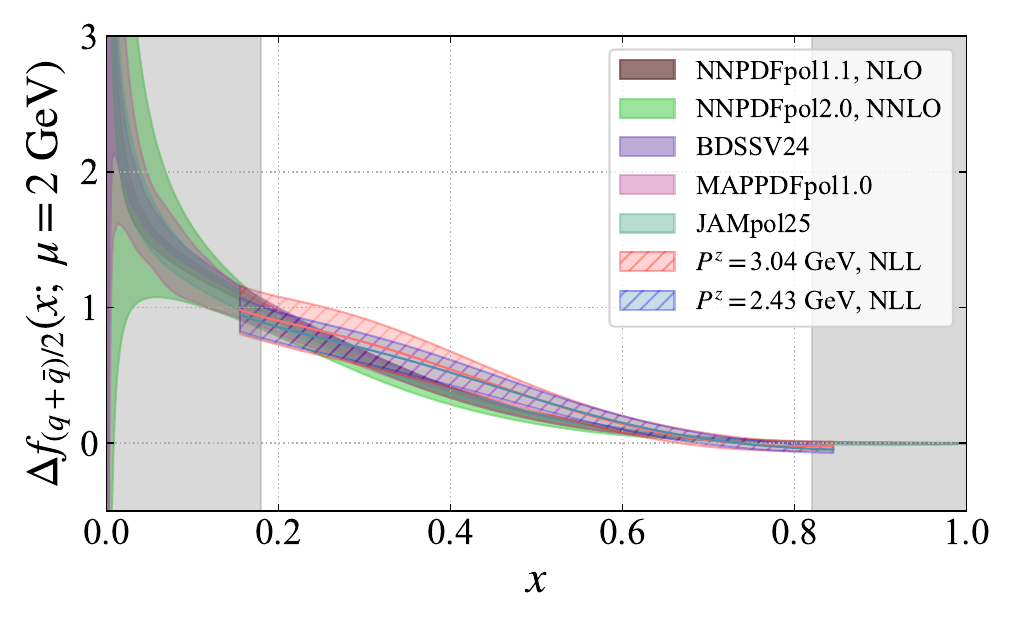}
    \includegraphics[width=0.49\linewidth]{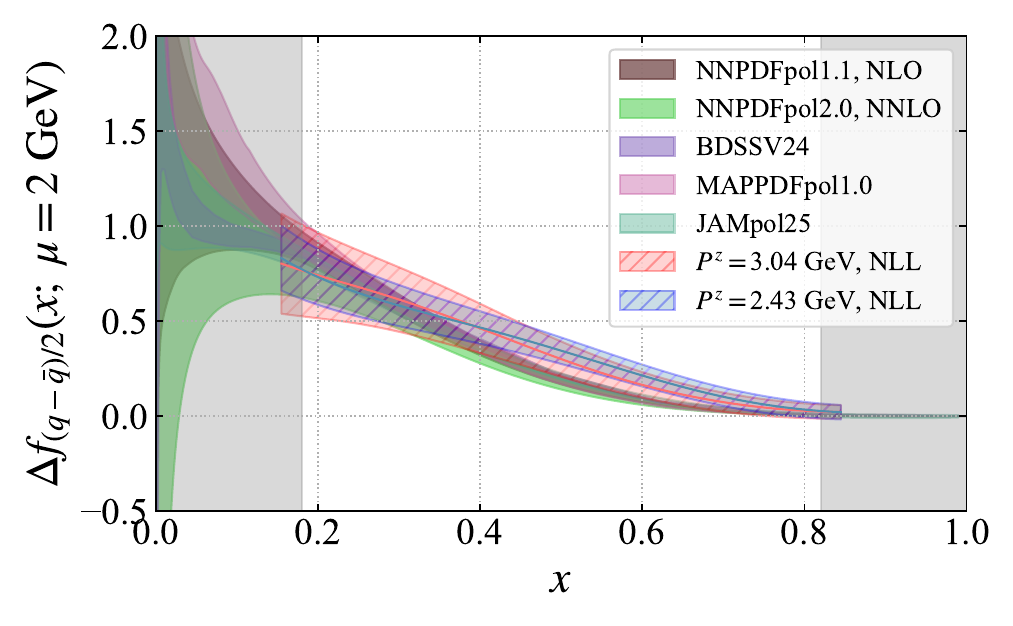}
    \caption{Helicity quark PDFs of the nucleon at momenta $P^z = 2.43$ and $P^z = 3.04$~GeV, compared with phenomenological determinations. The left and right panels show the full $(q+\bar q)/2$ and valence $(q-\bar q)/2$ components, respectively. The lattice results are presented at NLL accuracy, while the phenomenological bands include the NLO NNPDFpol1.1~\cite{Nocera:2014gqa}, NNLO NNPDFpol2.0~\cite{Cruz-Martinez:2025ahf}, BDSSV24~\cite{Borsa:2024mss}, MAPPDFpol1.0~\cite{Bertone:2024taw}, and JAMpol25~\cite{Cocuzza:2025qvf} analyses.}
    \label{fig:heli_xdep_sep}
\end{figure}

Similar to the unpolarized case, the full $(q + \bar q)/2$ and valence $(q - \bar q)/2$ components of the helicity PDFs are shown in the left and right panels of Fig.~\ref{fig:heli_xdep_sep}, respectively. Note that the full helicity distribution is obtained from the real part of the matrix elements, since the helicity quasi-distribution is even under charge conjugation. The helicity nucleon PDFs at hadron momenta $P^z = 2.43$ and $3.04$~GeV are compared with phenomenological determinations, including the NLO results from NNPDFpol1.1~\cite{Nocera:2014gqa}, the NNLO results from NNPDFpol2.0~\cite{Cruz-Martinez:2025ahf}, as well as the BDSSV24~\cite{Borsa:2024mss}, MAPPDFpol1.0~\cite{Bertone:2024taw}, and JAMpol25~\cite{Cocuzza:2025qvf} analyses. All phenomenological results are normalized by their respective axial charges $g_A$, while the valence $(q - \bar q)/2$ components of the lattice results are normalized using the $g_A$ value from JAMpol25~\cite{Cocuzza:2025qvf}. 

\begin{figure}[th!]
    \centering
    \includegraphics[width=0.49\linewidth]{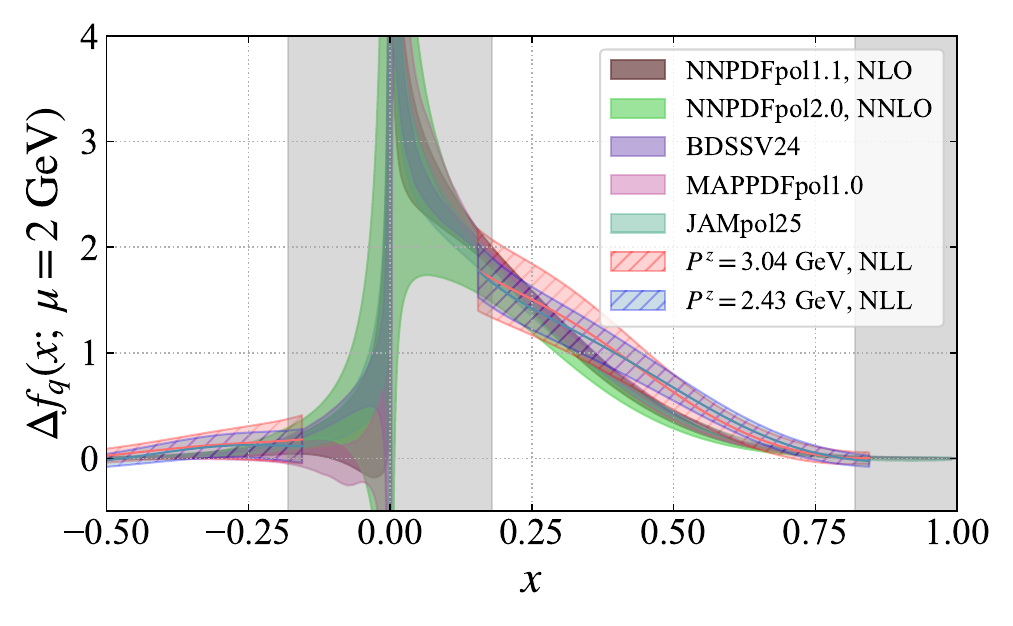}
    \caption{Helicity quark PDFs of the nucleon at momenta $P^z = 2.43$ and $P^z = 3.04$~GeV. The lattice-extracted distributions are shown together with phenomenological determinations for comparison.}
    \label{fig:heli_complete}
\end{figure}

From the plots in Fig.~\ref{fig:heli_xdep_sep}, we find that our results at $P^z = 3.04$ GeV are consistent with the phenomenological ones within one standard deviation for both the full and valence parts. Moreover, as in the unpolarized case, the discrepancy between the two momenta is more pronounced in the contribution from the imaginary part of the matrix elements. Combining the full and valence parts, the helicity nucleon PDFs are presented in Fig.~\ref{fig:heli_complete}.

\subsection{Transversity PDF}

\begin{figure}[th!]
    \centering
    \includegraphics[width=0.49\linewidth]{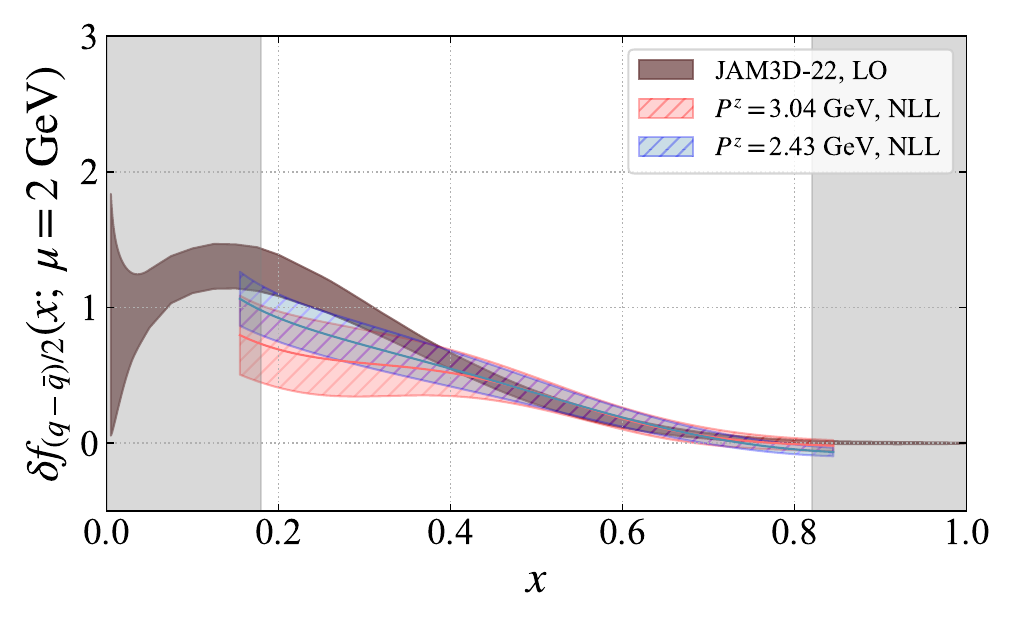}
    \includegraphics[width=0.49\linewidth]{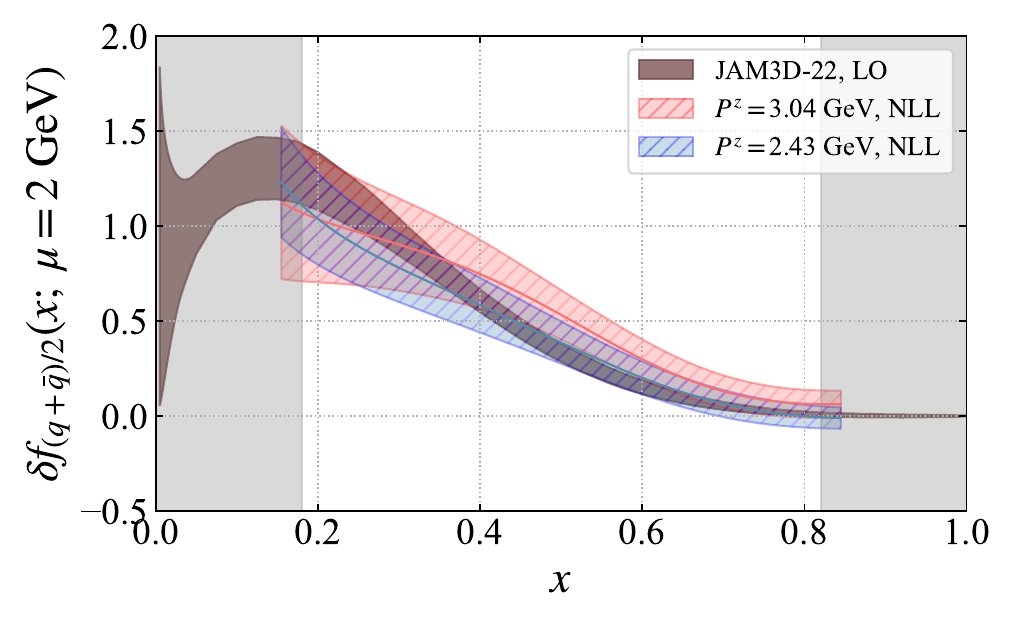}
    \caption{Transversity quark PDFs of the nucleon at momenta $P^z = 2.43$ and $P^z = 3.04$~GeV, compared with phenomenological and lattice determinations. The left and right panels show the valence $(q-\bar q)/2$ and full $(q+\bar q)/2$ components, respectively. The lattice results at NLL accuracy are compared with the leading-order phenomenological extraction from JAM3D-22~\cite{Gamberg:2022kdb}.}
    \label{fig:trans_xdep_sep}
\end{figure}

In analogy with the unpolarized and helicity cases, we also present our results for the transversity PDFs. The valence $(q - \bar q)/2$ and full $(q + \bar q)/2$ contributions are shown in the left and right panels of Fig.~\ref{fig:trans_xdep_sep}, respectively. The transversity nucleon PDFs at hadron momenta $P^z = 2.43$ and $3.04$~GeV are compared with the phenomenological extraction from JAM3D-22~\cite{Gamberg:2022kdb} at leading order (LO). The JAM3D-22 distributions are normalized by their own tensor charge $g_T$, and the full $(q + \bar q)/2$ contribution of the lattice results are normalized using the same value to ensure a consistent comparison. 

\begin{figure}[th!]
    \centering
    \includegraphics[width=0.49\linewidth]{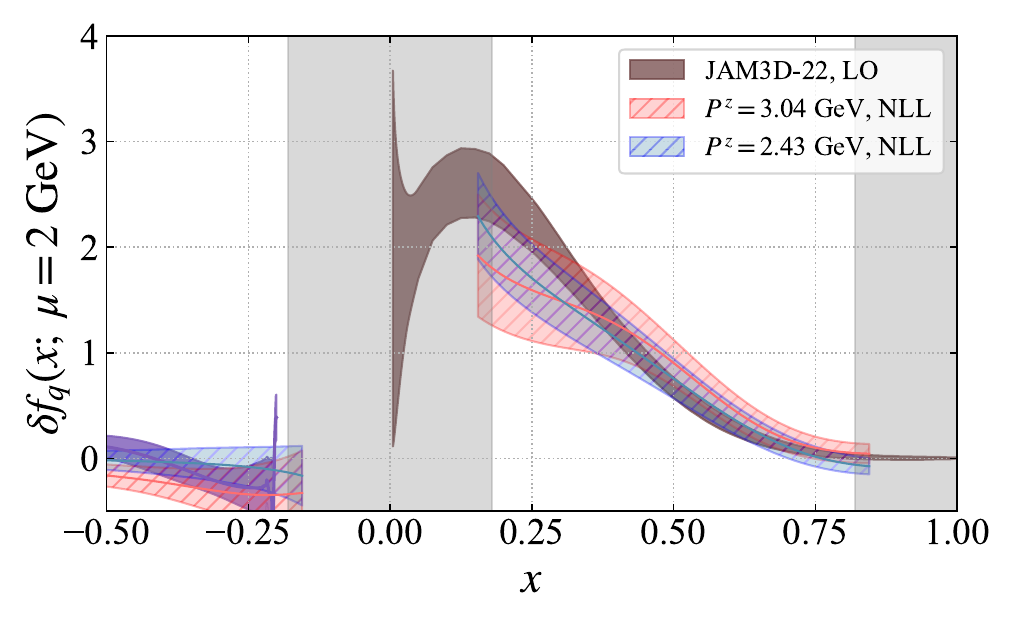}
    \caption{Transversity quark PDFs of the nucleon at momenta $P^z = 2.43$ and $P^z = 3.04$~GeV, shown together with phenomenological results for comparison. The lattice-extracted PDFs are compared with the JAM3D-22 extraction at LO.}
    \label{fig:trans_complete}
\end{figure}

As illustrated in Fig.~\ref{fig:trans_xdep_sep}, we observe that our lattice results show good agreement with the JAM3D-22 phenomenological extractions in the moderate-$x$ region. The level of consistency is even more evident in the comparison of the full part of the transversity PDFs. Similar to the unpolarized and helicity cases, the discrepancy between the two momenta is more pronounced in the contribution from the imaginary part of the matrix elements, underlining once again the need to carefully control excited-state contamination in this sector. The combined valence and full parts of the transversity nucleon PDFs are presented in Fig.~\ref{fig:trans_complete}. After all, since the phenomenological extraction of the transversity PDF still carries substantial uncertainties, caution is required in our comparison.

\section{Systematics}
\label{sec:sys}

In this section, we investigate different renormalization schemes and power corrections within the CG method. Besides, we also compare the CG results in this work to the PDFs calculated with the GI approach in the literature in Appendix~\ref{sec:gi_compare}. 

\subsection{Renormalization schemes}

\begin{figure}[th!]
    \centering
    \includegraphics[width=0.32\linewidth]{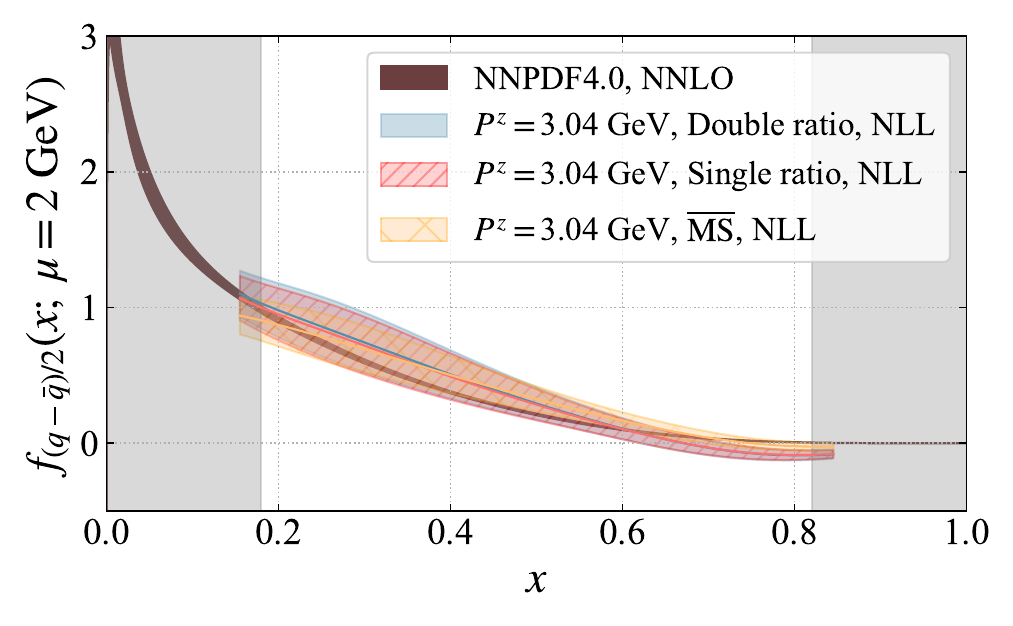}
    \includegraphics[width=0.32\linewidth]{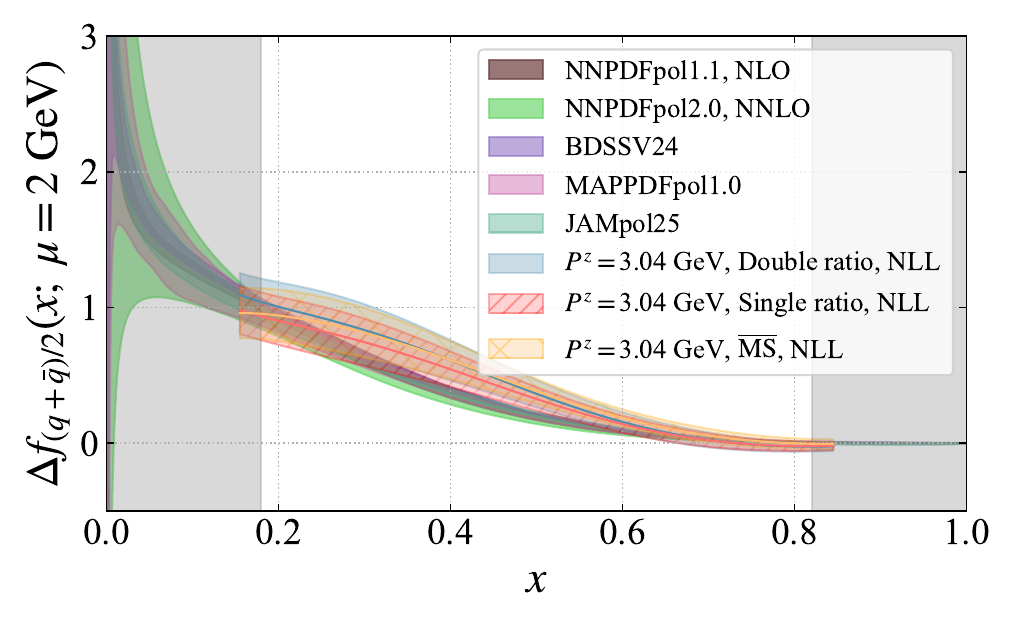}
    \includegraphics[width=0.32\linewidth]{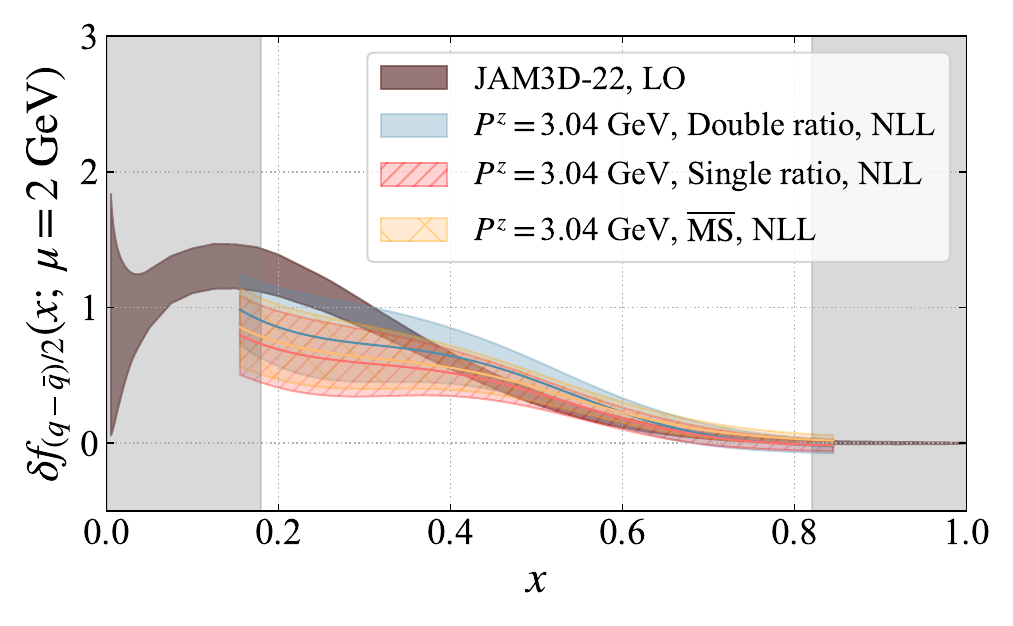}
    \caption{Comparison of different renormalization schemes for the contributions from the real part of the matrix elements of unpolarized (left), helicity (middle), and transversity (right) quasi-PDFs at $P^z = 3.04$ GeV. The results are obtained using the $\overline{\mathrm{MS}}$ scheme and the hybrid scheme with both double- and single-ratio methods. Note that the NLL labels indicate the perturbative accuracy of matching coefficients. }
    \label{fig:PDF_scheme_real}
\end{figure}

\begin{figure}[th!]
    \centering
    \includegraphics[width=0.32\linewidth]{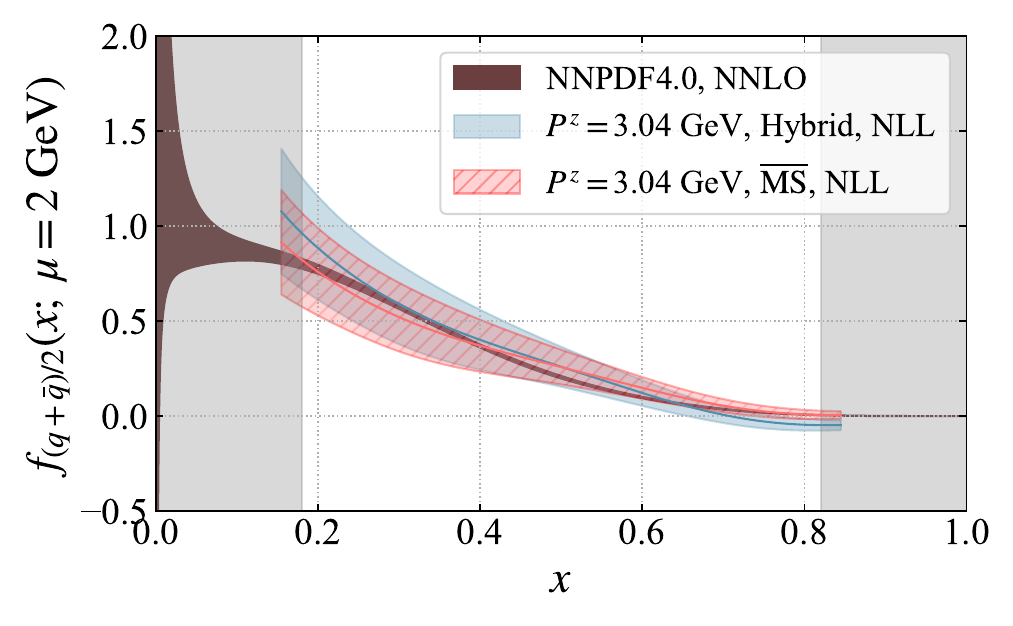}
    \includegraphics[width=0.32\linewidth]{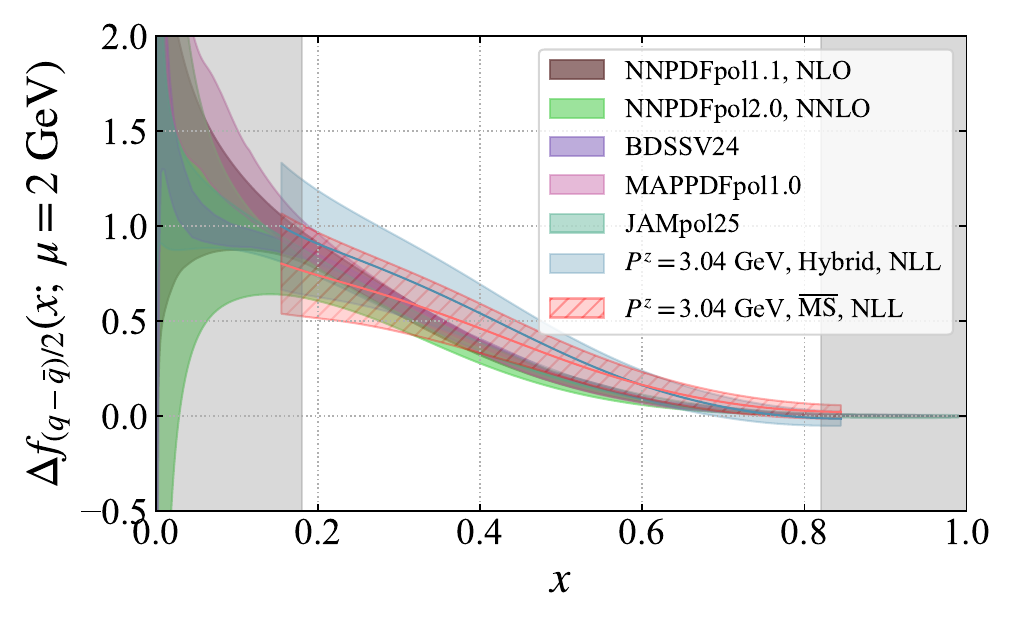}
    \includegraphics[width=0.32\linewidth]{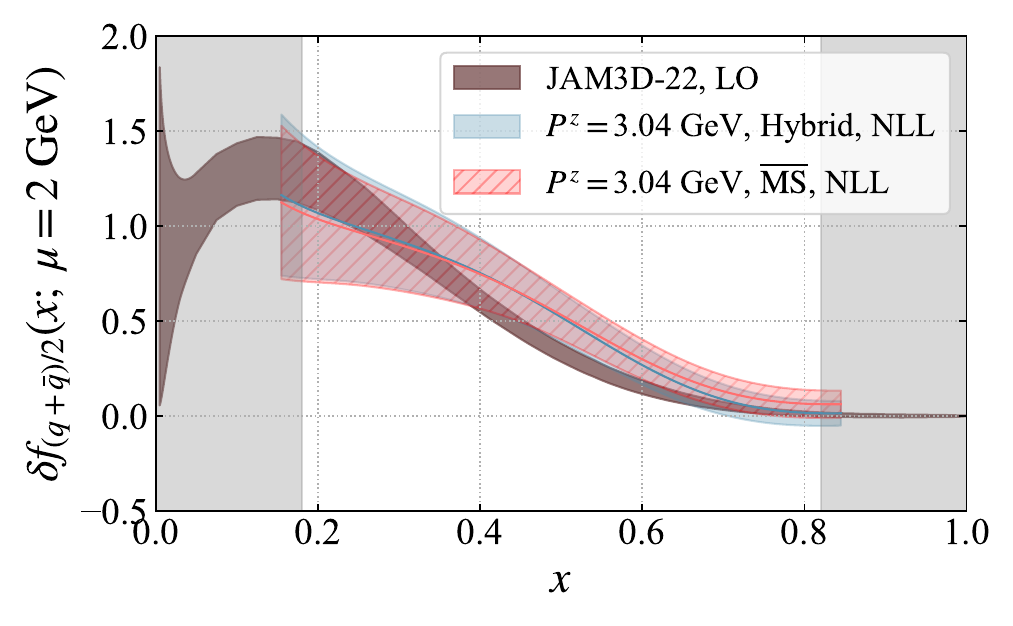}
    \caption{Comparison of different renormalization schemes for the contributions from the imaginary part of the matrix elements of unpolarized (left), helicity (middle), and transversity (right) quasi-PDFs at $P^z = 3.04$ GeV. The results are obtained using the renormalization constant $Z_\psi^{\overline{\mathrm{MS}}}$ at LL accuracy ($\overline{\mathrm{MS}}$) and the hybrid scheme with the single ratio (Hybrid).}
    \label{fig:PDF_scheme_imag}
\end{figure}

In this work, we propose using different renormalization schemes for the real and imaginary parts of the quasi-PDF matrix elements. To illustrate the comparison between these schemes, we take the results at $P^z = 3.04$ GeV as an example. The contributions from the real part of the matrix elements are extracted using the hybrid scheme with both double- (as shown in \Eq{hybrid_renorm_norm}) and single-ratio (as shown in \Eq{hybrid_renorm_no_norm}) methods, as presented in Fig.~\ref{fig:PDF_scheme_real}. For the imaginary part, the contributions are obtained using the renormalization constant $Z_\psi^{\overline{\mathrm{MS}}}$ (described in \Cref{ss:srip}) and the hybrid scheme with the single ratio, with the results presented in 
Fig.~\ref{fig:PDF_scheme_imag}.

In Fig.~\ref{fig:PDF_scheme_real}, the difference between the single- and double-ratio schemes is negligible for the unpolarized PDFs, while it becomes more pronounced in the helicity and transversity cases. This larger difference originates from the deviation of $N$ in \Eq{hybrid_renorm_norm} from unity, which is due to the lattice artifacts at $z=0$. In general, we find that the results obtained using the $\overline{\mathrm{MS}}$ scheme are consistent with those of the hybrid scheme with both single- and double-ratio methods.
In the helicity case, while we observe that the single-ratio scheme yields results that are closer to the phenomenological extractions, the large deviation of $N$ from unity suggests that further study of the lattice artifacts is necessary. Looking ahead, these issues could be systematically improved through calculations at multiple lattice spacings and with higher statistics for correlators at large time separations. In Fig.~\ref{fig:PDF_scheme_imag}, we observe that in the helicity case, the results obtained in the $\overline{\mathrm{MS}}$ scheme have a better consistency with the existing determinations than those from the hybrid scheme. {The systematic uncertainty associated with the renormalization schemes is incorporated into the error budget presented in Appendix~\ref{app:error}.}
However, since discretization effects and excited-state contamination are not disentangled in the present analysis, a more detailed and quantitative comparison between different renormalization schemes is left for future studies.

\subsection{Power corrections}
\label{ss:power_corr}

\begin{figure}[th!]
    \centering
    \includegraphics[width=0.49\linewidth]{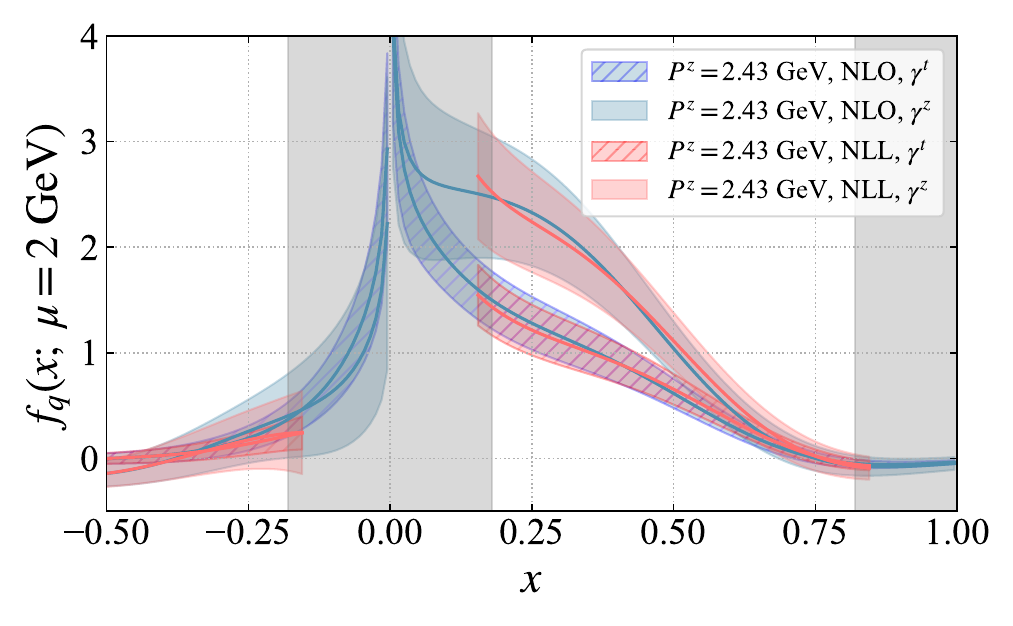}
    \includegraphics[width=0.49\linewidth]{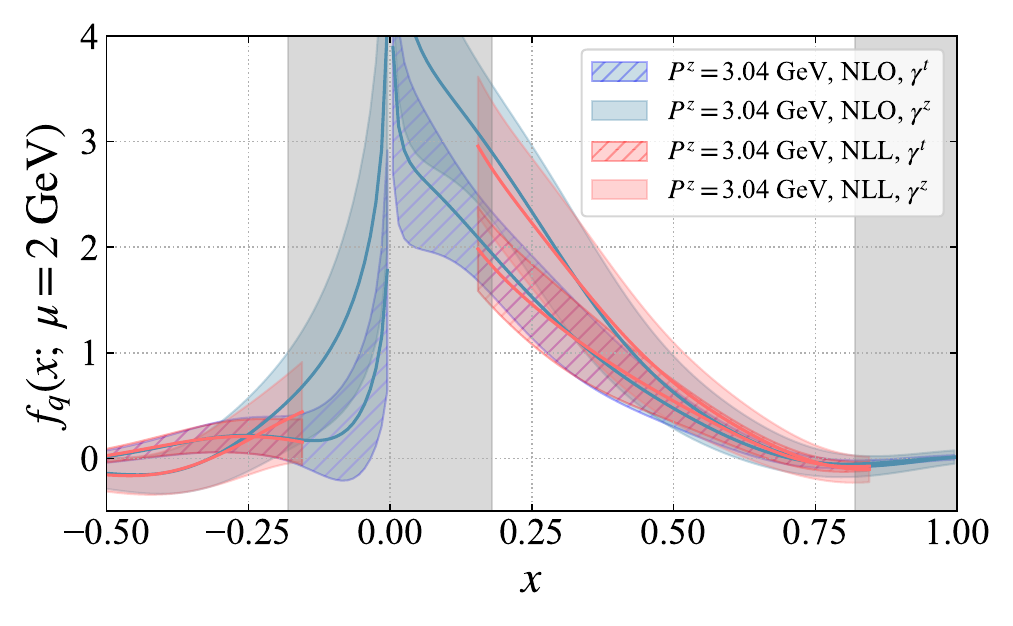}
    \caption{Comparison of unpolarized PDFs obtained with different choices of the Dirac structure $\tilde\Gamma$ at hadron momenta $P^z = 2.43$ and $3.04$ GeV.}
    \label{fig:unpol_gamma_comparison}
\end{figure}

\begin{figure}[th!]
    \centering
    \includegraphics[width=0.49\linewidth]{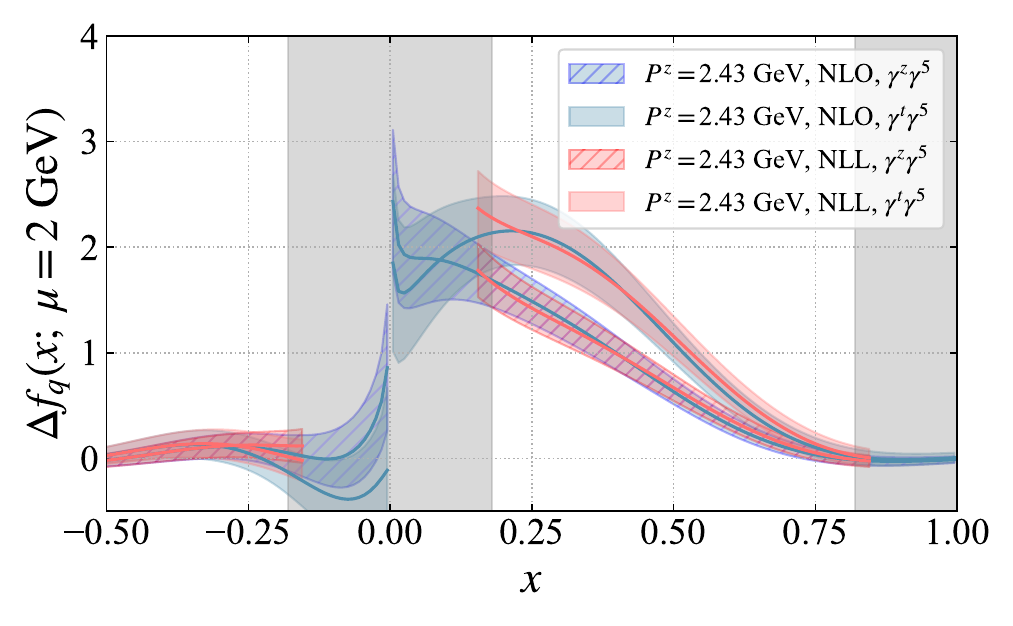}
    \includegraphics[width=0.49\linewidth]{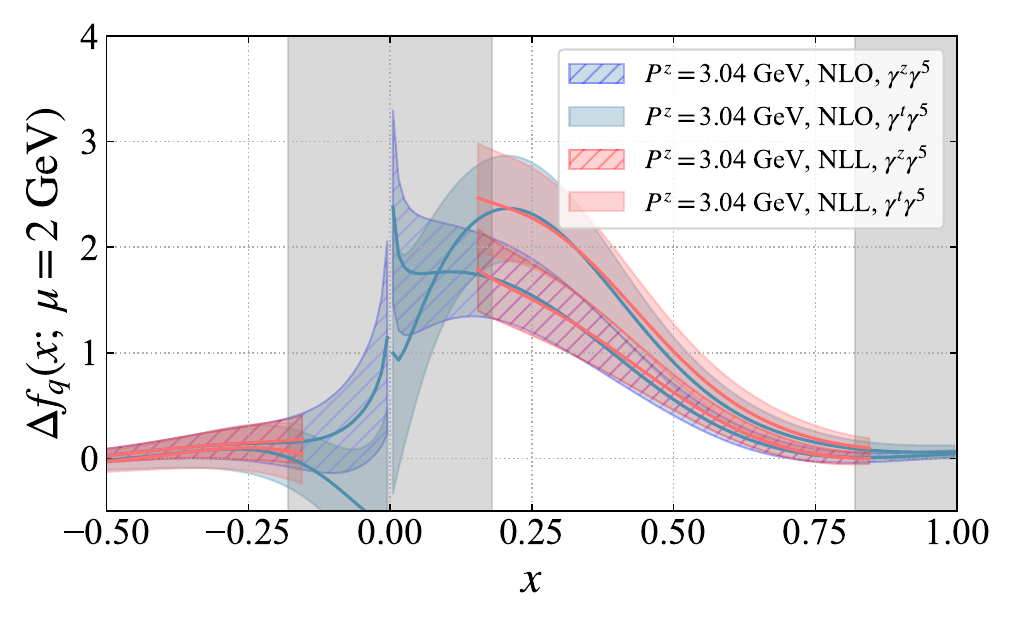}
    \caption{Comparison of helicity PDFs obtained with different choices of the Dirac structure $\tilde\Gamma$ at hadron momenta $P^z = 2.43$ and $3.04$ GeV.}
    \label{fig:heli_gamma_comparison}
\end{figure}

\begin{figure}[th!]
    \centering
    \includegraphics[width=0.49\linewidth]{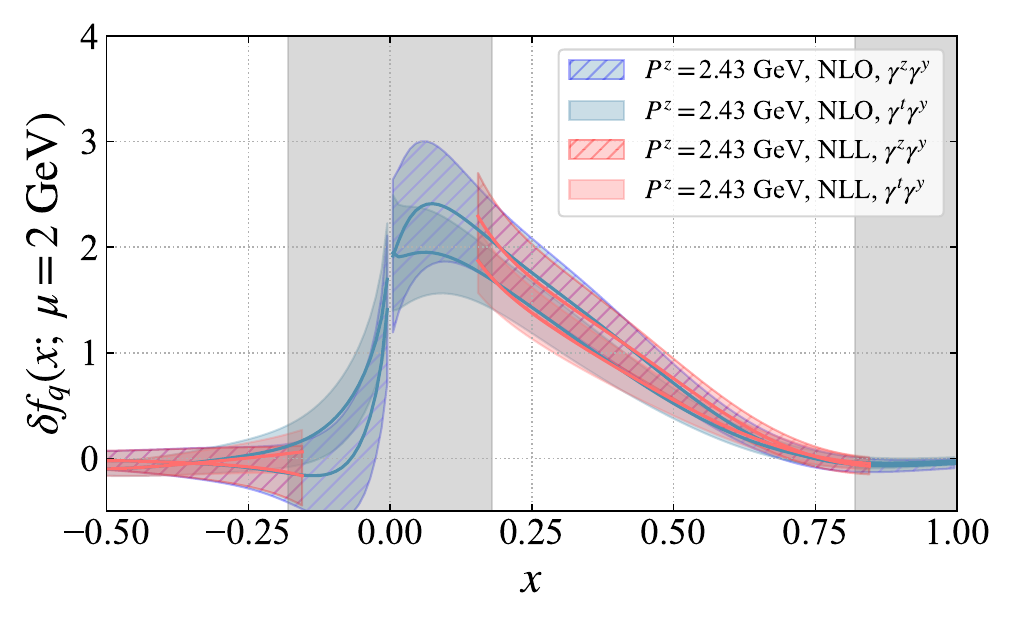}
    \includegraphics[width=0.49\linewidth]{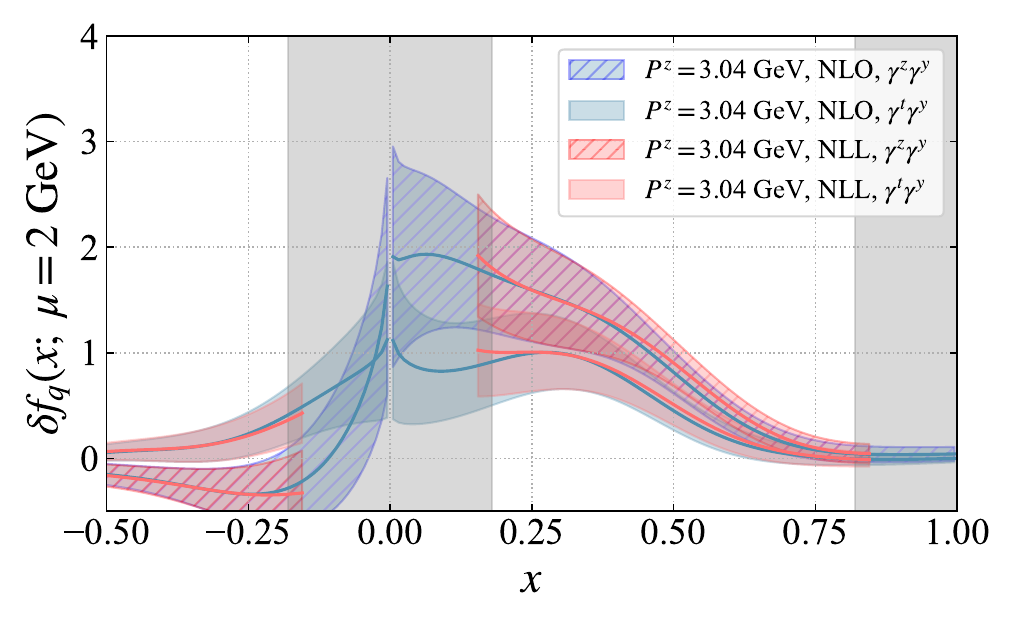}
    \caption{Comparison of transversity PDFs obtained with different choices of the Dirac structure $\tilde\Gamma$ at hadron momenta $P^z = 2.43$ and $3.04$ GeV.}
    \label{fig:trans_gamma_comparison}
\end{figure}

In Figs.~\ref{fig:unpol_gamma_comparison}, \ref{fig:heli_gamma_comparison}, and \ref{fig:trans_gamma_comparison}, we present a comparison of the unpolarized, helicity, and transversity PDFs obtained with different choices of the Dirac structure $\tilde\Gamma$ at hadron momenta $P^z = 2.43$ and $3.04$ GeV. For each polarization channel, multiple $\tilde \Gamma$ choices are available in the operator definition of the quasi-distributions, and while all of them can in principle be matched to the same light-cone PDF, the associated large-momentum expansion introduces different levels of power corrections. Importantly, in most cases, the discrepancies between the different choices of $\tilde\Gamma$ are reduced when the hadron momentum increases from $P^z=2.43$ to $3.04$ GeV, demonstrating that larger momenta suppress power corrections and bring the results of different operator structures closer to agreement. This behavior is consistent with theoretical expectations from the large-momentum effective expansion, providing further evidence that our lattice calculations capture the correct asymptotic trend.

For the gauge-invariant quasi-PDFs, the preferred choices of $\tilde\Gamma$ are $\gamma^t$, $\gamma^z\gamma_5$, and $\gamma^t\gamma_\perp^\alpha\gamma_5$ for the unpolarized, helicity, and transversity cases~\cite{Constantinou:2017sej,Chen:2017mie,Green:2017xeu}, as they are free from operator mixings at $O(a^0)$ induced by the breaking of chiral symmetry. As for the CG definitions, a systematic study of operator mixing under lattice discretization is not yet available. Nevertheless, there are indications that the same set of operators may be preferred. For example, the operator $\bar{\psi}\gamma^t\psi$ admits a natural number-density interpretation in free field theory. We leave a detailed investigation of this issue to future work.

\section{Conclusion}
\label{sec:conclusion}

In this work, we have carried out the first lattice QCD calculation of isovector nucleon PDFs using the CG method, including all three polarization channels. Employing an off-axis momentum setup, we achieved large hadron momenta up to $P^z > 3$ GeV, which allows for a controlled application of the LaMET framework. A systematic treatment of excited-state contamination was performed and we demonstrated that the CG method produces nucleon PDFs in the moderate-$x$ region that are broadly consistent with phenomenological determinations. These findings demonstrate the CG method as a promising tool for extracting precision-controlled nucleon parton distributions from lattice QCD.

More specifically, at the largest momentum in this work, $P^z=3.04$ GeV, we observed an encouraging agreement between our lattice results and the existing phenomenological extractions for all three polarizations of nucleon PDFs: unpolarized, helicity, and transversity. The consistency is more evident in the contributions from the real part of the matrix elements. 
The comparison with lattice calculations using the GI method is also presented in Appendix~\ref{sec:gi_compare}, where we find broad consistency with small deviations mainly caused by the power corrections.
In contrast, the contributions from the imaginary part of the matrix elements display larger discrepancies across different hadron momenta. This suggests that the imaginary sector is more sensitive to residual excited-state contamination and may require even more careful analyses in future studies.

The renormalization structure of the Coulomb-gauge quasi-PDF operators is significantly simpler than that of the GI operator due to the lack of the Wilson line (and hence, lack of linear divergence and associated renormalon). 
At short separations $z \sim a$, the operator $\bar{\psi}_0(z)\tilde\Gamma \psi_0(0)$ picks up large discretization effects. The hybrid scheme in \Cref{ss:hybrid} should remove the leading artifacts for the real part of the matrix elements, which, however, does not necessarily work for the imaginary part. Therefore, we propose to renormalize the latter using the S\ripmom \ scheme presented in \Cref{ss:srip}, which is a simple method for renormalizing the Coulomb-gauge quark fields. The flatness of the window shown in \Cref{fig:ZMS} is an encouraging sign that the relevant systematics are under control. 
To further understand the lattice artifacts, a full study of these schemes is planned by studying the quark mass and lattice spacing dependence of the renormalization constants. 

Together, these observations highlight both the promise and the challenges of extracting partonic information from the lattice at large momentum. The path forward is clear. With access to greater computational resources, it will become possible to carry out simulations at multiple lattice spacings, thereby enabling controlled continuum extrapolations that systematically reduce discretization effects. In parallel, accumulating higher statistics for correlators at large source–sink separations will improve the stability of the extracted observables and enhance the overall robustness of the final results. Further progress can be achieved by adopting more refined and effective strategies for treating excited-state contamination, such as the Lanczos algorithm~\cite{Wagman:2024rid,Hackett:2024xnx,Hackett:2024nbe} and the current-inspired GEVP~\cite{Barca:2024hrl,Barca:2025det}, which is expected to be particularly important for reliably resolving the imaginary components of matrix elements. Finally, the application of new lattice technologies aimed at improving the SNR, such as kinematically enhanced interpolating operators~\cite{Zhang:2025hyo} and more optimized smearing schemes, will be essential to extend these studies to higher momenta and increased precision. Combining these improvements will place the CG method on a firmer foundation and allow precise lattice determinations of nucleon PDFs that can stand alongside phenomenological and experimental results. In this way, our present work provides not only a first demonstration of the CG method for nucleon PDFs, but also a roadmap for systematically improving the approach toward a future of quantitatively precise lattice QCD parton distributions.

\begin{acknowledgments}

We thank Yushan Su for valuable communications.

This material is based upon work supported by the U.S. Department of Energy, Office of Science, Office of Nuclear Physics through Contract No.~DE-SC0012704, No.~DE-AC02-06CH11357, within the framework of Scientific Discovery through Advanced Computing (SciDAC) award Fundamental Nuclear Physics at the Exascale and Beyond, and under the umbrella of the Quark-Gluon Tomography (QGT) Topical Collaboration with Award DE-SC0023646. YZ and JL acknowledge support by the U.S. Department of Energy, Office of Science, Office of Nuclear Physics, Early Career Award through Contract No.~DE-SCL0000017. 

This research used awards of computer time provided by the INCITE program at Argonne Leadership Computing Facility, a DOE Office of Science User Facility operated under Contract DE-AC02-06CH11357, the ALCC program at the Oak Ridge Leadership Computing Facility, which is a DOE Office of Science User Facility supported under Contract DE-AC05-00OR22725, and the National Energy Research Scientific Computing Center, a DOE Office of Science User Facility supported by the Office of Science of the U.S. Department of Energy under Contract DE-AC02-05CH11231 using NERSC awards NP-ERCAP0028137 and NP-ERCAP0032114. Computations for this work were carried out in part on facilities of the USQCD Collaboration, which is funded by the Office of Science of the U.S. Department of Energy. We gratefully acknowledge the computing resources provided on Swing, a high-performance computing cluster operated by the Laboratory Computing Resource Center at Argonne National Laboratory. This research also used resources of the Argonne Leadership Computing Facility, a U.S. Department of Energy (DOE) Office of Science user facility at Argonne National Laboratory and is based on research supported by the U.S. DOE Office of Science-Advanced Scientific Computing Research Program, under Contract No. DE-AC02-06CH11357.

The measurement of the correlators was carried out with the \texttt{Qlua} software suite~\cite{qlua}, which utilized the multigrid solver in \texttt{QUDA}~\cite{Clark:2009wm,Babich:2011np}. 
Data analysis was performed using the Python package LaMETLat, which is specifically implemented to support the analysis of Lattice QCD data within the LaMET framework. The package is available at \href{https://github.com/Greyyy-HJC/LaMETLat}{https://github.com/Greyyy-HJC/LaMETLat} and is expected to undergo further development for prospective research applications.

\end{acknowledgments}

\appendix

\section{S\ripmom \ scheme}\label{sec:RIMOM}
\begin{figure}[t]
    \centering
    \includegraphics[width=0.49\linewidth]{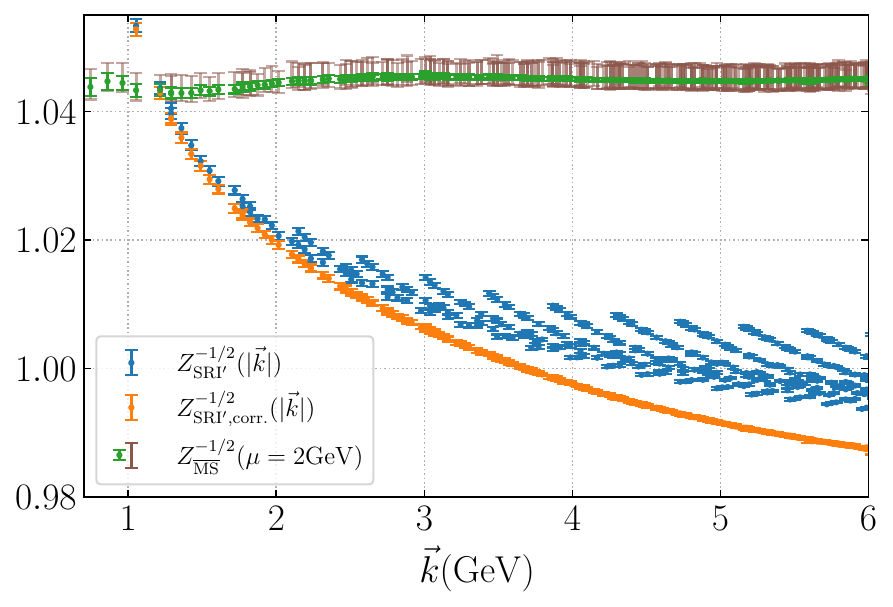}
    \caption{$Z_{\mathrm{SRI}'}^{-1/2}(|\vec{k}|)$ (shown in blue) is calculated from \Cref{eq:Zused} on 50 configurations, with 16 point-sources per configuration evenly spaced in Euclidean time and all located at the same spatial position. $Z_{\mathrm{SRI}',\mathrm{corr.}}^{-1/2}(\vec{|k|})$ (shown in orange) is corrected by using \Cref{eq:d1d2d3}, which depends on three fitted parameters $d_1,d_2,d_3$. Running and matching is then performed according to \Cref{eq:run,eq:match} to give $Z_{\overline{\mathrm{MS}}}^{-1/2}(\mu=2~\mathrm{GeV})$, which shows a nice window where it is constant.
    {Inner error bars (green) for $Z_\MS^{-1/2}(\mu=2~\mathrm{GeV})$ show the purely statistical error, whilst outer error bars (brown) include an estimate of the LL-running perturbative error. }
    $Z_{\overline{\mathrm{MS}}}^{-1/2}(\mu=2~\mathrm{GeV})$ is the factor which, when multiplied by the bare lattice quark field, renormalizes it into the $\overline{\mathrm{MS}}$ scheme at $\mu = 2~\mathrm{GeV}$. }
    \label{fig:ZMS}
\end{figure}

In principle it is possible to utilize the continuum form of $S_\mathrm{tree}(\vec{p})$ in the renormalization condition \Cref{eq:Zdef}, however that choice would lead to large lattice artifacts that begin at tree-level. 
Instead, we use the free static lattice propagator for the Wilson action to eliminate tree-level artifacts. 
For 3-momenta living in the cubic Brillouin zone $\vec{k}_i \in [-\frac{\pi}{a},\frac{\pi}{a})$, we define lattice momenta:  
\begin{equation}
\big(\widehat{\vec{k}}_1\big)_i = \mathrm{sin}(a \vec{k}_i), \quad \big(\widehat{\vec{k}}_2\big)_i = 2\mathrm{sin}(a \vec{k}_i/2), \quad \widehat{\vec{k}}_1^2 = \sum_{i=1}^3 
\big(\widehat{\vec{k}}_1\big)_i^2, \quad 
\widehat{\vec{k}}_2^2 = 
\big(\widehat{\vec{k}}_2\big)_i^2
\end{equation}
where both $\widehat{\vec{k}}_1^2$ and $\widehat{\vec{k}}_2^2$ approach the continuum $O(3)$-invariant $a^2\vec{k}^2$ in the continuum limit. 
The massless free static lattice propagator is given by (note that the $c_{\mathrm{SW}}$ improvement term does not affect this propagator~\cite{Capitani:2002mp,Sheikholeslami:1985ij}):
\begin{equation}
S_{\mathrm{tree}}^{\mathrm{lat.}}(\vec{k}) = \frac{-i \sum_{i=1}^3(\widehat{\vec{k}}_1)_i\gamma_i  + f(\vec{k}) \mathbbm{1}}{\left[\widehat{\vec{k}}_1^2 + \left(\frac{1}{2}\widehat{\vec{k}}_2^2\right)^2 \right]^{\frac12} \left[ \widehat{\vec{k}}_1^2 + \left(2 + \frac{1}{2}\widehat{\vec{k}}_2^2\right)^2\right]^{\frac12}},
\end{equation}
which in the continuum limit approaches $
S_\mathrm{cont}(\vec{k}) = \frac{-i \vec{k}_i \gamma_i}{2\sqrt{\vec{k}^2}}$ up to tree-level discretization artifacts. The component proportional to $ f(\vec{k}) \mathbbm{1}$ is unused in our renormalization conditions, as it is more sensitive to finite quark mass corrections (though the analytic form for $f(\vec{k})$ can be derived).
The renormalization condition used is explicitly given by:
\begin{equation}\label{eq:Zused}
Z_{\mathrm{SRI'}}(|\vec{k}|) =  
\frac{ 
\mathrm{Tr}\left[\sum_{i=1}^3 \gamma_i \big(\hat{\vec{k}}_1\big)_i \ S_0(\vec{k}) \right] }
{ \mathrm{Tr}\left[\sum_{i=1}^3 \gamma_i \big(\hat{\vec{k}}_1\big)_i \ S_\mathrm{tree}^{\mathrm{lat.}}(\vec{k})\right]}
\end{equation}
where the chiral limit has not been taken. 
Even after cancelling the tree-level artifacts, the next leading lattice artifacts arise from rotation-breaking effects proportional to the following cubic invariant:
\begin{equation}
{\vec{k}}^{[4]} = \sum_{i=1}^3 \vec{k}_i^4,
\end{equation}
which cause the `fishbone' structure shown in plot of $Z_{\mathrm{SRI'}}^{-1/2}(\vec{k})$ in \Cref{fig:ZMS}. 
One common strategy to suppress these artifacts has been to employ a `cylinder cut'~\cite{Bonnet:2002ih} and only use momenta close to the diagonal, for example restricting to $\vec{k} \sim (\frac{2n\pi}{a L}\pm1,\frac{2n\pi}{aL}\pm1,\frac{2n\pi}{aL}\pm1) $ for integer $n \in [-\frac{L}{2},\frac{L}{2})$. 
Due to the small statistical errorbars on our data, we instead keep all the data, and perform an analysis to subtract these leading artifacts using a variant of the analysis performed in Ref.~\cite{Blossier:2010vt}.

Firstly, for a given value of $\vec{k}^2$, if there are multiple corresponding values of $\vec{k}^{[4]}$, a linear fit is performed to the following form:
\begin{equation}
Z_{\mathrm{SRI'}}^{-1/2}(\vec{k}) = c_1(\vec{k}^2) + c_2(\vec{k}^2) a^4 \vec{k}^{[4]} .
\end{equation}
The correction coefficient $c_2$ is found to be well-described by the following form:
\begin{equation}\label{eq:d1d2d3_sub}
c_2(\vec{k}^2) = d_1\frac{1}{a^2\vec{k}^2} + d_2  + d_3 a^2\vec{k}^2 ,
\end{equation}
for constants $d_1,d_2,d_3$ such that the corrected renormalization factor is defined by:
\begin{equation}\label{eq:d1d2d3}
Z_{\mathrm{SRI',corr.}}^{-1/2}(\vec{k}) = Z_{\mathrm{SRI'}}^{-1/2}(\vec{k})- \left(d_1\frac{1}{a^2 \vec{k}^2} + d_2 + d_3 a^2 \vec{k}^2\right)  a^4 \vec{k}^{[4]}
\end{equation}
Note that the $d_1$ term in \Cref{eq:d1d2d3} is the leading $O(a^2)$ artifact proportional to $\vec{k}^{[4]}$, and is the cubic analogue of the $k^{[4]}/k^2$ term found at $O(g^2)$ in 4D lattice-perturbation theory~\cite{Constantinou:2009tr}.
The corrected renormalization factors are also plotted in \Cref{fig:ZMS} in orange, and evidently the subtraction procedure works remarkably well over a large range of momenta. 

Matching to $\overline{\mathrm{MS}}$ then requires dimensionally regulated perturbative calculations. 
The $O(\alpha_S)$ matching between S\ripmom~and $\overline{\mathrm{MS}}$ can be given by converting the $O(\alpha_S)$ four-dimensional Coulomb gauge quark propagator presented in Ref.~\cite{Popovici:2008ty} to the static limit:
\begin{align}\label{eq:match}
\sqrt{\frac{Z^{\mathrm{SRI}'}}{Z^{\mathrm{\overline{\mathrm{MS}}}}}}
&= 1 - \frac{\alpha_S C_F}{2\cdot(4 \pi)} \left[ \left(13 - \frac{4 \pi^2}{3} - 2\log 2 \right) + \log \frac{\mu^2}{\vec{k}^2} \right]
\end{align}
The anomalous dimension of the Coulomb gauge-fixed quark field is however only known at one-loop (leading logarithmic accuracy), and is provided by:
\begin{align}\label{eq:run}
&Z^{-1/2}_{\mathrm{SRI'}}(\vec{k}) = \left( \frac{\alpha_S(\vec{k})}{\alpha_S(\vec{k}_0)}\right)^{C_F/[{2\cdot(11-2N_f/3})]}Z^{-1/2}_{\mathrm{SRI'}}(\vec{k}_0). 
\end{align}
where $N_f = 3$ due to there being 3 light flavours in the sea. 
As such, the running and matching operations are not commutative, {in the sense that the following two pathways to obtaining $Z_\MS(\mu=2~\mathrm{GeV})$ are different:

\begin{equation}
\begin{tikzcd}[ampersand replacement=\&]
Z_\mathrm{SRI'}(|\vec{k}|) \arrow[r,"\mathrm{LL}"] \arrow[d,"O(\alpha_S)"] \& Z_\mathrm{SRI'}(2~\mathrm{GeV}) \arrow[d,"O(\alpha_S)"] \\
Z_\MS(\mu=|\vec{k}|) \arrow[r,"\mathrm{LL}"]                 \& Z_\MS(2~\mathrm{GeV})    
\end{tikzcd}
\end{equation}
To estimate the uncertainty due to this lack of commutativity as well as the perturbation theory uncertainty itself, we choose to run the operators in SRI'-MOM to a scale $|\vec{k}| = c \cdot 2~\mathrm{GeV}$, where $c$ is varied in the range $c \in [\frac{1}{\sqrt{2}},\sqrt{2}]$, before performing the $O(\alpha_S)$ matching between $Z_\mathrm{SRI'}(c \cdot 2~\mathrm{GeV})$ and $Z_\MS(2~\mathrm{GeV})$. 
Including this variation in the matching scale inflates the error-bars to the outer error bars shown in \Cref{fig:ZMS}, which is still much smaller than the other uncertainties. 
Note that this perturbation theory truncation error scales with $\sim\alpha_S(\mu = 2~\mathrm{GeV})$, and is largely $|\vec{k}|$-independent. 
}

Due to the inconsistent orders used between the running and matching procedures, there is no clear preference to use a consistent order for running of $\alpha_S$, and we subsequently use the PDG value for $\alpha_S(m_Z)$~\cite{ParticleDataGroup:2024cfk} with five-loop running~\cite{Herzog:2017ohr} and four-loop decoupling relations~\cite{Schroder:2005hy,Chetyrkin:2005ia}
to decouple the heavy flavours (bottom and charm). 

After running and matching, the complete factor required to multiply the bare lattice operator to renormalize it into $\overline{\mathrm{MS}}$ is shown in green in \Cref{fig:ZMS}. 
It is plotted as a function of $|\vec{k}|$, where it is understood that the operator is first renormalized in S\ripmom$(\vec{k})$, before running to $c \cdot 2~\mathrm{GeV}$ and finally matching to $\overline{\mathrm{MS}}$. 
One expects that at small $\vec{k}$ one has large perturbative errors (which can be improved by higher order perturbation theory) and non-perturbative effects, and at large $\vec{k}$ there are errors due to lattice artifacts. 
{Furthermore as noted in \Cref{ss:srip}, there is a systematic error due to the lack of chiral extrapolation which is expected to scale with $m_q/|\vec{k}|$, where $m_q$ is the renormalized quark mass.}
This leads to a window problem where results are only trustworthy within some intermediate window $ \Lambda_\mathrm{QCD} \ll |\vec{k}| \ll\frac{\pi}{a} $. The flatness of $Z_{\overline{\mathrm{MS}}}(\mu = 2~\mathrm{GeV})$ shown in \Cref{fig:ZMS} is encouraging and suggests that all of these effects are under control. 
The final result reported in \Cref{eq:finalresult} uses data in the range $2~\mathrm{GeV} \leq |\vec{k}| \leq 6~\mathrm{GeV}$, where the central value is given by a correlated average over those datapoints, and errorbars chosen conservatively so that every individual point lies within the quoted errors. 
{Beyond $|\vec{k}| = 6~\mathrm{GeV}$ there are noticeable lattice artifacts which should be expected as $a\cdot(6~\mathrm{GeV}) \simeq 1.9$, and below $|\vec{k}| = 2~\mathrm{GeV}$ one should expect large perturbative errors and non-perturbative effects - thus our window choice is quite conservative. }

\section{Comparison with lattice calculations using the GI method}
\label{sec:gi_compare}

\begin{figure}[th!]
    \centering
    \includegraphics[width=0.32\linewidth]{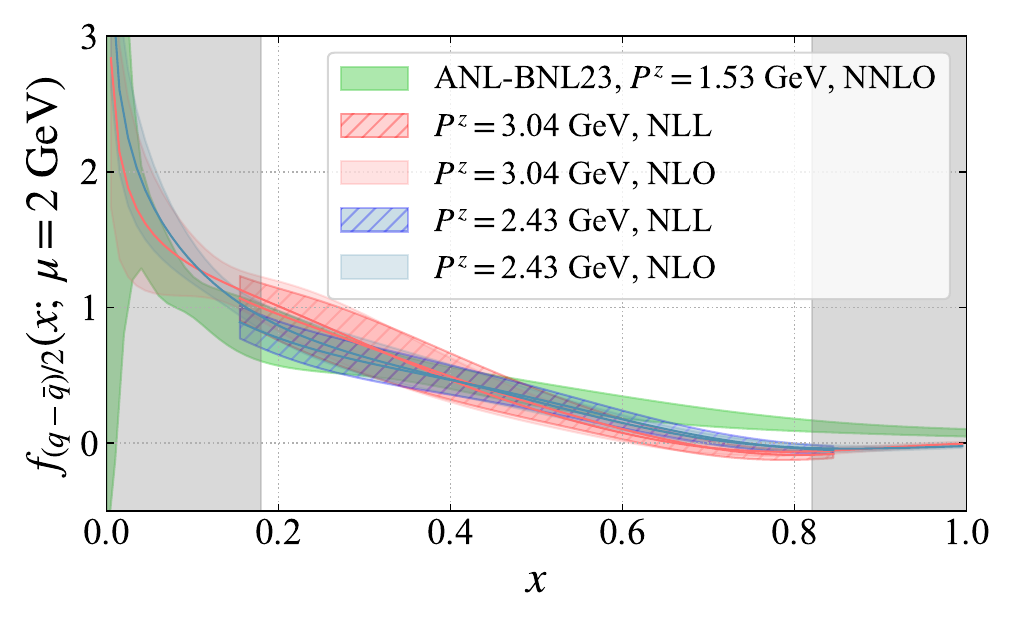}
    \includegraphics[width=0.32\linewidth]{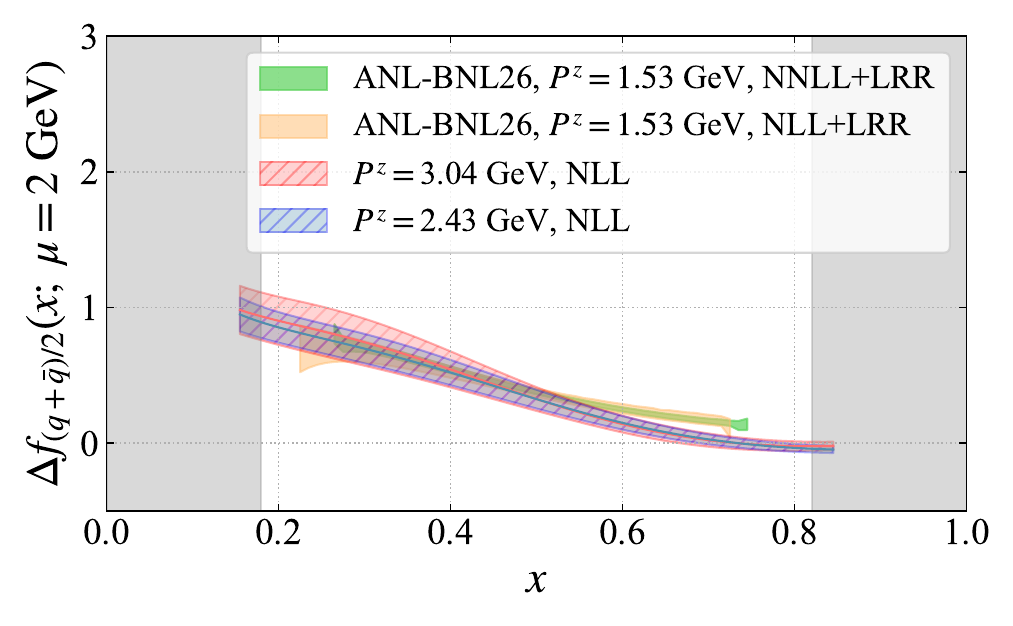}
    \includegraphics[width=0.32\linewidth]{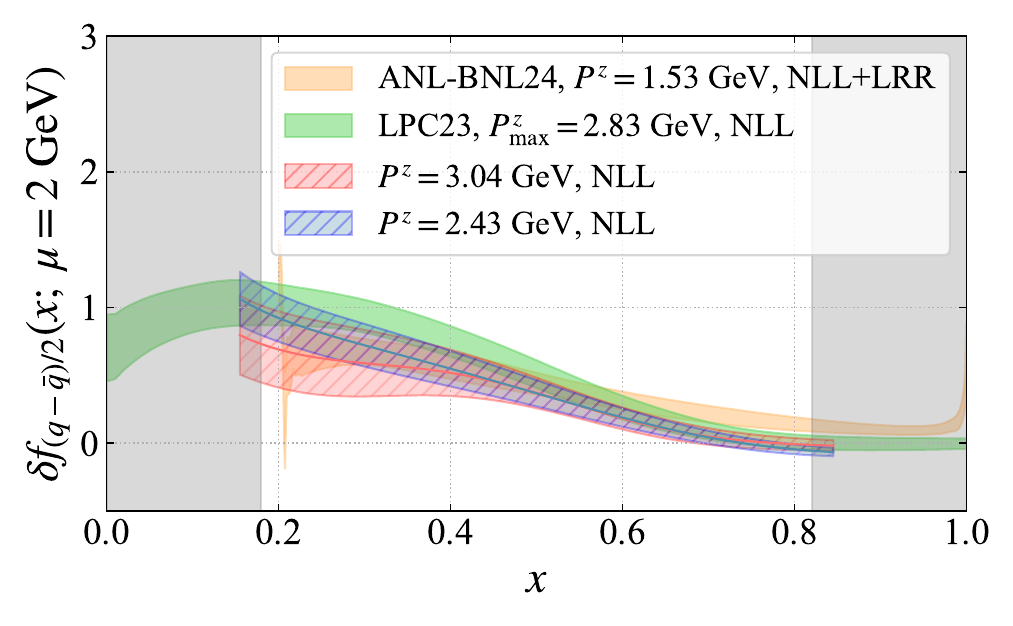}
    \caption{Comparison of the nucleon unpolarized (left), helicity (middle), and transversity (right) PDFs from the real part of the quasi-PDF matrix elements using the CG and GI methods. The GI results are from ANL-BNL23~\cite{Gao:2022uhg} at NNLO, {ANL-BNL26~\cite{Gao:2026wlz}} at NLL and NNLL with leading-renormalon resummation (LRR), ANL-BNL24~\cite{Gao:2023ktu} at NLL with LRR, and LPC23~\cite{LatticeParton:2022xsd} at NLL.
    }
    \label{fig:compare_gi}
\end{figure}

For completeness, we compare our results for the nucleon isovector unpolarized (left), helicity (middle), and transversity (right) PDFs from the real part of the quasi-PDF matrix elements to those computed using the GI method, as shown in Fig.~\ref{fig:compare_gi}.
The GI results include the unpolarized PDF from ANL-BNL23~\cite{Gao:2022uhg} at NNLO, helicity PDF from {ANL-BNL26~\cite{Gao:2026wlz}} at NLL and NNLL with leading-renormalon resummation (LRR), and the transversity PDF from ANL-BNL24~\cite{Gao:2023ktu} at NLL with LRR and LPC23~\cite{LatticeParton:2022xsd} at NLL. All the ANL-BNL23, ANL-BNL24, and ANL-BNL26 results are computed from the same HotQCD~\cite{HotQCD:2014kol} ensemble with lattice spacing $a=0.076$ fm, volume $L_s^3 \times L_t=64^3\times 64$, and physical valence pion mass $m_\pi$, at nucleon momentum $P^z = 1.53~\mathrm{GeV}$. The LPC23 transversity PDF was obtained from six different lattice ensembles generated by the CLS collaboration~\cite{Bruno:2014jqa} with lattice spacing $a = \{0.098, 0.085, 0.064, 0.049\}$ fm in the physical limits, with maximal momentum $P^z=2.83$ GeV at valence pion masses ranging between 354 and 222 MeV. Both the helicity and transversity PDFs are normalized with their corresponding charges---$g_A$ and $g_T$, respectively---calculated on the same lattice.

It should be emphasized that these lattice calculations are performed on different ensembles with varying spacings, pion masses, and nucleon momenta, so their systematic uncertainties are not the same. A clean comparison between the CG and GI methods should be conducted on the same ensemble with identical setups, such as the one in Ref.~\cite{Gao:2023lny}. We have also left out the comparison of contributions from the imaginary part of the quasi-PDF matrix elements, as they appear to be more sensitive to systematics such as excited-state contamination that challenge all methods.
Nevertheless, the comparison of the real-part contributions show that CG and GI results are broadly consistent at moderate $x$ but differ near the endpoint regions. Note that all the ANL-BNL results were obtained at a noticeably smaller momentum $P^z=1.53$ GeV. While other lattice systematics are still present, this discrepancy can be attributed to power corrections, which differ by the quasi-observables and are expected to be enhanced near the endpoints within the LaMET framework.

{
\section{Error Budget}
\label{app:error}

\begin{figure}[th!]
    \centering
    \includegraphics[width=0.49\linewidth]{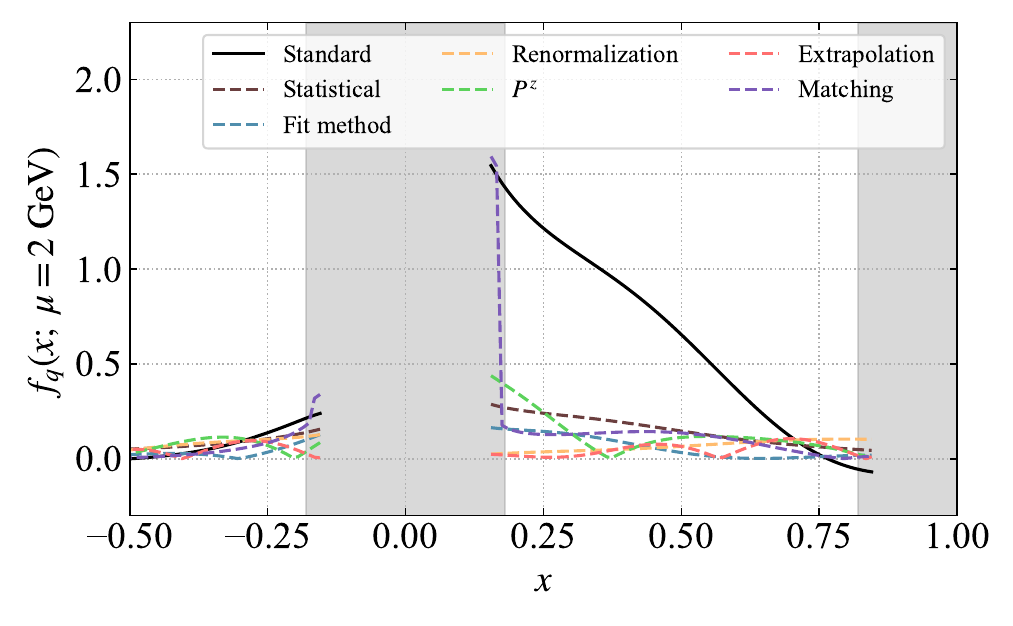}
    \caption{
    {
    Error budget for the nucleon unpolarized quark PDFs, shown here as an illustrative example. The black curve (``Standard") denotes the mean value of the unpolarized quark PDFs at $P^z = 2.43~\mathrm{GeV}$ at NLL accuracy, corresponding to the result shown in Fig.~\ref{fig:unpol_complete}, and is taken as the reference. The brown dashed curve (``Statistical") corresponds to the statistical uncertainty. The other dashed curves represent estimates of different sources of systematic uncertainty, defined as the differences between the final unpolarized quark PDFs obtained using different analysis strategies. Specifically, ``Fit method" corresponds to the difference between the joint fits with $\tau_{\mathrm{cut}}=2$ and $\tau_{\mathrm{cut}}=3$ (see Table~\ref{tb:gsfit}); ``Renormalization" denotes the difference between the hybrid and $\overline{\mathrm{MS}}$ renormalization schemes; ``Extrapolation" represents the difference between asymptotic extrapolations using fit ranges $z/a \in [6,14] \sqrt{2}$ and $z/a \in [10,20] \sqrt{2}$; and ``$P^z$" corresponds to the difference between results obtained at $P^z=2.43~\mathrm{GeV}$ and $P^z=3.04~\mathrm{GeV}$.
    }
    }
    \label{fig:error_budget}
\end{figure}

To summarize the systematic uncertainties in our analysis, we present the error budget for the nucleon unpolarized quark PDFs in Fig.~\ref{fig:error_budget}. Here the unpolarized quark PDFs are taken as a representative example, while the same analysis strategy is applied to the other observables. The black curve (``Standard") corresponds to our central result at $P^z=2.43~\mathrm{GeV}$ with NLL matching, the brown dashed curve (``Statistical") corresponds to the statistical uncertainty, and the remaining dashed curves show the differences between the final PDFs obtained using alternative analysis strategies. These differences are interpreted as estimates of the associated systematic uncertainties.

The systematic uncertainties considered in this work include the fitting strategy, renormalization scheme dependence, asymptotic extrapolation procedure, momentum dependence, and scale variation in the NLL matching. In particular, the ``Fit method" uncertainty is estimated from the difference between the joint fits with $\tau_{\mathrm{cut}}=2$ and $\tau_{\mathrm{cut}}=3$ (see Table~\ref{tb:gsfit}); the ``Renormalization" uncertainty is obtained from the difference between the hybrid and $\overline{\mathrm{MS}}$ schemes; the ``Extrapolation" uncertainty is estimated by varying the asymptotic fit range from $z/a \in [6,14] \sqrt{2}$ to $z/a \in [10,20] \sqrt{2}$; the ``$P^z$" uncertainty is determined from the difference between the results at $P^z=2.43~\mathrm{GeV}$ and $P^z=3.04~\mathrm{GeV}$; and the ``Matching" uncertainty is estimated by varying the initial scale of the resummation in NLL matching by a factor of $\sqrt{2}$.

We do not include an additional systematic uncertainty associated with different Dirac structures $\tilde{\Gamma}$, since the preferred choices adopted in this work are motivated in Sec.~\ref{ss:power_corr}. In addition, discretization effects are not included in the present error budget because only a single lattice spacing is available in this study. A systematic investigation of lattice-spacing dependence will be deferred to future work.
}

\bibliographystyle{JHEP}
\bibliography{revised}

\end{document}